\documentclass[12pt]{article}
\pdfoutput=1

\usepackage{epsfig}
\usepackage{psfrag}
\usepackage{latexsym}
\usepackage{indentfirst}
\usepackage{fancyhdr}
\usepackage{dsfont}
\usepackage{amssymb}
\usepackage{amsmath}
\usepackage{amsfonts}
\usepackage{pifont}
\usepackage{cite}
\usepackage{bbold}
\usepackage{color}
\usepackage{colordvi}
\usepackage{fancybox}
\usepackage[footnotesize]{caption2}
\usepackage{graphicx}
\usepackage[center,footnotesize,hang]{subfigure}
\usepackage{bbm}
\usepackage{url}
\usepackage{multirow}
\usepackage{array}
\usepackage{arydshln}
\usepackage{ulem}
\usepackage{enumitem}
\newcommand{\PreserveBackslash}[1]{\let\temp=\\#1\let\\=\temp}
\newcolumntype{C}[1]{>{\PreserveBackslash\centering}p{#1}}
\newcolumntype{R}[1]{>{\PreserveBackslash\raggedleft}p{#1}}
\newcolumntype{L}[1]{>{\PreserveBackslash\raggedright}p{#1}}
\allowdisplaybreaks  

\newcommand{\cleqn}{\setcounter{equation}{0}}

\newcommand{\bq}{\begin{eqnarray}}
\newcommand{\nq}{\end{eqnarray}}

\begin{document}

\title{\hfill ~\\[0mm]
        \textbf{Deviation from Bimaximal Mixing and Leptonic CP Phases in $S_4$ Family Symmetry and Generalized CP}}
\date{}

\author{\\[1mm]Cai-Chang Li\footnote{E-mail: {\tt lcc0915@mail.ustc.edu.cn}}~,~Gui-Jun Ding\footnote{E-mail: {\tt dinggj@ustc.edu.cn}}\\ \\
\it{\small Department of Modern Physics, University of Science and
    Technology of China,}\\
  \it{\small Hefei, Anhui 230026, China}\\[4mm] }
\maketitle

\begin{abstract}

The lepton flavor mixing matrix having one row or one column in common with the bimaximal mixing up to permutations is still compatible with the present neutrino oscillation data. We provide a thorough exploration of generating such a mixing matrix from $S_4$ family symmetry and generalized CP symmetry $H_{CP}$. Supposing that $S_4\rtimes H_{CP}$ is broken down to $Z^{ST^2SU}_2\times H^{\nu}_{CP}$ in the neutrino sector and $Z^{TST^{2}U}_4\rtimes H^{l}_{CP}$ in the charged lepton sector, one column of the PMNS matrix would be of the form $\left(1/2, 1/\sqrt{2}, 1/2\right)^{T}$ up to permutations, both Dirac CP phase and Majorana CP phases are trivial in order to accommodate the observed lepton mixing angles. The phenomenological implications of the remnant symmetry $K^{(TST^2, T^2U)}_4\rtimes H^{\nu}_{CP}$ in the neutrino sector and $Z^{SU}_{2}\times H^{l}_{CP}$ in the charged lepton sector are studied.  One row of PMNS matrix is determined to be $\left(1/2, 1/2, -i/\sqrt{2}\right)$, and all the three leptonic CP phases can only be trivial to fit the measured values of the mixing angles. Two models based on $S_4$ family symmetry and generalized CP are constructed to implement these model independent predictions enforced by remnant symmetry. The correct mass hierarchy among the charged leptons is achieved. The vacuum alignment and higher order corrections are discussed.

\end{abstract}
\thispagestyle{empty}
\vfill

\newpage
\setcounter{page}{1}

\section{\label{sec:introduction} Introduction}
\cleqn

The neutrino flavor mixing and neutrino oscillation have been firmly established so far. The standard three flavor neutrino oscillation relates the favor eigenstates of neutrinos to the mass eigenstates through the Pontecorvo-Maki-Nakagawa-Sakata (PMNS) mixing matrix. This matrix is a $3\times3$ unitary matrix and can be parameterized by three mixing angles $\theta_{12},\theta_{13},\theta_{23}$, one Dirac type CP violating phase $\delta_{CP}$ similar to the quark sector and two additional Majorana phases $\alpha_{21},\alpha_{31}$ if neutrinos are Majorana particles. Recently the last lepton mixing angle $\theta_{13}$ has been precisely measured to be about $9^{\circ}$~\cite{Abe:2011sj,Adamson:2011qu,Abe:2011fz,An:2012eh,Ahn:2012nd}. This discovery pushes  neutrino oscillation experiments into a new era  of precise determination of the lepton mixing angles and neutrino mass squared differences, and it also opens up new windows to probe leptonic CP violation. Although we still don't have convincing evidence for lepton CP violation, the current global fit to the neutrino oscillation data indicates nontrivial values of the Dirac type CP phase~\cite{GonzalezGarcia:2012sz,Capozzi:2013csa,Forero:2014bxa}.
The present T2K data already exclude values of $\delta_{CP}$ between $0.14\pi\sim0.87\pi$ at the 90\% confidence level~\cite{Abe:2013hdq,Abe:2014ugx}. Furthermore, several long-baseline neutrino oscillation experiments such as LBNE~\cite{Adams:2013qkq}, LBNO~\cite{Autiero:2007zj,Rubbia:2010fm,Angus:2010sz,Rubbia:2010zz,::2013kaa} and Hyper-Kamiokande~\cite{Abe:2011ts} are proposed to measure CP violation. Study of neutrino mixing including the CP violating phase would allow us to distinguish different flavor models.

In view of the fantastic experimental program of observing lepton CP violation and the fundamental role played by CP violation, it is crucial to be able to predict CP phases. The idea of combining flavor symmetry with generalized CP symmetry is a very interesting approach to predict both flavor mixing angles and CP phases from symmetry principle. The concept of generalized CP transformations has been put forward about thirty years ago. CP invariance at high energy scale
and its subsequent breaking lead to nontrivial constraints on the fermion mass matrices~\cite{Ecker:1981wv,Grimus:1995zi}. It is somewhat tricky to include the generalized CP symmetry in the presence of a family symmetry. Generally the generalized CP transformation must be subject to the so-called consistency condition, which implies that the generalized CP transformation corresponds to an automorphism of the family symmetry group~\cite{Feruglio:2012cw,Holthausen:2012dk}. Furthermore, it is shown that that physical CP transformations always have to be class-inverting automorphisms of family symmetry group~\cite{Chen:2014tpa}. As a result, the conventional CP transformation $\varphi\mapsto\varphi^{*}$ can not be consistently defined, but rather a non-trivial transformation in flavor space is needed.

Generalized CP symmetry together with family symmetry can give us interesting phenomenological predictions. The simplest example is the so-called $\mu-\tau$ reflection symmetry which is a combination of the canonical CP transformation and the $\mu-\tau$ exchange symmetry. The invariance of the light neutrino mass matrix under $\mu-\tau$ reflection in the charged lepton diagonal basis leads to maximal atmospheric mixing angle $\theta_{23}$ and maximal Dirac CP phase $\delta_{CP}$ with $\delta_{CP}=\pm\pi/2$~\cite{Harrison:2002kp,Grimus:2003yn,Farzan:2006vj}.
The phenomenological implications of the generalized CP symmetry has been analyzed within the context of popular $A_4$~\cite{Feruglio:2012cw,Ding:2013bpa}, $S_4$~\cite{Feruglio:2012cw,Ding:2013hpa,Feruglio:2013hia,Luhn:2013lkn,Li:2013jya} and $T^{\prime}$~\cite{Girardi:2013sza} family symmetries. By breaking the full symmetry down to $Z_2\times CP$ in the neutrino sector, the $\text{TM}_1$ and $\text{TM}_2$ mixing patterns in which the first and the second columns of the tri-bimaximal mixing is kept respectively, can be exactly produced. The Dirac CP phase $\delta_{CP}$ is predicted to be conserved or maximally broken. Concrete models in which these symmetry breaking patterns are achieved dynamically have been proposed. Furthermore, the generalized CP has been extended to $\Delta(48)$~\cite{Ding:2013nsa}, $\Delta(96)$~\cite{Ding:2014ssa} and $\Delta(6n^2)$ series~\cite{King:2014rwa,Hagedorn:2014wha,Ding:2014ora} family symmetries as well. Some new mixing textures compatible with the experimental data are found, in particular CP phases can be neither vanishing nor maximal. A number of interesting models with definite predictions for CP phases have been constructed. There are other approaches to dealing with family symmetry and CP violation~\cite{Branco:1983tn,Babu:2004tn,Chen:2009gf}.

Besides the well-known tri-bimaximal mixing, the bimaximal (BM) mixing can also be naturally derived from the $S_4$ family symmetry~\cite{Altarelli:2009gn,Meloni:2011fx,Ding:2013eca}. Since $\theta_{13}$ is not so small as expected and $\theta_{12}$ is not maximal, the BM pattern has been ruled out. However, the scheme with only one row or one column of the BM mixing preserved is still viable. In the present work, we shall assume $S_4$ family symmetry and generalized CP symmetry which is then spontaneously broken down to $Z_2\times CP$ in the neutrino sector or the charged lepton sector. As a consequence, only one column or one row of the BM mixing is preserved and the PMNS matrix deviates from BM pattern. Moreover, the concrete forms of the deviation from the BM mixing are constrained by the remnant symmetry, the corresponding predictions for the mixing angles and CP phases are investigated in a model independent way. Furthermore two models realizing these scenarios are built.

The paper is organized as follows. In section~\ref{sec:framework}, we present the basic concept of generalized CP symmetry and model independent approach of predicting lepton flavor mixing from remnant symmetry. The deviation from BM mixing induced by a rotation between two generation neutrinos or a rotation between two generation charged lepton fields is investigated in section~\ref{sec:2_derivation}, and the corresponding phenomenological predictions for the lepton mixing parameters are discussed. In section~\ref{sec:general_analysis_one_row}, the phenomenological implications of the symmetry breaking pattern of $S_4\rtimes H_{CP}$ into $K^{(TST^2, T^2U)}_4\rtimes H^{\nu}_{CP}$ in the neutrino sector and $Z^{SU}_{2}\times H^{l}_{CP}$ in the charged lepton sector are studied in a model-independent way. The resulting PMNS matrix has a row of form $\left(1/2, 1/2, 1/\sqrt{2}\right)$, and all the the three leptonic CP phases are conserved to fit the measured values of the mixing angles. In section~\ref{sec:model_column}, we construct an $S_4$ model with generalized CP symmetry, where the mixing pattern with one column $\left(1/2, 1/\sqrt{2}, 1/2\right)^{T}$ and conserved CP found in Ref.~\cite{Feruglio:2012cw} are produced exactly at leading order. Agreement with experimental data can be achieved after subleading order contributions are considered. The model reproducing all aspects of the general results of section~\ref{sec:general_analysis_one_row} is presented in section~\ref{sec:model_row}. Section~\ref{sec:conclusion} is devoted to our conclusion. The group theory of $S_4$ and the Clebsch-Gordan coefficients in our basis are collected in Appendix~\ref{sec:appA}. Finally the scenario of residual symmetry $Z_2\times H^{\nu}_{CP}$ in the neutrino sector and $K_4\rtimes H^{l}_{CP}$ in the charged lepton sector is discussed in Appendix~\ref{sec:appB}.

\section{\label{sec:framework}Basic framework}

\cleqn

We now consider a theory which is invariant under both family symmetry $S_4$ and generalized CP at high energy scale. For a field multiplet $\varphi(x)$ in a irreducible representation $\mathbf{r}$ of $S_4$, it transforms under the action of $S_4$ as
\begin{equation}
\label{eq:flavor_trans}\varphi(x)\stackrel{g}{\longmapsto}\rho_{\mathbf{r}}(g)\varphi(x),\quad g\in S_4\,,
\end{equation}
where $\rho_{\mathbf{r}}(g)$ is the representation matrix for the element $g$ in the representation $\mathbf{r}$. The generalized CP transformation on $\varphi$ is defined as
\begin{equation}
\label{eq:GCP_trans_def}\hskip-0.5in\varphi(x)\stackrel{CP}{\longmapsto}X_{\mathbf{r}}\varphi^{*}(t,-\mathbf{x})\,,
\end{equation}
where $X_{\mathbf{r}}$ is the generalized CP transformation matrix, and it is a unitary matrix to keep the kinetic term invariant. In Ref.~\cite{Feruglio:2012cw} the authors assumed that $X_{\mathbf{r}}$ is a unitary symmetric matrix. However,
it is not necessary to require $X_{\mathbf{r}}$ symmetric if there is a family symmetry~\cite{Holthausen:2012dk,Chen:2014tpa}, since performing two CP transformations in succession could be equivalent to a family symmetry transformation. Furthermore, given a well-defined CP transformation $X_{\mathbf{r}}$, Eqs.~(\ref{eq:flavor_trans}, \ref{eq:GCP_trans_def}) imply that $\rho_{\mathbf{r}}(g)X_{\mathbf{r}}$ with $g\in S_4$ is a CP symmetry of the theory. However $\rho_{\mathbf{r}}(g)X_{\mathbf{r}}$ is not always a symmetric matrix even if $X_{\mathbf{r}}$ is a symmetric matrix. As a result, we don't need to require the CP transformation $X_{\mathbf{r}}$ imposed at high energy scale is symmetric, although the residual CP transformation should be symmetric to avoid partially degenerate lepton masses, as will be shown below. Note that the obvious action of CP on the possible spinor indices has been suppressed in Eq.~\eqref{eq:GCP_trans_def}. One subtle point that we should treat with care is that the family symmetry and the generalized CP must be compatible with each other. The following consistency condition has to be fulfilled~\cite{Grimus:1995zi,Holthausen:2012dk,Feruglio:2012cw},
\begin{equation}
\label{eq:consistency_equ}X_{\mathbf{r}}\rho^{*}_{\mathbf{r}}(g)X^{-1}_{\mathbf{r}}=\rho_{\mathbf{r}}(g^{\prime}),\qquad g,g^{\prime}\in S_4\,,
\end{equation}
which maps one element $g$ into another element $g^{\prime}$. For the faithful representation $\mathbf{r}=\mathbf{3}, \mathbf{3^{\prime}}$, the representation matrices of no two elements are identical. As a consequence, the mapping of $g\rightarrow g^{\prime}$ is bijective, and then the consistency equation of Eq.~\eqref{eq:consistency_equ} will define a unique mapping of the family symmetry group $S_4$ to itself. Hence the generalized CP transformation $X_{\mathbf{r}}$ corresponds an automorphism of $S_4$. Generally the element $g$ is distinct from $g^{\prime}$ in Eq.~\eqref{eq:consistency_equ}. Hence the mathematical structure of the full symmetry group comprising family symmetry $S_4$ and generalized CP symmetry is in general a semi-direct product~\cite{Feruglio:2012cw}. Consequently, the imposed symmetry at high energy scale is $S_4\rtimes H_{CP}$.

Since the outer automorphism group of $S_4$ is trivial~\cite{Holthausen:2012dk,Ding:2014ora}, all the automorphisms of $S_4$ are inner automorphisms, and can be generated by group conjugation. Consequently the generalized CP transformation compatible with $S_4$ family symmetry is defined by the inner  automorphism of $S_4$ through the consistency condition. Now we determine the explicit form of these CP transformation matrices in our working basis. We consider the representative inner automorphism $\sigma_{TST^2}:(S,T,U)\rightarrow(S,ST,SU)$, where $\sigma_{h}$ is defined as $\sigma_{h}:$ $g\rightarrow hgh^{-1}$ for any $h, g\in S_4$. The corresponding generalized CP transformation denoted by $X^{0}_{\mathbf{r}}$ should satisfy the following consistency equations:
\begin{eqnarray}
\nonumber&&X^{0}_{\mathbf{r}}\rho_{\mathbf{r}}^{*}(S)\left(X^{0}_{\mathbf{r}}\right)^{-1}=\rho_{\mathbf{r}}\left({\sigma_{TST^2}(S)}\right)=\rho_{\mathbf{r}}(S),\\
\nonumber&&X^{0}_{\mathbf{r}}\rho_{\mathbf{r}}^{*}(T)\left(X_{\mathbf{r}}^{0}\right)^{-1}=\rho_{\mathbf{r}}\left({\sigma_{TST^2}(T)}\right)=\rho_{\mathbf{r}}(ST),\\
&&X^{0}_{\mathbf{r}}\rho_{\mathbf{r}}^{*}(U)\left(X_{\mathbf{r}}^{0}\right)^{-1}=\rho_{\mathbf{r}}\left({\sigma_{TST^2}(U)}\right)=\rho_{\mathbf{r}}(SU)\;.
\end{eqnarray}
Given the representation matrices listed in Table~\ref{tab:representation}, we see that the following relations are satisfied for any irreducible representations $\mathbf{r}$ of $S_4$,
\begin{equation}
\rho^{*}_{\mathbf{r}}(S)=\rho_{\mathbf{r}}(S),\qquad  \rho^{*}_{\mathbf{r}}(T)=\rho_{\mathbf{r}}(ST),\qquad \rho^{*}_{\mathbf{r}}(U)=\rho_{\mathbf{r}}(SU)\,.
\end{equation}
Therefore $X^{0}_{\mathbf{r}}$ is determined to be a unity matrix up to an overall phase,
\begin{equation}
X^{0}_{\mathbf{r}}=1\,.
\end{equation}
For a given solution $X_{\mathbf{r}}$ of Eq.~\eqref{eq:consistency_equ}, we can easily check that $\rho_{\mathbf{r}}(h)X_{\mathbf{r}}$ is also a solution for any $h\in S_4$. Since $\rho_{\mathbf{r}}(h)X_{\mathbf{r}}$ maps one group element $g$ into $hg^{\prime}h^{-1}\equiv\sigma_{h}(g^{\prime})$~\footnote{We have $\big(\rho_{\mathbf{r}}(h)X_{\mathbf{r}}\big)\rho^{*}_{\mathbf{r}}(g)\big(\rho_{\mathbf{r}}(h)X_{\mathbf{r}}\big)^{-1}=\rho_{\mathbf{r}}(h)\big(X_{\mathbf{r}}\rho^{*}(g)X^{-1}_{\mathbf{r}}\big)\rho^{-1}_{\mathbf{r}}(h)=
\rho_{\mathbf{r}}(h)\rho_{\mathbf{r}}(g^{\prime})\rho^{-1}_{\mathbf{r}}(h)=\rho_{\mathbf{r}}(hg^{\prime}h^{-1})$.}, the inner automorphism is equivalent to a family symmetry transformation. As a consequence, the generalized CP transformation compatible with the $S_4$ family symmetry is of the form
\begin{equation}
\label{eq:GCP_trans}\rho_{\mathbf{r}}(h)X^{0}_{\mathbf{r}}=\rho_{\mathbf{r}}(h),\qquad h\in S_4\,,
\end{equation}
where $h$ can be any of the 24 group elements of $S_4$. In particular we see that the canonical CP transformation with $\rho_{\mathbf{r}}(1)=X^{0}_{\mathbf{r}}=1$ is allowed. Therefore all coupling constants would be constrained to be real in a $S_4$ model with imposed CP symmetry.

Being similar to the paradigm of family symmetry, the imposed symmetry is $S_4\rtimes H_{CP}$ at high energy in the present work, where $H_{CP}$ is the CP transformation consistent with $S_4$ family symmetry and its elements is given in Eq.~\eqref{eq:GCP_trans}. Subsequently $S_4\rtimes H_{CP}$ is broken down to different residual symmetry subgroups $G_{\nu}\rtimes H^{\nu}_{CP}$ and $G_{l}\rtimes H^{l}_{CP}$ in the neutrino and the charged lepton sectors respectively. The misalignment between $G_{\nu}\rtimes H^{\nu}_{CP}$  and $G_{l}\rtimes H^{l}_{CP}$ leads to particular predictions for mixing angles and CP phases. The basic procedure of predicting lepton flavor mixing from remnant symmetries in a model independent way has been stated clearly in Refs.~\cite{Ding:2013bpa,Ding:2013hpa,Ding:2013nsa,Ding:2014ssa}. In the following, we briefly review the most important points which will be exploited later. Without loss of generality, the three generations of left-handed lepton doublets are assigned to be a $S_4$ triplet $\mathbf{3}$. The irreducible representation $\mathbf{3^{\prime}}$ is distinct from $\mathbf{3}$ in the overall sign of the generator $U$, therefore the same results are obtained if the lepton doublet fields are embedded into $\mathbf{3^{\prime}}$ instead of $\mathbf{3}$. Firstly, invariance under the residual symmetries $G_l$ and $G_{\nu}$ implies
\begin{eqnarray}
\nonumber&&\rho^{\dagger}_{\mathbf{3}}(g_l)m^{\dagger}_{l}m_{l}\rho_{\mathbf{3}}(g_l)=m^{\dagger}_{l}m_{l},\qquad g_{l}\in G_{l}\\
&&\label{eq:invariance1}\rho^{T}_{\mathbf{3}}(g_{\nu})m_{\nu}\rho_{\mathbf{3}}(g_{\nu})=m_{\nu},\qquad g_{\nu}\in G_{\nu} \,,
\end{eqnarray}
where the charged lepton mass matrix $m_l$ is given in the convention in which the right-handed (left-handed) fields are on the left-hand (right-hand) side of $m_{l}$. Furthermore, the neutrino and the charged lepton mass matrices are subject to the constraint of residual CP symmetry as follows,
\begin{eqnarray}
\nonumber&&X^{T}_{\nu\mathbf{3}}m_{\nu}X_{\nu\mathbf{3}}=m^{*}_{\nu},\qquad
\quad X_{\nu\mathbf{3}}\in H^{\nu}_{CP}\,,\\
\label{eq:inv_CP}&&X^{\dagger}_{l\mathbf{3}}m^{\dagger}_{l}m_{l}X_{l\mathbf{3}}=\left(m^{\dagger}_{l}m_{l}\right)^{*},\quad
X_{l\mathbf{3}}\in H^{l}_{CP}\,.
\end{eqnarray}
The remnant family symmetry should be consistent with remnant CP symmetry, and hence the following consistency conditions should be fulfilled,
\begin{eqnarray}
\nonumber&&X_{\nu\mathbf{r}}\rho^{*}_{\mathbf{r}}(g_{\nu})X^{-1}_{\nu\mathbf{r}}=\rho_{\mathbf{r}}(g^{\prime}_{\nu}),\qquad
g_{\nu},g^{\prime}_{\nu}\in G_{\nu},\\
\label{eq:invariance2}&&X_{l\mathbf{r}}\rho^{*}_{\mathbf{r}}(g_{l})X^{-1}_{l\mathbf{r}}=\rho_{\mathbf{r}}(g^{\prime}_{l}),\qquad
g_{l},g^{\prime}_{l}\in G_{l}\,.
\end{eqnarray}
Given a set of solutions $X_{\nu\mathbf{r}}$ and $X_{l\mathbf{r}}$, we can
straightforwardly check that the CP transformations $\rho_{\mathbf{r}}(g_{\nu})X_{\nu\mathbf{r}}$
and $\rho_{\mathbf{r}}(g_{l})X_{l\mathbf{r}}$ with $g_{\nu}\in G_{\nu}$, $g_{l}\in G_{l}$ are admissible as well, and they lead to the same constraints shown in Eq.~\eqref{eq:inv_CP} on the lepton mass matrices as $X_{\nu\mathbf{r}}$ and $X_{l\mathbf{r}}$.
Furthermore, from the invariant condition of Eq.~\eqref{eq:inv_CP}, we can derive that
\begin{equation}
U^{\dagger}_{\nu}X_{\nu\mathbf{3}}U^{*}_{\nu}=\text{diag}(\pm1, \pm1, \pm1),\qquad U^{\dagger}_{l}X_{l\mathbf{3}}U^{*}_{l}=\text{diag}(e^{i\rho_1}, e^{i\rho_2}, e^{i\rho_3})\,,
\end{equation}
where $\rho_{i}(i=1,2,3)$ is an arbitrary real phase, $U_{\nu}$ and $U_{l}$ are the unitary diagonalization matrices of $m_{\nu}$ and $m^{\dagger}_{l}m_{l}$ respectively with $U^{T}_{\nu}m_{\nu}U_{\nu}=\text{diag}(m_1,m_2,m_3)$ and $U^{\dagger}_{l}m^{\dagger}_{l}m_{l}U_{l}=\text{diag}(m^2_e, m^2_{\mu}, m^2_{\tau})$. As a consequence, the residual CP transformations $X_{\nu\mathbf{3}}$ and $X_{l\mathbf{3}}$ should be symmetric otherwise the neutrino or the charged lepton masses would be constrained to be partially degenerate which is not compatible with experimental data. Note that the conclusion that the remnant CP transformations in the neutrino sector have to be symmetric is also reached in Ref.~\cite{Feruglio:2012cw}. As $X_{\nu\mathbf{r}}$ and $\rho_{\mathbf{r}}(g_{\nu})X_{\nu\mathbf{r}}$ with $g_{\nu}\in G_{\nu}$ lead to the same constraint on the neutrino mass matrix $m_{\nu}$, the light neutrino masses would be partially degenerate if $\rho_{\mathbf{r}}(g_{\nu})X_{\nu\mathbf{r}}$ is non-symmetric even if $X_{\nu\mathbf{r}}$ is a symmetric matrix. Therefore that symmetric remnant CP transformation is necessary but not sufficient for non-degenerate lepton masses, and we shall require both  $X_{\nu\mathbf{r}}$ and $\rho_{\mathbf{r}}(g_{\nu})X_{\nu\mathbf{r}}$ are symmetric in the following. Similarly $X_{l\mathbf{r}}$ and $\rho_{\mathbf{r}}(g_{l})X_{l\mathbf{r}}$ with $g_{l}\in G_{l}$ are also required to be symmetric.

We can obtain the most general form of $m_{\nu}$ and $m^{\dagger}_{l}m_{l}$ from the invariant requirements of Eq.~\eqref{eq:invariance1} and Eq.~\eqref{eq:inv_CP}, then diagonalize them, and finally we can determine the lepton mixing matrix $U_{PMNS}$. Last but not least, generally we have many possible choices for the residual symmetry subgroups. However, if the residual family
symmetries are taken to be another pair of subgroups $G^{\prime}_{\nu}$ and $G^{\prime}_{l}$ which are conjugate to $G_{\nu}$ and $G_{l}$,
\begin{equation}
 G^{\prime}_{\nu}=hG_{\nu}h^{-1},\qquad G^{\prime}_{l}=hG_{l}h^{-1},\qquad h\in S_4\,.
\end{equation}
Solving the consistency condition, we find that the residual CP symmetries $H^{\nu'}_{CP}$ and $H^{l'}_{CP}$ compatible with $G^{\prime}_{\nu}$ and $G^{\prime}_{l}$ are of the form
\begin{equation}
\label{eq:CP_conju}H^{\nu'}_{CP}=\rho_{\mathbf{r}}(h)H^{\nu}_{CP}\rho^{T}_{\mathbf{r}}(h),\qquad
H^{l'}_{CP}=\rho_{\mathbf{r}}(h)H^{l}_{CP}\rho^{T}_{\mathbf{r}}(h)\,.
\end{equation}
This means that the elements of $H^{\nu'}_{CP}$ and $H^{l'}_{CP}$ are given by $\rho_{\mathbf{r}}(h)X_{\nu\mathbf{r}}\rho^{T}_{\mathbf{r}}(h)$ and $\rho_{\mathbf{r}}(h)X_{l\mathbf{r}}\rho^{T}_{\mathbf{r}}(h)$ respectively with $X_{\nu\mathbf{r}}\in H^{\nu}_{CP}$ and $X_{l\mathbf{r}}\in H^{l}_{CP}$. The neutrino and charged lepton mass matrices $m^{\prime}_{\nu}$ and $m^{\prime\dagger}_{l}m^{\prime}_{l}$ invariant under $G^{\prime}_{\nu}\rtimes H^{\nu'}_{CP}$ and $G^{\prime}_{l}\rtimes H^{l'}_{CP}$ respectively are determined to be
\begin{equation}
m^{\prime}_{\nu}=\rho^{*}_{\mathbf{3}}(h)m_{\nu}\rho^{\dagger}_{\mathbf{3}}(h), \qquad m^{\prime\dagger}_{l}m^{\prime}_{l}=\rho_{\mathbf{3}}(h)m^{\dagger}_{l}m_{l}\rho^\dagger_{\mathbf{3}}(h)\,.
\end{equation}
Obviously the lepton mixing matrix would be predicted to be of the same form as that in $G_{\nu}$, $G_{l}$ case~\cite{Ding:2013bpa,Ding:2013nsa,Ding:2014ssa}. As a result, we only need to analyze few independent residual family symmetries not related by group conjugation and the compatible remnant CP. We assume that the light neutrinos are Majorana particles, and hence the remnant family symmetry $G_{\nu}$ in the neutrino sector must be $K_4$ or $Z_2$ subgroups. The case that $S_4\rtimes H_{CP}$ is broken down to $Z_2\times H^{\nu}_{CP}$ in the neutrino sector and $Z_3\rtimes H^{l}_{CP}$ in the charged lepton sector has been comprehensively studied~\cite{Ding:2013hpa,Li:2013jya}, One column of the PMNS matrix is then determined to be proportional to $\left(2,-1,-1\right)^{T}$ or $(1,1,1)^{T}$, i.e. the so-called $\text{TM}_1$ and $\text{TM}_2$ mixing patterns can be produced exactly. Besides the $Z_3$ subgroup, the residual family symmetry $G_{l}$ in the charged lepton sector can be $Z_4$ or $K_4$ subgroups of $S_4$~\footnote{Choosing $G_{l}$ to be a non-abelian subgroup would lead to a degenerate mass spectrum.}. For example, the choice $G_{l}=Z^{TST^2U}_4$ (or $G_{l}=K^{\left(S,U\right)}_4$) and $G_{\nu}=K^{(TST^2,T^2U)}_4$ leads to BM mixing no matter whether the generalized CP is included or not. In order to be in accordance with experimental data, we degrade $G_{\nu}$ from $K_4$ to $Z_2$ or $G_{l}$ from $Z_4(K_4)$ to $Z_2$ such that only one column or one row of the BM mixing matrix is fixed. After the generalized CP transformation defined in Eq.~\eqref{eq:GCP_trans} is taken into account further, the resulting lepton mixing matrix $U_{PMNS}$ is found to depend on only one free real parameter. In the following, we shall firstly investigate the phenomenological predictions of preserving one column or one row of BM mixing, which may originate from a $2\times 2$ rotation in the neutrino or the charged lepton sector. Furthermore, the $S_4$ family symmetry together with the the generalized CP is imposed onto the theory, and then lepton flavor mixing arising from the symmetry breaking into different residual subgroups in the neutrino and the charged lepton sectors are discussed in section~\ref{sec:general_analysis_one_row} and section~\ref{sec:model_column}. We find that the PMNS matrix has one column or one row in common with BM mixing up to permutations, and moreover the CP phases are predicted to take definite values because of the constraint of generalized CP symmetry.

\section{\label{sec:2_derivation} Phenomenological analysis of deviation from bimaximal mixing}
\cleqn

In a particular phase convention, the BM mixing matrix $U_{BM}$ (ignoring possible Majorana phases) is of the following form~\cite{Barger:1998ta}
\begin{equation}
\label{eq:BM}U_{BM}=\left(\begin{array}{ccc}
  \frac{1}{\sqrt{2}}  &  -\frac{1}{\sqrt{2}}  &  0  \\
    \frac{1}{2}  &  \frac{1}{2}  &  -\frac{1}{\sqrt{2}}  \\
    \frac{1}{2}  &  \frac{1}{2}  &  \frac{1}{\sqrt{2}}
  \end{array}\right)\,,
\end{equation}
which leads to the three lepton mixing angles
\begin{equation}
\theta^{BM}_{12}=\theta^{BM}_{23}=45^{\circ},\qquad \theta^{BM}_{13}=0^{\circ}\,.
\end{equation}
Comparing with the latest global fitting results~\cite{GonzalezGarcia:2012sz,Capozzi:2013csa,Forero:2014bxa}, we see that rather large corrections are needed to be compatible with the experimental data. In the following, we shall consider the minimal modifications for simplicity. The additional rotation of the 1-2, 1-3 or 2-3 generation of charged leptons or neutrinos in the BM basis
would be considered. As a consequence, one column or one row of BM mixing would be retained. Similar deviation from tri-bimaximal mixing has been widely studied~\cite{Albright:2008rp,He:2011gb,Wang:2013wya,Shimizu:2014ria,Petcov:2014laa}. Notice that the Majorana CP violating phases are not constrained at all in this phenomenological approach, since they are indeterminant in the starting BM mixing. First of all, we discuss the case of an extra 1-2 rotation in the charged lepton sector. The PMNS mixing matrix is obtained by multiplying the BM matrix $U_{BM}$ by a 1-2 rotation matrix in the left-hand side as follows:
\begin{equation}
U_{PMNS}=\left(\begin{array}{ccc}
    \cos\theta  & -\sin\theta e^{-i\delta}  &  0  \\
    \sin\theta e^{i\delta}  &  \cos\theta  &  0  \\
    0  &  0  &  1
  \end{array}\right)U_{BM}\,,
\end{equation}
where $\theta$ and $\delta$ are real free parameters, and their values can be fitted by the experimental data. Then the three mixing angles read as
\begin{equation}
\sin^{2}\theta_{13}=\frac{1}{2}\sin^{2}\theta, \quad  \sin^{2}\theta_{12}=\frac{1}{2}+\frac{\sqrt{2}\sin2\theta\cos\delta}{3+\cos2\theta},\quad \sin^{2}\theta_{23}=1-\frac{2}{3+\cos2\theta}\,.
\end{equation}
We see that the atmospheric and reactor mixing angles are related with each other by
\begin{equation}
\label{eq:theta23_theta_13_1}\sin^{2}\theta_{23}=\frac{1}{2}-\frac{1}{2}\tan^{2}\theta_{13}\,.
\end{equation}
Hence $\theta_{23}$ is constrained to lie in the first octant, i.e. $\theta_{23}<\frac{\pi}{4}$. The Jarlskog invariant $J_{CP}$ is given by
\begin{eqnarray}
J_{CP}=\frac{\sin2\theta\sin\delta}{8\sqrt{2}}\,.
\end{eqnarray}
Then the Dirac CP phase $\delta_{CP}$ in the standard parameterization~\cite{pdg} is
\begin{eqnarray}
\nonumber  \sin\delta_{CP}=\frac{(3+\cos2\theta)\sin2\theta\sin\delta}{|\sin2\theta|\sqrt{(3+\cos2\theta)^{2}-8\sin^{2}2\theta\cos^{2}\delta}}\,.
\end{eqnarray}
For the value of $\delta=0$, the above mixing parameters are simplified into
\begin{eqnarray}
\sin^{2}\theta_{13}=\frac{1}{2}\sin^{2}\theta,~~ \sin^{2}\theta_{12}=\frac{1}{2}+\frac{\sqrt{2}\sin2\theta}{3+\cos2\theta},~~
\sin^{2}\theta_{23}=\frac{2\cos^{2}\theta}{3+\cos2\theta},~~ \sin\delta_{CP}=0\,,
\end{eqnarray}
where the Dirac CP is conserved. Since the rotation of 2-3 generation of charged leptons gives a vanishing $\theta_{13}$, we turn to investigate an additional rotation of 1-3 generations. We can obtain the PMNS mixing matrix by multiplying the BM matrix by a 1-3 rotation matrix in the left-hand side as
\begin{equation}
U_{PMNS}=\left(\begin{array}{ccc}
\cos\theta  & 0  &  ~-\sin\theta e^{-i\delta}  \\
   0  &  1  &  0  \\
\sin\theta e^{i\delta}  ~&  0  &  \cos\theta
\end{array}\right)U_{BM}\,.
\end{equation}
The lepton mixing angles can be straightforwardly extracted as follows,
\begin{equation}
\sin^{2}\theta_{13}=\frac{1}{2}\sin^{2}\theta,\quad \sin^{2}\theta_{12}=\frac{1}{2}+\frac{\sqrt{2}\sin2\theta\cos\delta}{3+\cos2\theta},\quad
\sin^{2}\theta_{23}=\frac{2}{3+\cos2\theta}\,.
\end{equation}
The atmospheric and reactor mixing angles are related by,
\begin{equation}
\label{eq:theta23_theta_13_2}\sin^{2}\theta_{23}=\frac{1}{2}+\frac{1}{2}\tan^{2}\theta_{13}\,.
\end{equation}
which implies $\theta_{23}>\pi/4$ and $\theta_{23}$ is in the second octant. The Jarlskog invariant reads as
\begin{equation}
J_{CP}=-\frac{\sin2\theta\sin\delta}{8\sqrt{2}}\,,
\end{equation}
and then the Dirac CP phase is given by
\begin{equation}
\sin\delta_{CP}=-\frac{(3+\cos2\theta)\sin2\theta\sin\delta}{|\sin2\theta|\sqrt{(3+\cos2\theta)^{2}-8\sin^{2}2\theta\cos^{2}\delta}}\;.
\end{equation}
\begin{figure}[t!]
\centering
\includegraphics[width=0.32\textwidth]{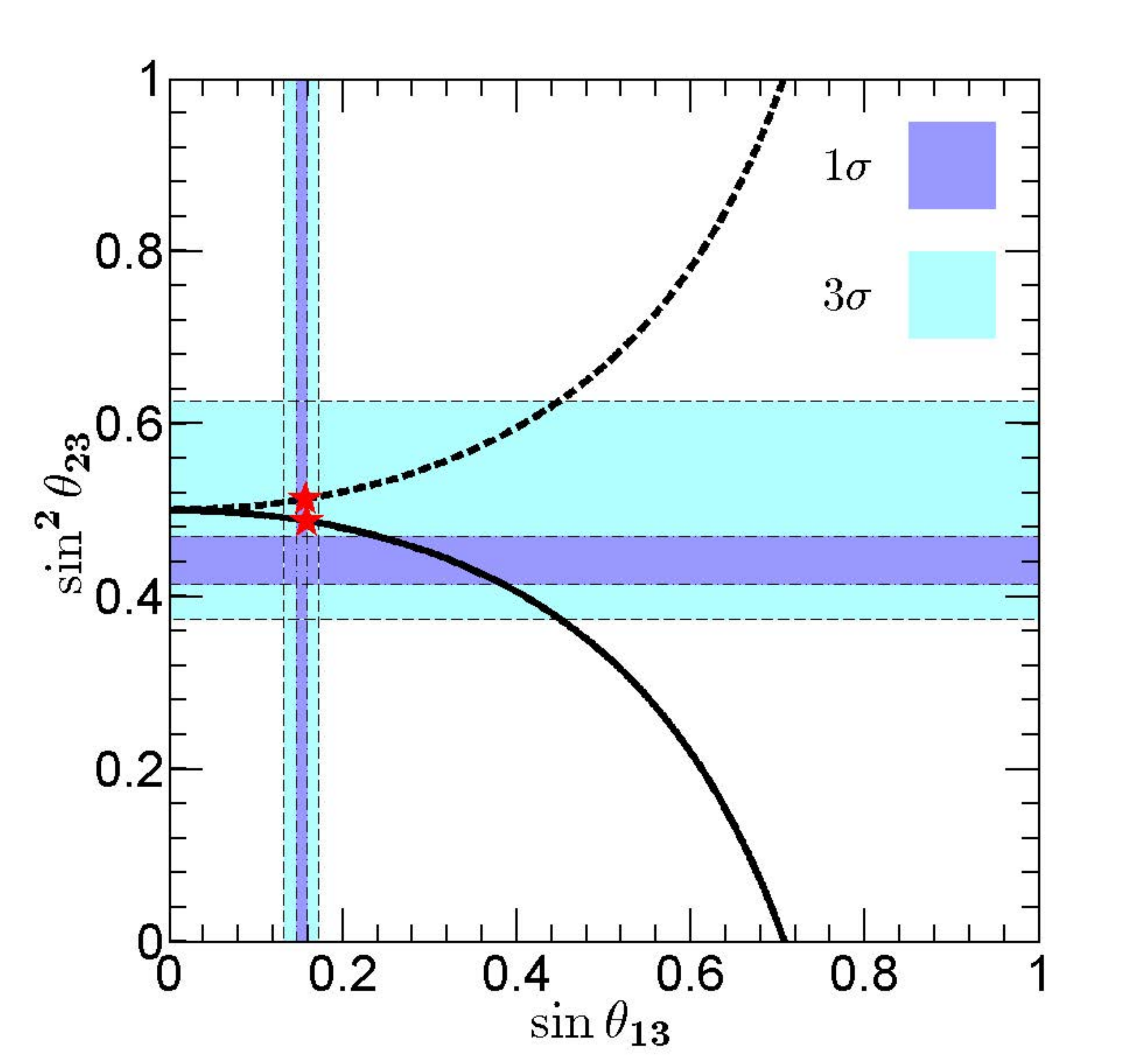}
\includegraphics[width=0.32\textwidth]{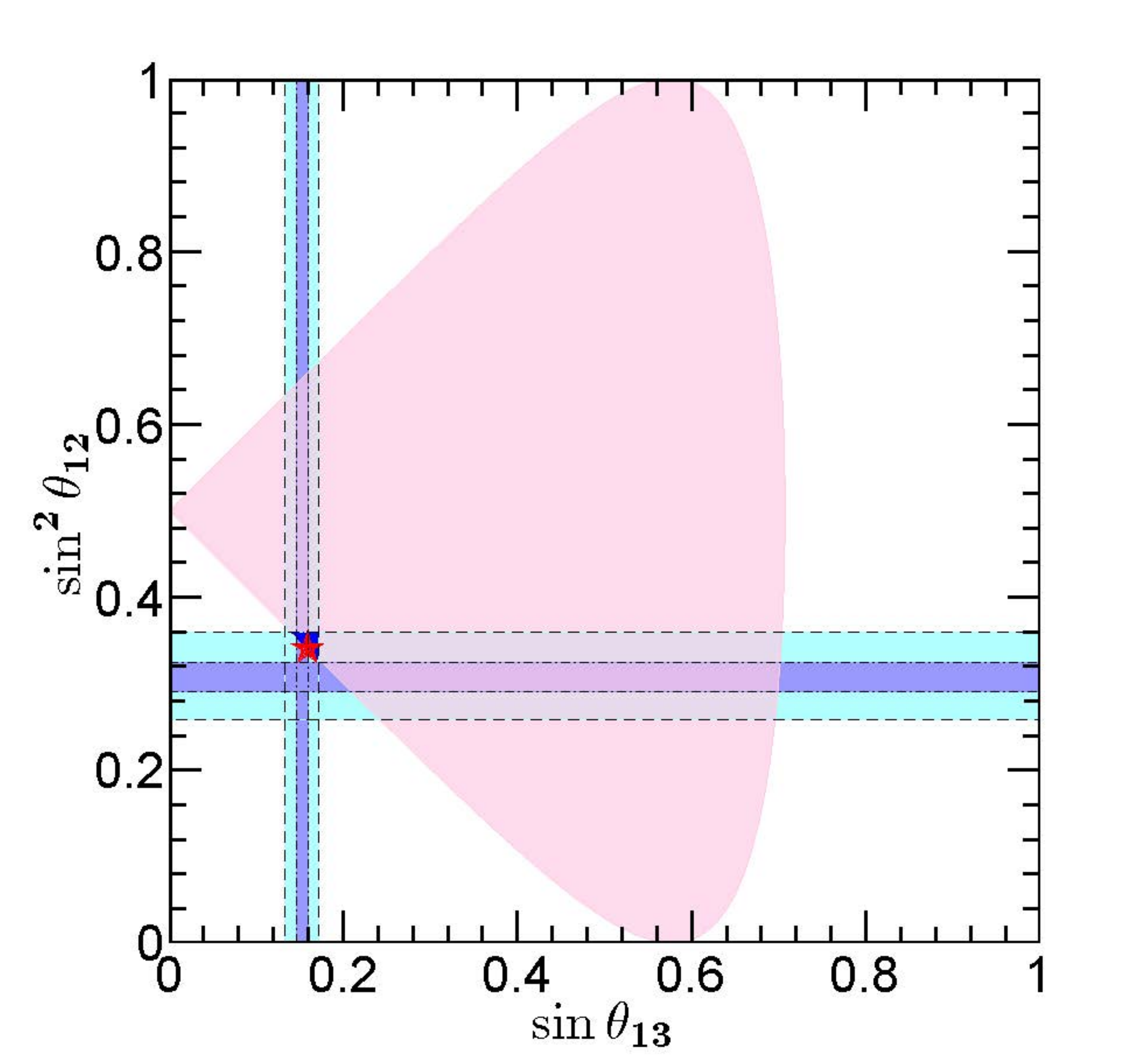}
\includegraphics[width=0.32\textwidth]{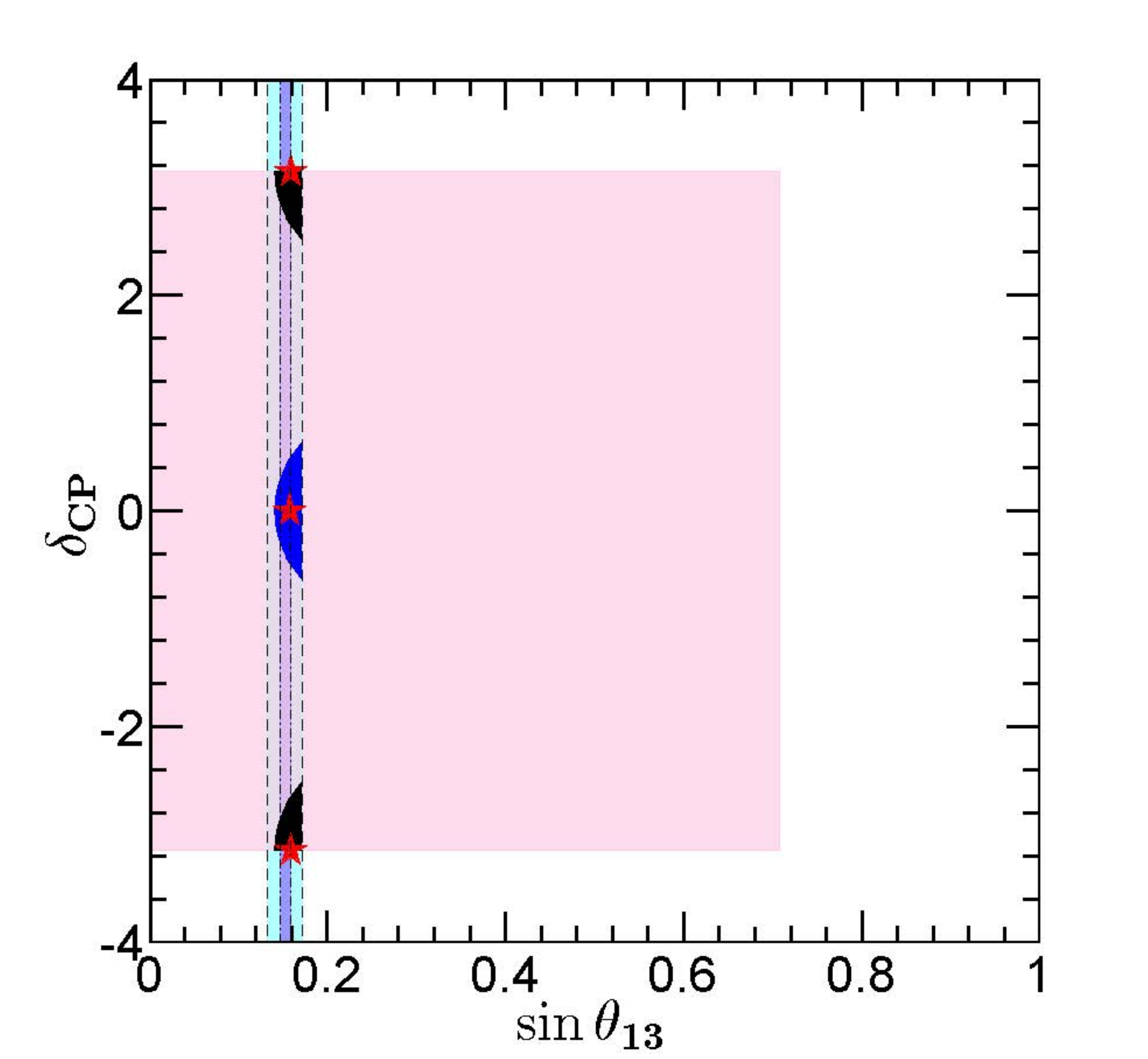}
\caption{\label{fig:deviation_one}Correlations among mixing angles ($\sin\theta_{13}, \sin^{2}\theta_{12}, \sin^{2}\theta_{23}$) and CP phase $\delta_{CP}$ for additional rotations of 1-2 and 1-3 generation of charged leptons in the BM basis. In the first panel, the results of $\sin^{2}\theta_{23}$ vs. $\sin\theta_{13}$ for 1-2 and 1-3 rotations are shown in solid line and dashed line respectively. The pink regions in the last two subfigures are the predictions for $\sin^{2}\theta_{12}$ and $\delta_{CP}$ with respect to $\sin\theta_{13}$ if both $\theta$ and $\delta$ vary in the range of $-\pi$ to $\pi$. The black areas in the third panel denote the allowed region by the experimental data of three mixing angles for 1-2 rotation and the blue areas for 1-3 rotation. In the second subfigure, the allowed regions for 1-2 and 1-3 rotations coincide. The red stars represent the best fit values in $S_4$ family symmetry combined with generalized CP.}
\end{figure}
We perform numerical analysis by scanning the free parameters $\theta$ and $\delta$ in the regions of $-\pi<\theta\leq\pi$ and $-\pi<\delta\leq\pi$. The correlations and the possible allowed values of the mixing parameters are obtained, as shown in Fig.~\ref{fig:deviation_one}. We see that there is a strong correlation between $\sin^2\theta_{23}$ and $\sin\theta_{13}$, which is given in Eq.~\eqref{eq:theta23_theta_13_1} and Eq.~\eqref{eq:theta23_theta_13_2}. Note that the allowed regions of the mixing parameters are rather large although only two free parameters $\theta$ and $\delta$ are involved. Furthermore, we take into account the current bounds for three neutrino mixing angles presented in Ref.~\cite{Capozzi:2013csa}, then the values of the mixing parameters would shrink to quite small areas. It is remarkable that the Dirac CP phase $\delta_{CP}$ is constrained to be in the range of $\pm\left[2.52,\pi\right]$ and $\left[-0.62, 0.62\right]$ for 1-2 and 1-3 rotations respectively. For comparison with the above phenomenological analysis, the theoretical predictions of the generalized CP symmetry discussed in section~\ref{sec:general_analysis_one_row} are also shown in Fig.~\ref{fig:deviation_one}.

Then we study the deviation from BM mixing induced by a rotation in the neutrino sector. Since the rotation of 1-2 generations leads to $\theta_{13}=0$, we do not discuss this scenario. Firstly we consider the case that the neutrino mass matrix is rotated between 1-3 generations in the BM basis. The PMNS matrix is obtained by multiplying the BM matrix $U_{BM}$ by a 1-3 rotation matrix in the right-hand side as follows:
\begin{equation}
U_{PMNS}=U_{BM}\begin{pmatrix}
\cos\theta  & 0  &  ~\sin\theta e^{-i\delta} \\
0   &  1   &   0 \\
-\sin\theta e^{i\delta}~   &  0  &  \cos\theta
\end{pmatrix}\,,
\end{equation}
which gives rise to the solar mixing angle $\sin^2\theta_{12}=\frac{2}{3+\cos2\theta}\geq\frac{1}{2}$. This mixing pattern is obviously not compatible with the experimental data~\cite{GonzalezGarcia:2012sz,Capozzi:2013csa,Forero:2014bxa}. Next we consider the rotation of 2-3 generation of neutrinos. The PMNS matrix is given by
\begin{equation}
U_{PMNS}=U_{BM}\begin{pmatrix}
    1  &  0  &  0  \\
    0  &  \cos\theta  &~  e^{-i\delta}\sin\theta  \\
    0  &~  -e^{i\delta}\sin\theta  &  \cos\theta
  \end{pmatrix}\,.
\end{equation}
The relation $2\cos^2\theta_{12}\cos^2\theta_{13}=1$ is found be fulfilled due to the fixed form of the first column. Using the $3\sigma$ range $1.76\times10^{-2}\leq\sin^2\theta_{13}\leq2.98\times10^{-2}$ as input, we obtain $0.485\leq\sin^2\theta_{12}\leq0.491$ which is outside of the experimentally preferred $3\sigma$ range~\cite{GonzalezGarcia:2012sz,Capozzi:2013csa,Forero:2014bxa}. Consequently this case doesn't agree with the experimental data as well. In short summary, simple perturbative rotation to the BM mixing in the neutrino sector is not viable because the observed values of $\theta_{12}$ and $\theta_{13}$ can not be produced simultaneously. It is notable that agreement with the experimental data could be achieved if permutations of rows and columns are allowed. If we perform both 2-3 rotation of neutrino and exchanges of rows and columns, the following PMNS matrix can be obtained
\begin{equation}
\label{eq:neutrino_perturbation} U_{PMNS}=\frac{1}{2}
\begin{pmatrix}
\sqrt{2}\cos\theta+\sin\theta e^{-i\delta} & 1  & \cos\theta-\sqrt{2}\sin\theta e^{i\delta}\\
-\sqrt{2}\sin\theta e^{-i\delta} & \sqrt{2} & -\sqrt{2}\cos\theta \\
-\sqrt{2}\cos\theta+\sin\theta e^{-i\delta}~ & 1 & ~\cos\theta+\sqrt{2}\sin\theta e^{i\delta}
\end{pmatrix}\,.
\end{equation}
\begin{figure}[t!]
\centering
\includegraphics[width=0.32\textwidth]{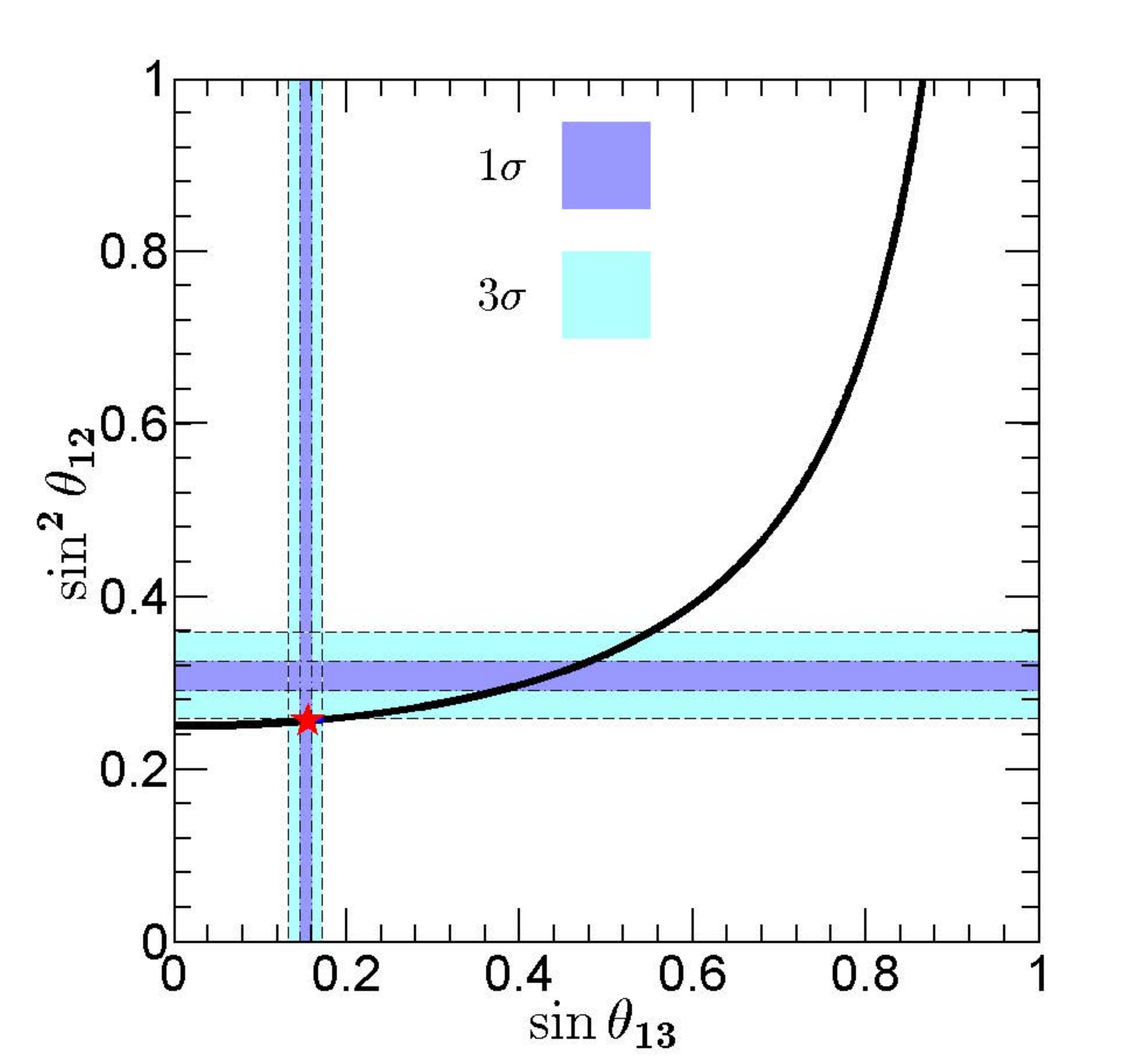}
\includegraphics[width=0.32\textwidth]{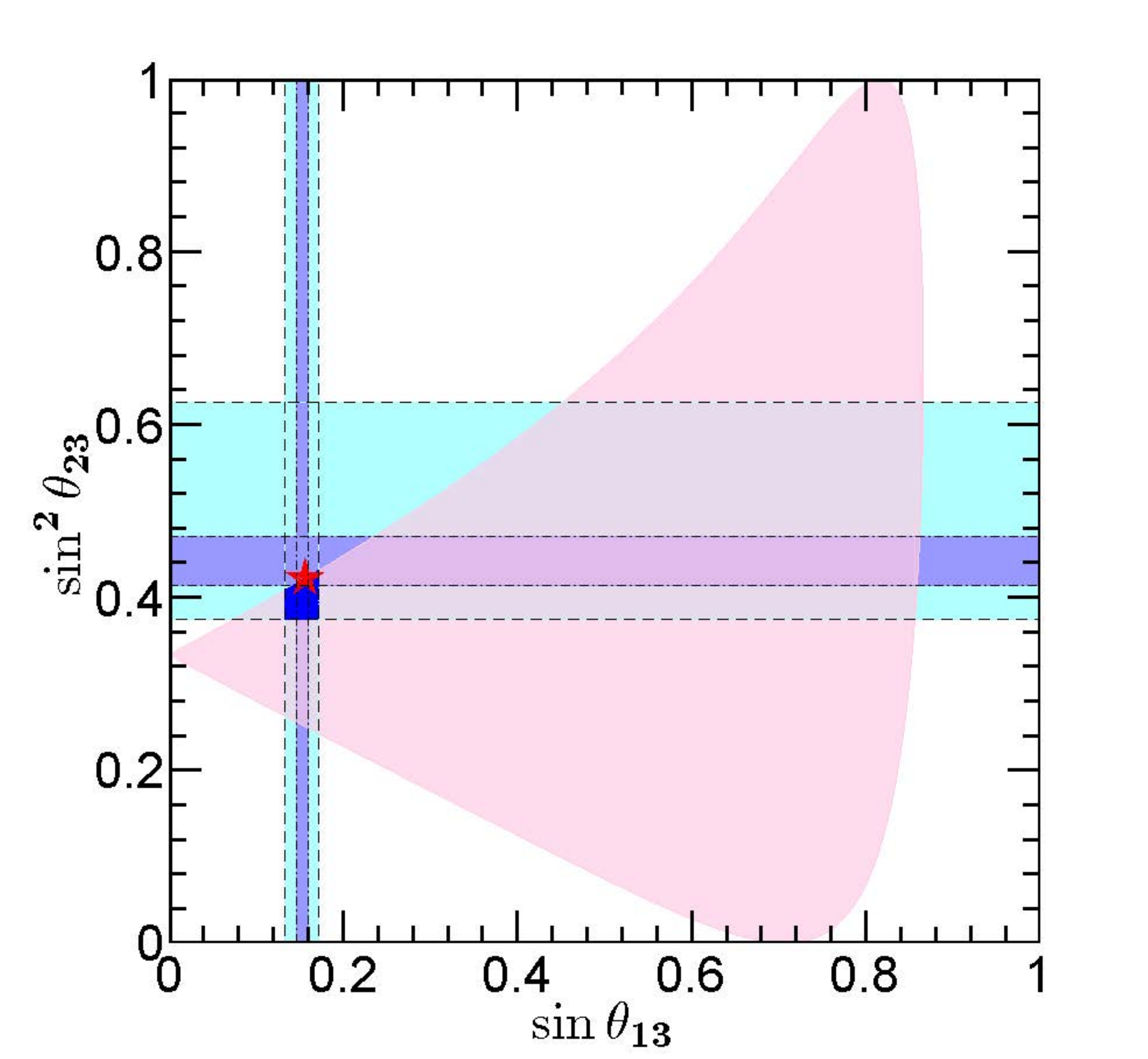}
\includegraphics[width=0.32\textwidth]{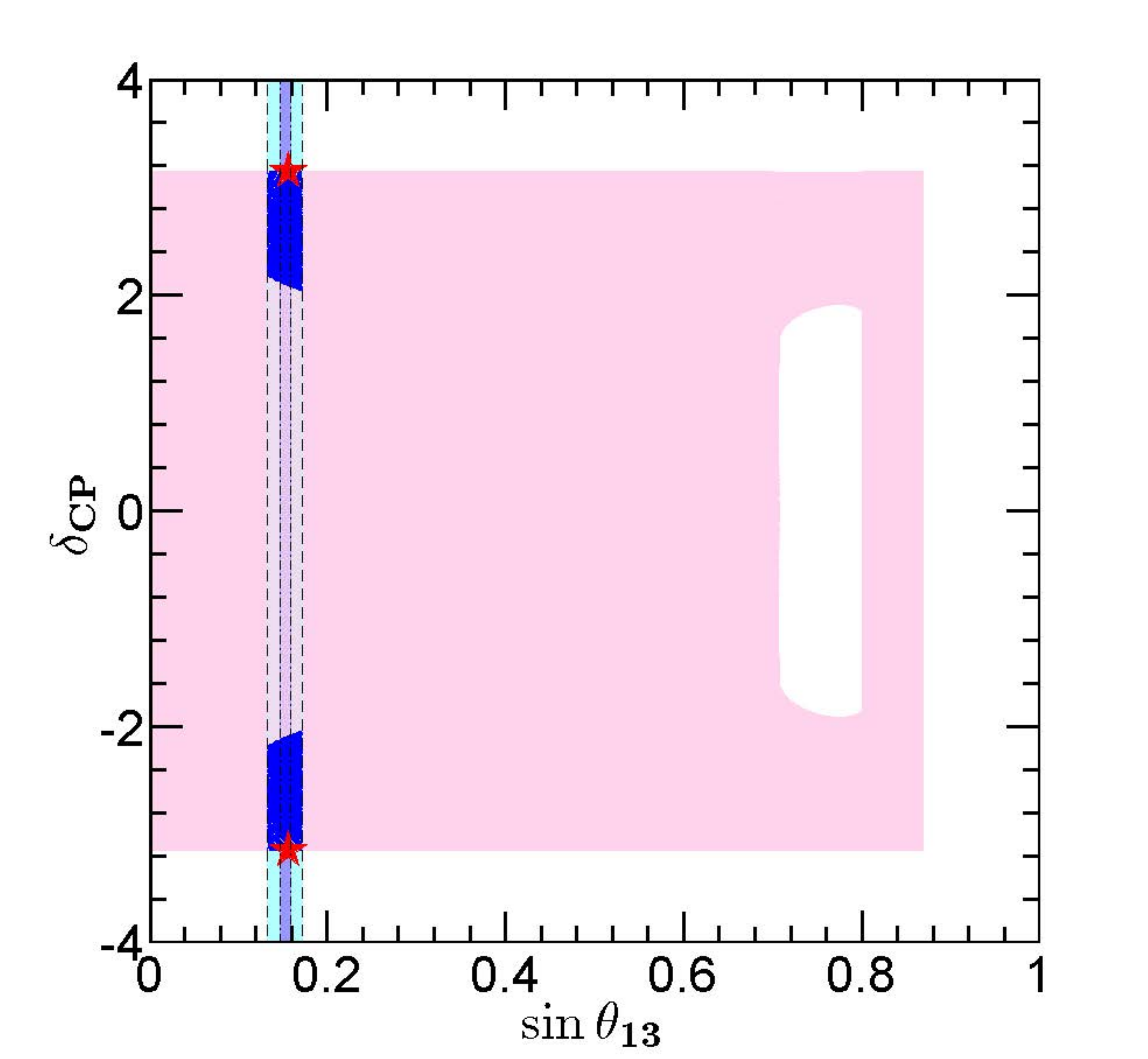}
\caption{\label{fig:deviation_two}Correlations among mixing angles ($\sin\theta_{13}, \sin^{2}\theta_{12}, \sin^{2}\theta_{23}$) and the Dirac CP phase $\delta_{CP}$ for the perturbation from the neutrino sector with permutations of rows and columns. The pink areas denote the allowed parameter values when both $\theta$ and $\delta$ vary in the range of $[-\pi,\pi]$. The blue ones are allowed regions if both $\theta_{13}$ and $\theta_{23}$ are required to lie in the experimentally preferred $3\sigma$ ranges~\cite{Capozzi:2013csa}. The red stars represent the best fit values in generalized CP which will be discussed at the beginning of section~\ref{sec:model_column}.}
\end{figure}
The three mixing angles read as
\begin{eqnarray}
\nonumber && \sin^{2}\theta_{13}=\frac{1}{8}(3-\cos2\theta-2\sqrt{2}\sin2\theta\cos\delta),\\ \nonumber&&\sin^{2}\theta_{12}=\frac{2}{5+\cos2\theta+2\sqrt{2}\sin2\theta\cos\delta},\\
\label{eq:neu_dev_mixing_angles}&&\sin^{2}\theta_{23}=\frac{2+2\cos2\theta}{5+\cos2\theta+2\sqrt{2}\sin2\theta\cos\delta}\,.
\end{eqnarray}
The following correlation is found
\begin{equation}
\label{eq:theta12_13_correlation}4\sin^2\theta_{12}\cos^2\theta_{13}=1\,.
\end{equation}
For the fitted $3\sigma$ range of $\theta_{13}$, the solar mixing angles is constrained to be in the interval of $0.254\leq\sin^2\theta_{12}\leq0.258$ which is rather close to its $3\sigma$ lower limit 0.259~\cite{Capozzi:2013csa}. As a result, we suggest this mixing pattern is a good leading order approximation since  accordance with experimental data should be easily achieved after subleading contributions are taken into account.
The Jarlskog invariant $J_{CP}$ is given by
\begin{eqnarray}
J_{CP}=-\frac{\sin2\theta\sin\delta}{8\sqrt{2}}\,,
\end{eqnarray}
The Dirac CP phase $\delta_{CP}$ is determined to be
\begin{eqnarray}
\label{eq:neu_dev_DCP}\hskip-0.20in\sin\delta_{CP}=-\frac{(5+\cos2\theta+2\sqrt{2}\sin2\theta\cos\delta)\sin2\theta\sin\delta}{\left|\cos\theta\right|\sqrt{2\left(3+\cos2\theta+2\sqrt{2}\sin2\theta\cos\delta\right)\left[(3-\cos2\theta)^{2}-8\sin^{2}2\theta\cos^{2}\delta\right]}}\;.
\end{eqnarray}
Similar to perturbative rotation from the charged lepton sector discussed above, the numerical results are presented in Fig.~\ref{fig:deviation_two}, where we demand that $\theta_{13}$ and $\theta_{23}$ are in their $3\sigma$ intervals~\cite{Capozzi:2013csa} while $\theta_{12}$ is fixed by the correlation of Eq.~\eqref{eq:theta12_13_correlation} and it is slightly beyond the present $3\sigma$ range. We see that $\theta_{23}$ is constrained to be smaller than $45^{\circ}$, and $\delta_{CP}$ is in the range of $\pm\left[2.04, \pi\right]$. The situation of $\theta_{23}$ in the second octant can be accounted for by exchanging the second and the third rows in Eq.~\eqref{eq:neutrino_perturbation}. Then $\theta_{23}$ would become $\pi/2-\theta_{23}$ and $\delta_{CP}$ becomes $\pi+\delta_{CP}$ while the predictions for $\theta_{12}$ and $\theta_{13}$ are the same as those in Eq.~\eqref{eq:neu_dev_mixing_angles}. It is straightforward to get numerical results for this case, as shown in Fig.~\ref{fig:deviation_three}. $\delta_{CP}$ is constrained to be in the range $\left[-1.10, 1.10\right]$. In the following, we shall show a lepton mixing matrix with one column or one row in common with BM mixing can be achieved from $S_4$ family symmetry, and $\delta_{CP}$ is predicted to take specific values 0 or $\pi$ after generalized CP is imposed.

\begin{figure}[t!]
\centering
\includegraphics[width=0.48\textwidth]{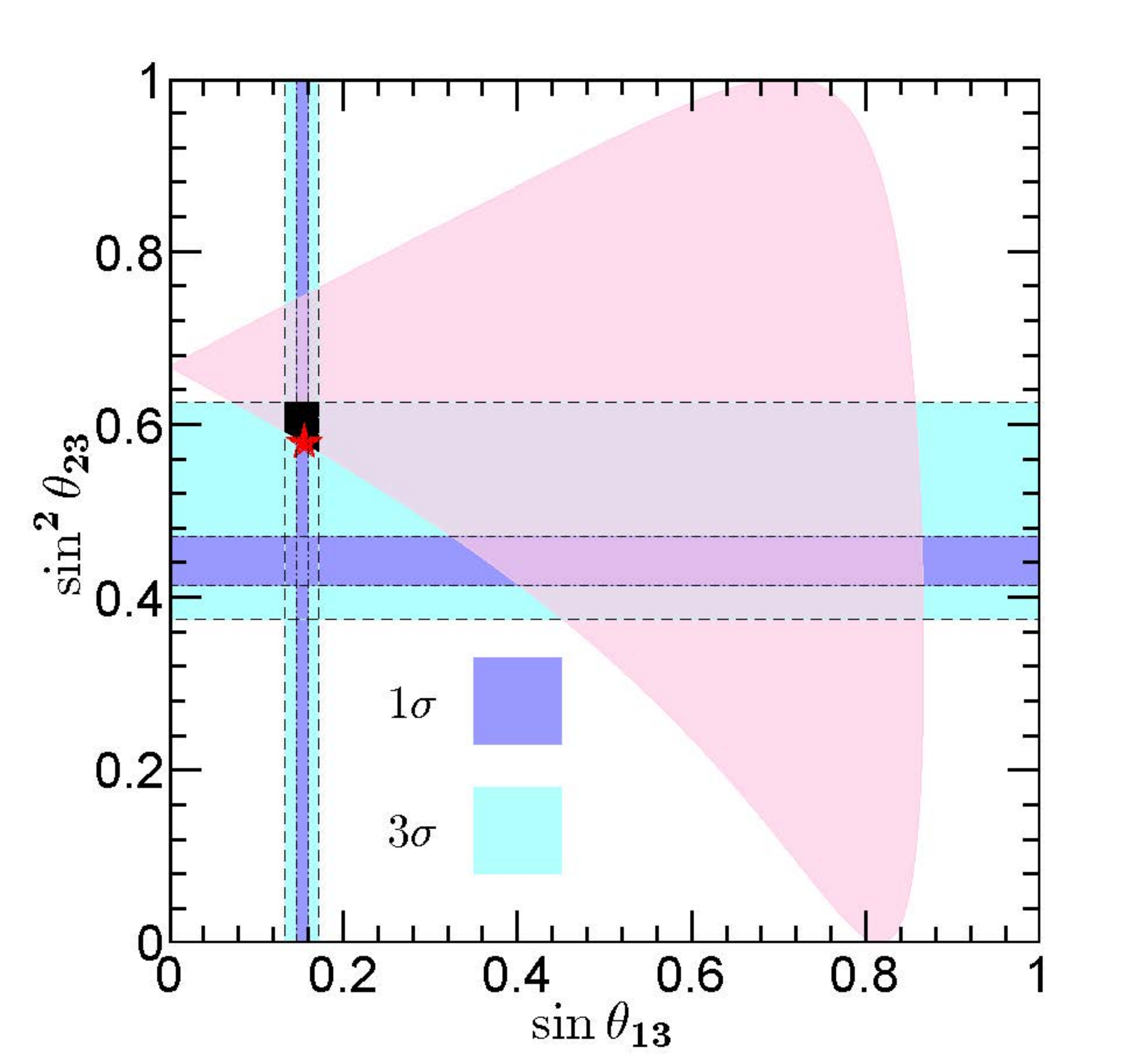}
\includegraphics[width=0.48\textwidth]{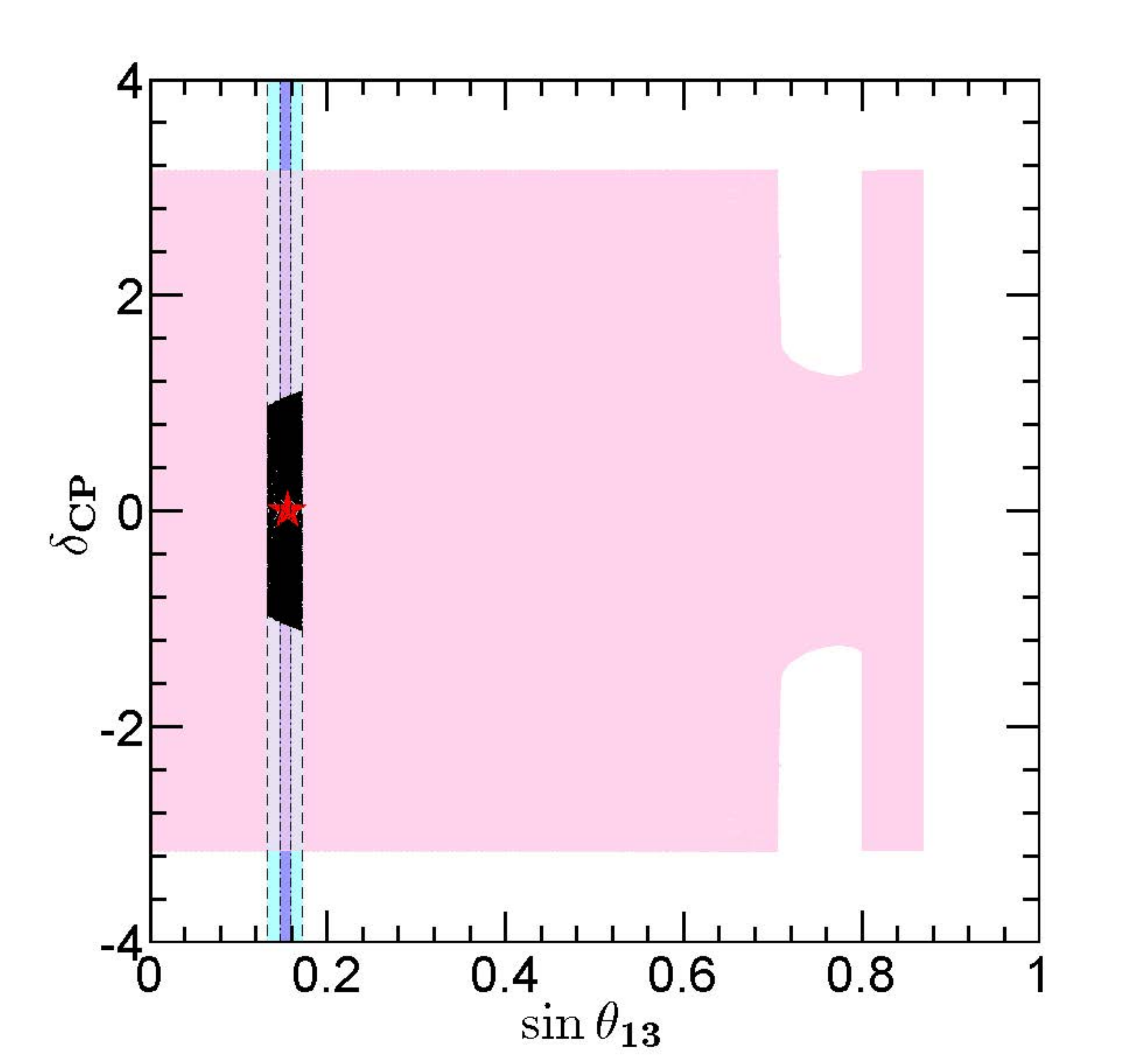}
\caption{\label{fig:deviation_three}Correlations among $\sin\theta_{13}$, $\sin^{2}\theta_{23}$ and Dirac CP phase $\delta_{CP}$ for the lepton mixing obtained by exchanging the second and the third row of Eq.~\eqref{eq:neutrino_perturbation}. The result for $\sin^{2}\theta_{12}$ with respect to $\sin\theta_{13}$ are not shown here since it is the same as the one in Fig.~\ref{fig:deviation_two}. The black areas denote the allowed regions after the measured $3\sigma$ bounds of $\theta_{13}$ and $\theta_{23}$ are imposed.}
\end{figure}

\section{\label{sec:general_analysis_one_row}Lepton flavor mixing from remnant symmetries $K^{(TST^2, T^2U)}_4\rtimes H^{\nu}_{CP}$ in the neutrino sector and $Z^{SU}_{2}\times H^{l}_{CP}$ in the charged lepton sector}
\cleqn

In this work, we shall extend the flavor symmetry to include additional CP symmetry. Analogous to the paradigm of flavor symmetry, lepton flavor mixing still arises from the mismatch between the remnant symmetries in the neutrino and the charged lepton sectors. The phenomenological implications of the breaking pattern of $S_4$ and generalized CP into $Z_2\times CP$ in the neutrino sector and an abelian subgroup of $S_4$ in the charged lepton sector have been investigated in Ref.~\cite{Feruglio:2012cw}. In this section, we shall study another scenario that $S_4$ is broken down to $K_4$ and $Z_2$ subgroups in the neutrino and the charged lepton sectors respectively. Including the generalized CP symmetry, the representative remnant symmetry considered here is $K^{(TST^2, T^2U)}_4\rtimes H^{\nu}_{CP}$ in the neutrino sector and $Z^{SU}_{2}\times H^{l}_{CP}$ in the charged lepton sector. Other possible choices of remnant symmetry are related to this one by similarity transformations or lead to a vanishing reactor mixing angle. In this case, only one row (instead of one column) of the PMNS matrix can be fixed because of the residual $Z^{SU}_2$ in the charged lepton sector. In this approach the remnant symmetries are assumed and we do not consider how the required vacuum alignment needed to achieve the remnant symmetries is dynamically realized, since the resulting lepton flavor mixing is independent of vacuum alignment mechanism although there are generally many possible symmetry breaking implementation schemes. Furthermore, we shall present dynamical models realizing the concerned symmetry breaking pattern in section~\ref{sec:model_row}.

Firstly we consider the neutrino sector. The full symmetry $S_4\rtimes H_{CP}$ is broken to $K^{(TST^2, T^2U)}_4\rtimes H^{\nu}_{CP}$. In order to consistently formulate such a setup, the element $X_{\nu\mathbf{r}}$ of $H^{\nu}_{CP}$ must satisfy the following consistence conditions:
\begin{equation}
\label{eq:consist_con_K4}X_{\nu\mathbf{r}}\rho^{*}_{\mathbf{r}}(h) X^{-1}_{\nu\mathbf{r}}=\rho_{\mathbf{r}}(h^{\prime}),\quad \mathrm{with}\quad  h,h^{\prime}\in K^{(TST^2, T^2U)}_{4}\,.
\end{equation}
We find that the residual CP transformation $X_{\nu\mathbf{r}}$ can take 4 possible values,
\begin{eqnarray}
H^{\nu}_{CP}=\{\rho_{\mathbf{r}}(1), \rho_{\mathbf{r}}(TST^{2}), \rho_{\mathbf{r}}(T^{2}U), \rho_{\mathbf{r}}(ST^{2}SU)\}\,.
\end{eqnarray}
Note that $X_{\nu\mathbf{r}}=\rho_{\mathbf{r}}(S)$, $\rho_{\mathbf{r}}(T^{2}ST)$, $\rho_{\mathbf{r}}(ST^{2}U)$, $\rho_{\mathbf{r}}(T^{2}SU)$ also fulfill the consistence condition of Eq.~\eqref{eq:consist_con_K4}, nevertheless they don't satisfy our symmetric requirement of section~\ref{sec:framework}, as both $\rho_{\mathbf{r}}(ST^{2}U)$ and $\rho_{\mathbf{r}}(T^{2}SU)$ are not symmetric matrices. The light neutrino mass matrix $m_{\nu}$ is constrained by the residual family symmetry $K^{(TST^2, T^2U)}_4$ and the residual CP symmetry $H^{\nu}_{CP}$ as
\begin{eqnarray}
\label{eq:neutrino_flavor_reK4}\rho^{T}_{\mathbf{3}}(h)m_{\nu}\rho_{\mathbf{3}}(h)&=m_{\nu}, \qquad &h\in K^{(TST^2, T^2U)}_4\;,\\
X^{T}_{\nu\mathbf{3}}m_{\nu}X_{\nu\mathbf{3}}&=m^{*}_{\nu}, \qquad \label{eq:neutrino_CP_reK4}&X_{\nu\mathbf{3}}\in H^{\nu}_{CP}\,.
\end{eqnarray}
Eq.~\eqref{eq:neutrino_flavor_reK4} constrains the light neutrino mass matrix to be of the form
\begin{equation}
\label{eq:neutrino_matrix_reK4}m_{\nu}=a\left(
\begin{array}{ccc}
 0 & 0 & 1 \\
 0 & 1 & 0 \\
 1 & 0 & 0
\end{array}
\right)+b\left(
\begin{array}{ccc}
 3 & 0 & -1 \\
 0 & 2 & 0 \\
 -1 & 0 & 3
\end{array}
\right)+c\left(
\begin{array}{ccc}
 0 & 1 & 0 \\
 1 & 0 & 1 \\
 0 & 1 & 0
\end{array}
\right)\,,
\end{equation}
which can be diagonalized by a unitary matrix $U_{\nu}$, i.e.
\begin{equation}
\label{eq:nu_mass_reK4}U^{T}_{\nu}m_{\nu}U_{\nu}=\mathrm{diag}\left(a+2b-\sqrt{2}c, a+2b+\sqrt{2}c, -a+4b \right)\,,
\end{equation}
where
\begin{equation}
\label{eq:Unu:reK4}U_{\nu}=\frac{1}{2}\left(
\begin{array}{ccc}
1  &  ~1   &~-\sqrt{2} \\
-\sqrt{2}  &~  \sqrt{2} & 0  \\
1  &  ~1  & ~ \sqrt{2}
\end{array}
\right)\,.
\end{equation}
Note that $U_{\nu}$ is fixed up to column permutations since the order of the eigenvalues of $m_{\nu}$ in Eq.~\eqref{eq:nu_mass_reK4} is not determined. Furthermore, the residual CP symmetry invariant condition of Eq.~\eqref{eq:neutrino_CP_reK4} implies that all the three parameters $a$, $b$ and $c$ are real for $X_{\nu\mathbf{r}}=\rho_{\mathbf{r}}(1), \rho_{\mathbf{r}}(TST^{2}), \rho_{\mathbf{r}}(T^{2}U), \rho_{\mathbf{r}}(ST^{2}SU)$. Then the light neutrino masses are determined by three real parameters $a$, $b$ and $c$. As a consequence, either normal ordering (NO) or inverted ordering (IO) neutrino mass spectrum can be accommodated.

\begin{figure}[t!]
\begin{center}
\includegraphics[width=0.99\textwidth]{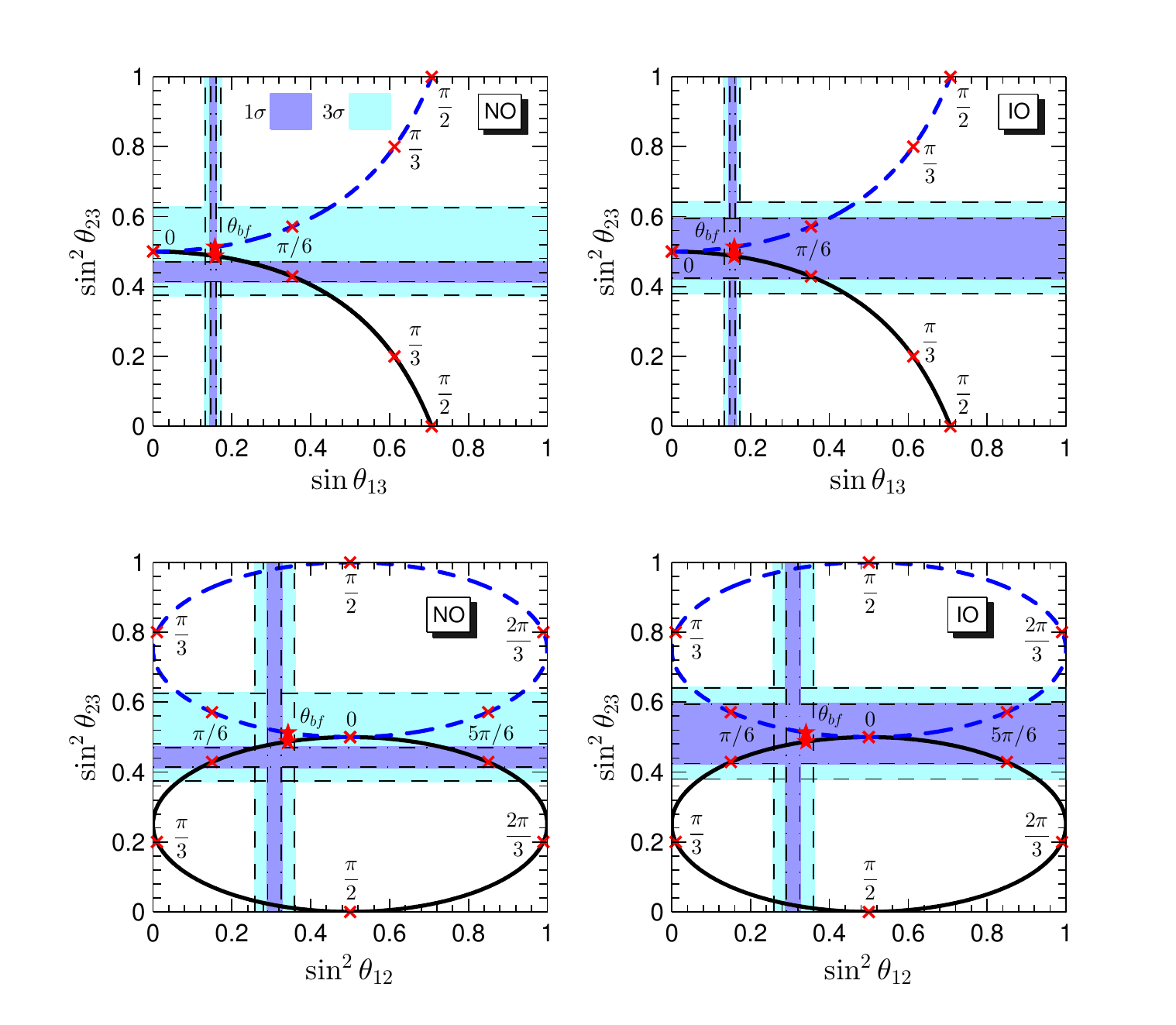}
\caption{\label{fig:best_fit_four}The relations between $\sin^{2}\theta_{23}$ and $\sin\theta_{13}$, $\sin^{2}\theta_{12}$ in the case of $Z_2\times CP$ preserved in the charged lepton sector. The solid lines and dashed lines represent the results for $\theta_{23}<\pi/4$ and $\theta_{23}>\pi/4$ respectively. The best fitting value is marked with a red star, and the points for $\theta=0$, $\pi/6$, $\pi/3$, $\pi/2$, $2\pi/3$ and $5\pi/6$ are labelled with a cross to guide the eye. The shown $1\sigma$ and $3\sigma$ ranges for the mixing angles are taken from Ref.~\cite{Capozzi:2013csa}.}
\end{center}
\end{figure}

Now we turn to the charged lepton sector. The $S_4$ flavor symmetry is broken down to $G_{l}=Z^{SU}_2$. The remnant CP symmetry $H^{l}_{CP}$ has to be consistent with the remnant family symmetry $Z^{SU}_2$. That is to say, its element $X_{l\mathbf{r}}$ should satisfy the consistency equation
\begin{eqnarray}
\label{eq:cons_eq_charl_row}
X_{l\mathbf{r}}\rho^{*}_{\mathbf{r}}(SU)X^{-1}_{l\mathbf{r}}=\rho_{\mathbf{r}}(SU)\,.
\end{eqnarray}
This restricted consistency equation can be derived from the general consistency condition of Eq.~\eqref{eq:consistency_equ} with $g,\, g^{\prime}\in Z^{SU}_2$. For $g=SU$, $g^{\prime}$ can only be $SU$ (can not be identity element) since it is the unique element which has the same order as $g=SU$. This implies that the remnant CP symmetry $H^{l}_{CP}$ is commutable with the remnant family symmetry $Z^{SU}_2$, and therefore the semidirect product between family and generalized CP symmetries will reduce to a direct product. As a consequence, the residual symmetry in the neutrino sector would be $Z^{SU}_2\times H^{l}_{CP}$ in this case. In fact, the reduction of the semidirect product structure to direct product holds true for a generic residual $Z_2$ family symmetry~\cite{Ding:2013hpa,Ding:2013bpa}. It is easy to check that only four generalized CP transformations are acceptable,
\begin{equation}
\label{eq:rem_GCP_reK4}H^{l}_{CP}=\{\rho_{\mathbf{r}}(TST^{2}), \rho_{\mathbf{r}}(TST^{2}U), \rho_{\mathbf{r}}(T^{2}ST), \rho_{\mathbf{r}}(T^{2}STU)\}\;.
\end{equation}
We are able to construct the hermitian combination $m^{\dagger}_{l}m_{l}$ of the charged lepton mass matrix from its invariance under the residual symmetry $Z^{SU}_{2}\times H^{l}_{CP}$,
\begin{equation}
\label{eq:constraint_reK4}\begin{array}{l}
\rho^{\dagger}_{\mathbf{3}}(SU)m^{\dagger}_{l}m_{l}\rho_{\mathbf{3}}(SU)=m^{\dagger}_{l}m_{l},\\
X^{\dagger}_{l\mathbf{3}}m^{\dagger}_{l}m_{l}X_{l\mathbf{3}}=\left(m^{\dagger}_{l}m_{l}\right)^{*}\,.
\end{array}
\end{equation}
Since $X_{l\mathbf{r}}$ and $\rho_{\mathbf{r}}(g_l)X_{l\mathbf{r}}$ with $g_{l}\in Z^{SU}_2$ lead to the same constraints on the charged lepton mass matrix, as shown in section~\ref{sec:framework}. Two distinct phenomenological predictions arise for the four possible generalized CP transformations in Eq.~\eqref{eq:rem_GCP_reK4}. Firstly we focus on the case of $X_{l\mathbf{r}}=\rho_{\mathbf{r}}(TST^{2}), \rho_{\mathbf{r}}(TST^{2}U)$. The most general $m^{\dagger}_{l}m_{l}$ satisfying Eq.~\eqref{eq:constraint_reK4} is of the following form
\begin{equation}
\label{eq:charged_mass_matrix_one}
 m^{\dagger}_{l}m_{l}=\left(
\begin{array}{ccc}
 \alpha  & (1+i)\beta  & i\epsilon \\
 (1-i)\beta  &  \gamma  &  (1+i)\beta  \\
 -i\epsilon  &  (1-i)\beta  & \alpha
\end{array}
\right)\,,
\end{equation}
where $\alpha$, $\beta$, $\gamma$ and $\epsilon$ are real. It can be diagonalized by the unitary transformation
\begin{equation}
U_{l}=\frac{1}{\sqrt{2}}\left(
\begin{array}{ccc}
e^{\frac{i\pi}{4}}\sin\theta  ~&~ e^{\frac{i\pi}{4}}\cos\theta ~&~ e^{-\frac{i\pi}{4}} \\
-\sqrt{2}\cos\theta ~&~ \sqrt{2}\sin\theta ~&~   0 \\
e^{-\frac{i\pi}{4}}\sin\theta ~&~ e^{-\frac{i\pi}{4}}\cos\theta ~&~ e^{\frac{i\pi}{4}} \\
\end{array}
\right)
\end{equation}
up to rephasings and column permutations, and the angle $\theta$ is specified by
\begin{equation}
\tan2\theta=\frac{4\beta}{\alpha+\epsilon-\gamma}\;.
\end{equation}
The charged lepton masses are
\begin{eqnarray}
\nonumber && m^{2}_{e}=\frac{1}{2}\left[\alpha+\epsilon+\gamma-\mathrm{sign}\left((\alpha+\epsilon-\gamma)\cos(2\theta)\right)\sqrt{16\beta^{2}+(\alpha+\epsilon-\gamma)^{2}}\right]\;,\\
\nonumber && m^{2}_{\mu}=\frac{1}{2}\left[\alpha+\epsilon+\gamma+\mathrm{sign}\left((\alpha+\epsilon-\gamma\right)\cos(2\theta))\sqrt{16\beta^{2}+(\alpha+\epsilon-\gamma)^{2}}\right]\;,\\
\label{eq:ch_mass_CaseIII}&& m^{2}_{\tau}=\alpha-\epsilon\;.
\end{eqnarray}
Combining the unitary transformations $U_{\nu}$ and $U_{l}$ from neutrino and charged lepton sectors, we obtain the predictions for the PMNS matrix:
\begin{equation}
\label{eq:PMNS_caseIII_1oct}U_{PMNS}=U^{\dagger}_{l}U_{\nu}=\frac{1}{2}\left(
\begin{array}{ccc}
\sin\theta+\sqrt{2}\cos\theta ~&~ \sin\theta-\sqrt{2}\cos\theta ~&~ i\sqrt{2}\sin\theta \\
\cos\theta-\sqrt{2}\sin\theta ~&~ \cos\theta+\sqrt{2}\sin\theta ~&~ i \sqrt{2}\cos\theta \\
 1 ~&~ 1 ~&~ -i \sqrt{2} \\
\end{array}
\right)\,,
\end{equation}
The lepton mixing parameters can be straightforwardly extracted as follows
\begin{eqnarray}
\nonumber&\sin\delta_{CP}=\sin\alpha_{21}=\sin\alpha_{31}=0, \\
\label{eq:mix_parameters_caseIII_1oct}&\sin^{2}\theta_{13}=\frac{1}{2}\sin^{2}\theta,\quad \sin^{2}\theta_{12}=\frac{1}{2}-\frac{\sqrt{2}\sin2\theta}{3+\cos2\theta},\quad\sin^{2}\theta_{23}=\frac{1+\cos2\theta}{3+\cos2\theta}\,,
\end{eqnarray}
\begin{figure}[t!]
\begin{center}
\includegraphics[width=0.5\textwidth]{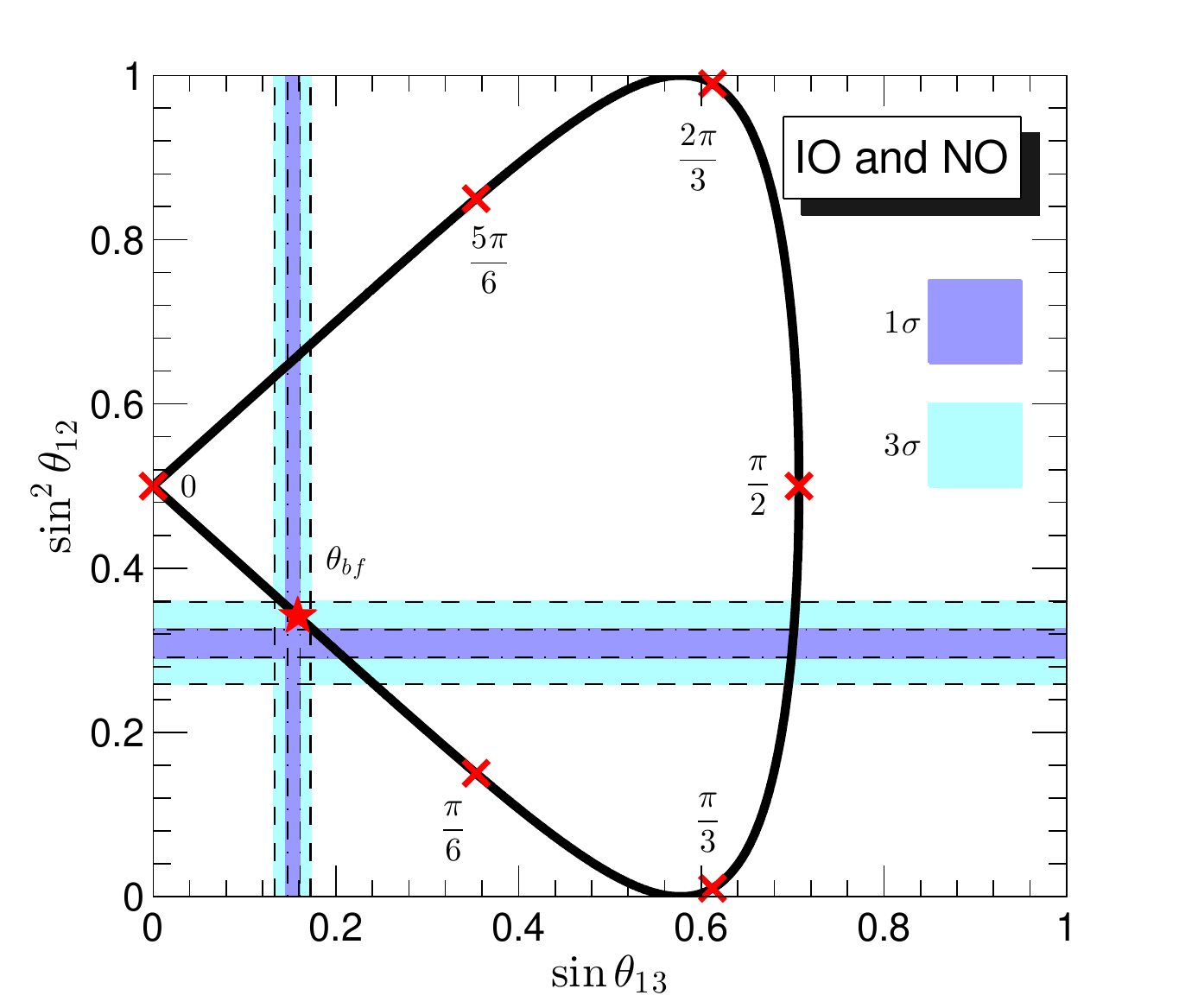}
\caption{\label{fig:best_fit_three}Correlations between $\sin\theta_{13}$ and $\sin^{2}\theta_{12}$ in case III. The best fitting value is marked with a red star, and the points for $\theta=0$, $\pi/6$, $\pi/3$, $\pi/2$, $2\pi/3$ and $5\pi/6$ are labelled with a cross to guide the eye. The shown $1\sigma$ and $3\sigma$ ranges for the mixing angles are taken from Ref.~\cite{Capozzi:2013csa}.}
\end{center}
\end{figure}
where the PDG convention for the lepton mixing angles and CP phases is adopted~\cite{pdg}, $\delta_{CP}$ is the Dirac CP phase, $\alpha_{21}$ and $\alpha_{31}$ stand for the Majorana CP phases. We see that all CP phases are trivial, this is because that a common CP transformation $X_{l\mathbf{r}}=X_{\nu\mathbf{r}}=\rho_{\mathbf{r}}(TST^2)$ is shared by the charged lepton and neutrino sectors. Note that the remaining twenty-three CP transformations in Eq.~\eqref{eq:GCP_trans} are broken by either the neutrino or the charged lepton mass terms. In contrast with the general phenomenological analysis of section~\ref{sec:2_derivation}, the CP phases are predicted to be of definite value $0$ or $\pi$ due to the imposed CP symmetry. Furthermore, the mixing angles are closely related with each other as follows,
\begin{equation}
\label{eq:correlation_caseIII}\sin^2\theta_{12}=\frac{1}{2}\pm\tan\theta_{13}\sqrt{1-\tan^2\theta_{13}},\qquad 2\cos^2\theta_{13}\cos^2\theta_{23}=1\,.
\end{equation}
The measured values of reactor mixing angle $\sin^2\theta_{13}=0.0234$ fixes the parameter $\theta\simeq12.494^{\circ}$, and then the other two mixing angles are determined to be $\sin^2\theta_{12}\simeq0.347$, $\sin^2\theta_{23}\simeq0.488$ which are in the experimentally allowed regions. The correlations among the mixing angles are plotted in Fig.~\ref{fig:best_fit_three} and Fig.~\ref{fig:best_fit_four}. We see that the predictions for the lepton mixing angles agree rather well with their measured values for certain values of the parameter $\theta$. The best fitting results of this mixing pattern for NO (IO) are:
\begin{eqnarray}
\nonumber&\hskip-0.3in\theta_{bf}=0.225 (0.227),\qquad \sin^2\theta_{12}(\theta_{bf})=0.342(0.341),\\
\label{eq:bf_caseIII_1oct}&\hskip-0.3in\sin^2\theta_{13}(\theta_{bf})=0.0250 (0.0253),~~ \sin^2\theta_{23}(\theta_{bf})=0.487(0.487),~~ \chi^2_{min}=6.938(4.288)\,.
\end{eqnarray}
Hence this mixing pattern can describe the experimental data very well, as the global minimum of the $\chi^2$ is quite small: 4.288 for IO and 6.938 for NO spectrum. From Eq.~\eqref{eq:correlation_caseIII}, we have $\sin^2\theta_{23}=1-1/(2\cos^2\theta_{13})<1/2$, namely $\theta_{23}$ is in the first octant, as can be seen from Fig.~\ref{fig:best_fit_four}. The present neutrino oscillation data can not tell us whether $\theta_{23}$ is larger or smaller than $45^{\circ}$. $\theta_{23}$ in the second octant can be achieved by exchanging the second and the thirds rows of the PMNS matrix in Eq.~\eqref{eq:PMNS_caseIII_1oct}. The observed values of the three mixing angles can also be accommodated. Results of the $\chi^2$ analysis are as follows:
\begin{eqnarray}
\nonumber&
\hskip-0.3in\theta_{bf}=0.224 (0.227),\qquad \sin^2\theta_{12}(\theta_{bf})=0.343(0.341),\\
\label{eq:bf_caseIII_2oct}&\hskip-0.3in\sin^2\theta_{13}(\theta_{bf})=0.0248(0.0253),~~\sin^2\theta_{23}(\theta_{bf})=0.513 (0.513 ),~~\chi^2_{min}=9.890(4.409)
\end{eqnarray}
for NO (IO) mass spectrum.

It is notable that the Dirac CP $\delta_{CP}$ is predicted to be conserved here. The present experiments have very low sensitivity to leptonic CP. T2K has recently reported a weak indication for $\delta_{CP}$ around $3\pi/2$~\cite{Abe:2013hdq}. Analysis of the SuperKamiokande atmospheric neutrino data gives preferable range $\left(1.2\pm0.5\right)\pi$~\cite{Himmel:2013jva}. The global analysis of all oscillation data gives $\delta_{CP}=1.39^{+0.38}_{-0.27}\pi (1\sigma)$ for NO and $\delta_{CP}=1.31^{+0.29}_{-0.33}\pi (1\sigma)$ for IO and no restriction appears at $3\sigma$ level~\cite{Capozzi:2013csa}. Hence conserved CP with $\delta_{CP}=0, \pi$ can be accommodated by both present experimental data and global analysis. Future long baseline neutrino experiments LBNE~\cite{Adams:2013qkq}, LBNO~\cite{Autiero:2007zj,Rubbia:2010fm,Angus:2010sz,Rubbia:2010zz,::2013kaa} and Hyper-Kamiokande~\cite{Abe:2011ts} are designed to measure the Dirac phase. If the signal of leptonic CP violation is discovered, our proposal would be ruled out. In addition, the predictions for the atmospheric mixing angle $\theta_{23}$ can be tested by future atmospheric neutrino oscillation experiments such as the India-based Neutrino Observatory.

\begin{figure}[t!]
\begin{center}
\includegraphics[width=0.66\textwidth]{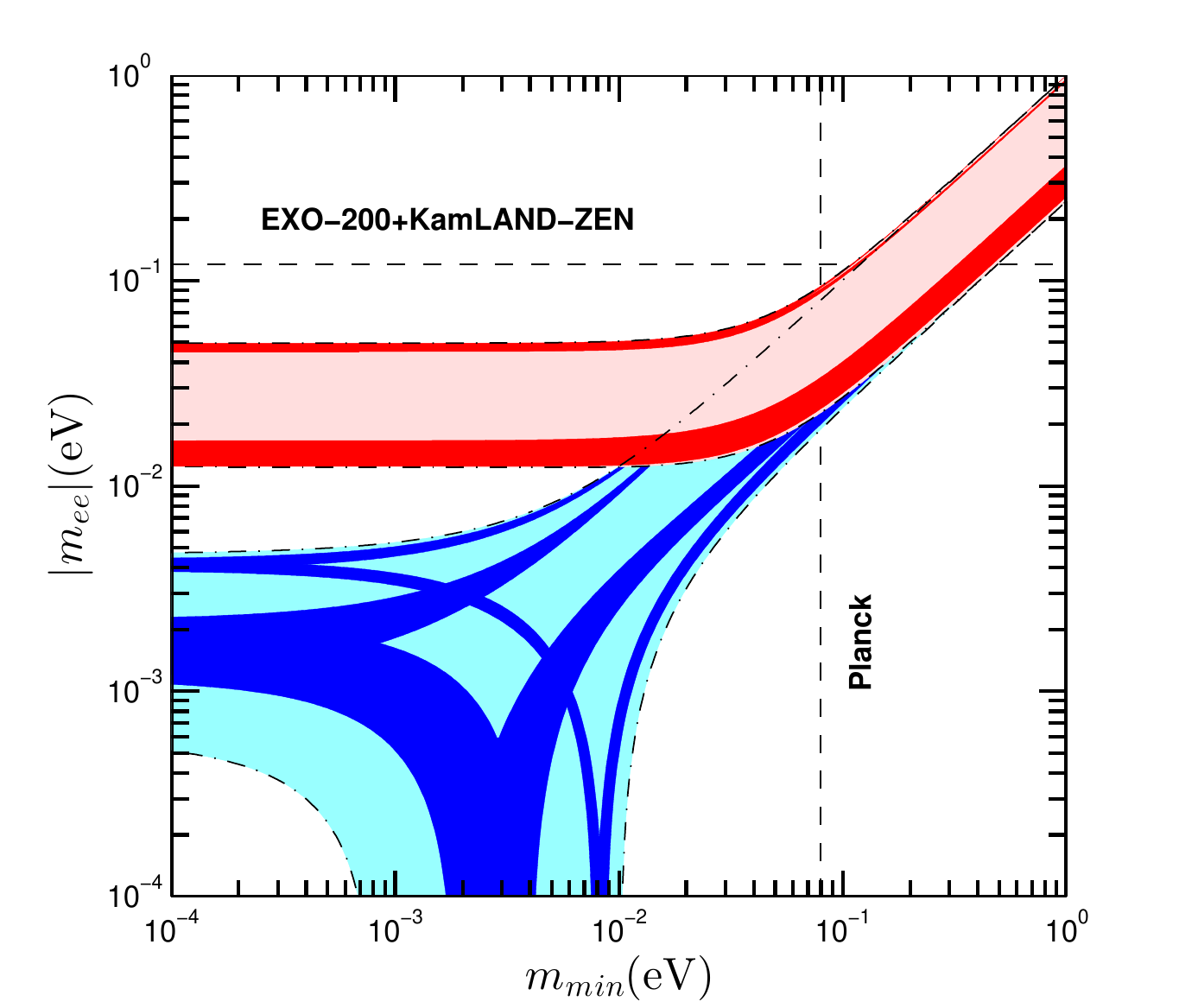}
\caption{\label{fig:onubetabeta_caseIII} The allowed values of the effective mass $\left|m_{ee}\right|$ with respect to the lightest neutrino mass in the case of $Z_2\times CP$ preserved in the charged lepton sector, where the light red and light blue bands denote the regions for the $3\sigma$ ranges of the oscillation parameters in the inverted and normal neutrino mass spectrum respectively~\cite{Capozzi:2013csa}. The red and blue regions are the predictions for inverted hierarchy and normal hierarchy with the PMNS matrix given in Eq.~\eqref{eq:PMNS_caseIII_1oct}. The upper bound $|m_{ee}|<(0.120-0.250)$ eV comes from the combination of the EXO-200~\cite{Auger:2012ar, Albert:2014awa} and KamLAND-ZEN experiments~\cite{Gando:2012zm}. The upper limit on the mass of the lightest neutrino is derived from the latest Planck result $m_1+m_2+m_3<0.230$ eV at $95\%$ confidence level~\cite{Ade:2013zuv}.}
\end{center}
\end{figure}

Furthermore, the predictions for the conserved Dirac and Majorana CP phases in Eq.~\eqref{eq:mix_parameters_caseIII_1oct} can be checked by the neutrinoless double beta ($0\nu\beta\beta$) decay experiment which is an important probe for the Majorana nature of neutrino and lepton number violation. It is well-known that the $0\nu\beta\beta$-decay amplitude depends on the following effective Majorana mass:
\begin{equation}
\left|m_{ee}\right|=\left|(m_1 c_{12}^2+m_2s^2_{12}e^{i \alpha_{21}})c_{13}^2+m_3 s_{13}^2 e^{i (\alpha_{31}-2\delta_{CP})}\right|\,,
\end{equation}
where $c_{ij}\equiv\cos\theta_{ij}$ and $s_{ij}\equiv\sin\theta_{ij}$. The predictions for the effective mass are plotted in Fig.~\ref{fig:onubetabeta_caseIII}. We see that $\left|m_{ee}\right|$ is determined to be around the $3\sigma$ upper limit (0.049eV) or lower limit (0.013eV) for inverted hierarchy, which is within the reach of the forthcoming $0\nu\beta\beta$ experiments. A large region of possible values of $|m_{ee}|$ is allowed in case of NO, and $|m_{ee}|$ could be rather small depending on the value of the lightest neutrino mass. Consequently this mixing pattern would be preferred if the $\left|m_{ee}\right|$ is measured to be close to 0.049eV or 0.013eV in future. Note that the effective mass $\left|m_{ee}\right|$ doesn't depend on $\theta_{23}$, and therefore it is invariant under the exchange of the 2nd and the 3rd rows of the PMNS matrix.

The phenomenological implications for the remaining two remnant CP transformations $X_{l\mathbf{r}}=\rho_{\mathbf{r}}(T^{2}ST), \rho_{\mathbf{r}}(T^{2}STU)$ can be studied in the same way. However, we find that the observed values of the three lepton mixing angles can not be fitted simultaneously. Hence this case will not be discussed in detail.

In short, the perturbative rotations to the BM mixing from the charged lepton sector, which is discussed in section~\ref{sec:2_derivation}, can be realized by breaking the $S_4$ family symmetry to a $Z_2$ subgroup in the charged lepton and to $K_4$ in the neutrino sector.  By further extending the $S_4$ family symmetry to consistently include generalized CP symmetry, the phase $\delta$ of the perturbative rotation can not take arbitrary value anymore. We have definite predictions for the leptonic CP phases: both Dirac CP phase and Majorana CP phases are trivial in order to fit the data of mixing angles.

\section{\label{sec:model_column}Model predicting one column of BM mixing with $S_4$ and generalized CP}
\cleqn

\begin{figure}[t!]
\begin{center}
\includegraphics[width=0.66\textwidth]{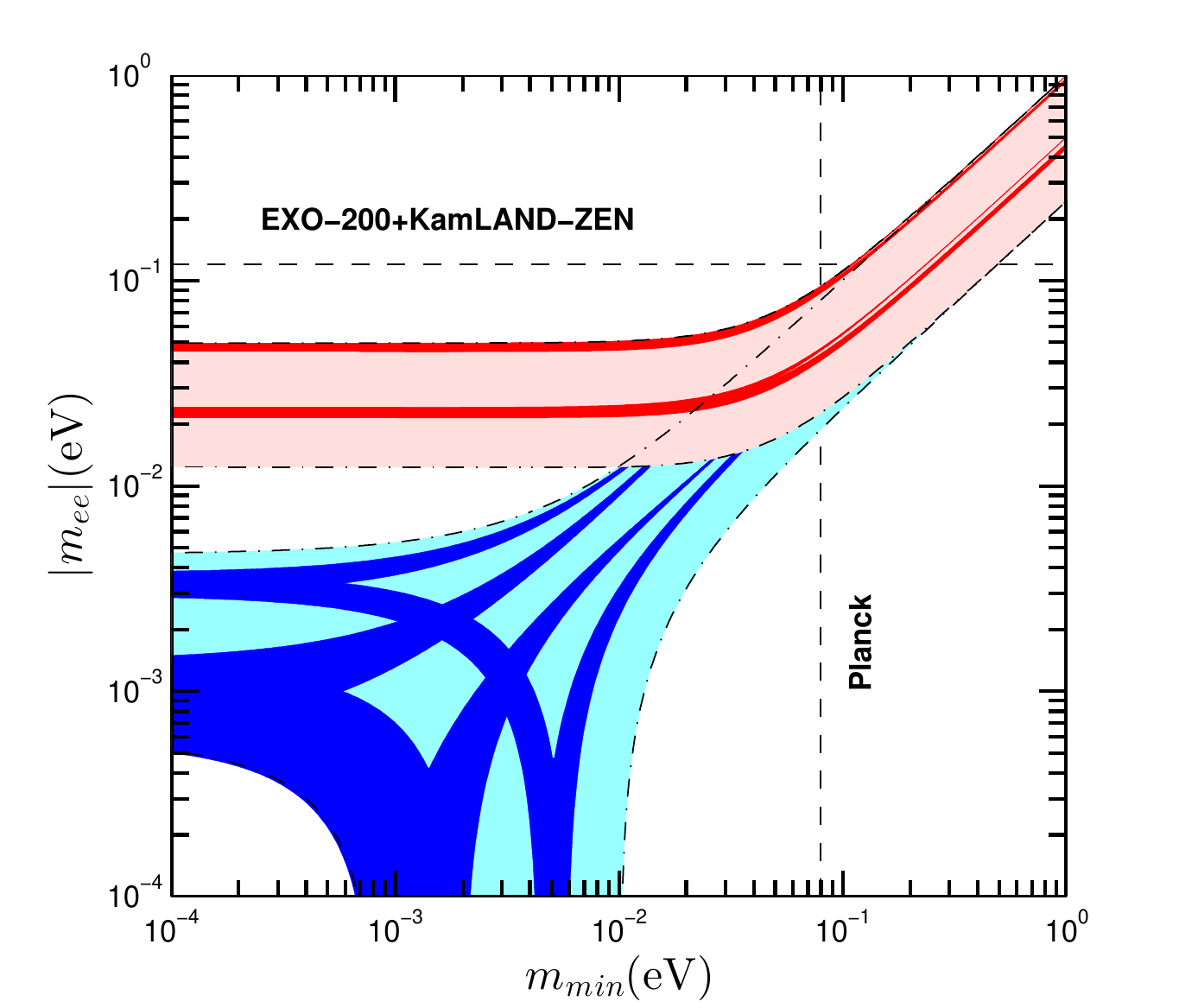}
\caption{\label{fig:onubetabeta_caseI_new} The allowed values of the effective mass $\left|m_{ee}\right|$ in the case of $Z_2\times CP$ preserved in the neutrino sector, where the light red and light blue bands denote the regions for the $3\sigma$ ranges of the oscillation parameters in the inverted and normal neutrino mass spectrum respectively~\cite{Capozzi:2013csa}. The red and blue regions are the predictions for inverted hierarchy and normal hierarchy with the PMNS matrix given in Eq.~\eqref{eq:UPMNS_one}. The upper bound $|m_{ee}|<(0.120-0.250)$ eV comes from the combination of the EXO-200~\cite{Auger:2012ar,Albert:2014awa} and KamLAND-ZEN experiments~\cite{Gando:2012zm}. The upper limit on the mass of the lightest neutrino is derived from the latest Planck result $m_1+m_2+m_3<0.230$ eV at $95\%$ confidence level~\cite{Ade:2013zuv}.}
\end{center}
\end{figure}

The scenario of the $S_4$ flavor symmetry breaking to $Z_2$ and $Z_4$ subgroups in the neutrino and charged lepton sectors respectively with generalized CP symmetry has been investigated in Ref.~\cite{Feruglio:2012cw}.
In terms of the notation of present work, the representative residual symmetries can be chosen to be $Z^{ST^2SU}_2\times H^{\nu}_{CP}$ in the neutrino sector and $Z^{TST^{2}U}_4\rtimes H^{l}_{CP}$ in the charged lepton sector, where $H^{\nu}_{CP}=\{\rho_{\mathbf{r}}(1), \rho_{\mathbf{r}}(ST^{2}SU)\}$ and $H^{l}_{CP}=\{\rho_{\mathbf{r}}(1),\rho_{\mathbf{r}}(TST^2U),\rho_{\mathbf{r}}(S), \rho_{\mathbf{r}}(T^{2}STU)\}$. Before presenting the model, we shall firstly review the lepton mixing arising from this breaking pattern. Note that the residual CP transformations $H^\nu_{CP}=\{ \rho_{\mathbf{r}}(T^{2}U) ,\rho_{\mathbf{r}}(TST^2)\}$ are also compatible with the remnant flavor symmetry $Z^{ST^2SU}_2$. However the measured values of the three mixing angles can not be accommodated simultaneously in that case. As the representation matrix of $TST^2U$ is diagonal in all irreducible representations of $S_4$, the hermitian combination $m^{\dagger}_{l}m_{l}$ is diagonal. Hence lepton flavor mixing completely arises from the neutrino sector. Straightforward calculations demonstrate that the neutrino mass matrix preserving $Z^{ST^2SU}_2\times H^{\nu}_{CP}$ is of the following form:
\begin{eqnarray}
\label{eq:gen_mass_nu}
m_{\nu}=\alpha\left(
\begin{array}{ccc}
 0 & 0 & 1 \\
 0 & 1 & 0 \\
 1 & 0 & 0
\end{array}
\right)+\beta \left(
\begin{array}{ccc}
 -3 & 0 & 1 \\
 0 & -2 & 0 \\
 1 & 0 & -3
\end{array}
\right)+\gamma\left(
\begin{array}{ccc}
 0 & 1 & 0 \\
 1 & 0 & 1 \\
 0 & 1 & 0
\end{array}
\right)+\epsilon\left(
\begin{array}{ccc}
 \sqrt{2} &~ -1 & 0 \\
 -1 &~ 0 & 1 \\
 0 &~ 1 & -\sqrt{2}
\end{array}
\right)\,,
\end{eqnarray}
where the all four parameters $\alpha$, $\beta$, $\gamma$ and $\epsilon$ are real. The lepton mixing matrix $U_{PMNS}$, which diagonalizes the neutrino mass matrix in Eq.~\eqref{eq:gen_mass_nu}, is determined to be of the form
\begin{equation}
\label{eq:UPMNS_one}
U_{PMNS}=\frac{1}{2}\left(
\begin{array}{ccc}
\sin\theta+\sqrt{2}\cos\theta & ~1~ & \cos\theta-\sqrt{2}\sin\theta\\
-\sqrt{2}\sin\theta & ~\sqrt{2}~ & -\sqrt{2}\cos\theta \\
\sin\theta-\sqrt{2}\cos\theta & ~1~ & \cos\theta+\sqrt{2}\sin\theta
\end{array}
\right)K_{\nu}\,,
\end{equation}
up to row and column permutations, where $K_{\nu}$ is a unitary diagonal matrix with entries $\pm1$ or $\pm i$ which renders the light neutrino masses positive. The rotation angle $\theta$ is given by
\begin{eqnarray}
\tan2\theta=\frac{-4\epsilon}{2\alpha+2\beta-\sqrt{2}\,\gamma}\,.
\end{eqnarray}
The light neutrino masses are
\begin{eqnarray}
\nonumber &&\hskip-0.3in m_1=\frac{1}{2}\left|6\beta+\sqrt{2}\gamma+\text{sign}((2\alpha+2\beta-\sqrt{2}\gamma)\cos2\theta)\sqrt{16\epsilon^2+(2\alpha+2\beta-\sqrt{2}\gamma)^2}\right|,\\
\nonumber &&\hskip-0.3in m_2=|\alpha -2 \beta +\sqrt{2} \gamma|,  \\
\label{eq:neutrino_mass_caseI}&&\hskip-0.3in m_3=\frac{1}{2}\left|6\beta+\sqrt{2}\gamma-\text{sign}((2\alpha+2\beta-\sqrt{2}\gamma)\cos2\theta)\sqrt{16\epsilon^2+(2\alpha+2\beta-\sqrt{2}\gamma)^2}\right|\,.
\end{eqnarray}
Notice that the mixing pattern with one column $(1/2, 1/\sqrt{2}, 1/2)^{T}$ has been proposed in Ref.~\cite{Hernandez:2012ra}, where the scenario of only one $Z_2$ symmetry imposed in the neutrino sector was analyzed in a general way. The lepton mixing angles and CP phases can be read out from Eq.~\eqref{eq:UPMNS_one} as follow
\begin{eqnarray}
\nonumber \sin^{2}\theta_{13}&&=\frac{1}{8}(3-\cos2\theta-2\sqrt{2}\sin2\theta)\,,\qquad \sin^{2}\theta_{12}=\frac{2}{5+\cos2\theta+2\sqrt{2}\sin2\theta}\,,\\ \label{eq:mixing_paramaters_nu1}\sin^{2}\theta_{23}&&=\frac{4\cos^{2}\theta}{5+\cos2\theta+2\sqrt{2}\sin2\theta}\,,\quad \sin\delta_{CP}=\sin\alpha_{21}=\sin\alpha_{31}=0\,,
\end{eqnarray}
which match with the expressions of mixing parameters in Ref.~\cite{Feruglio:2012cw} after parameter redefinition $\theta\rightarrow\pi/4+\theta$. We see that the three mixing angles are strongly correlated with each other
\begin{equation}
\label{eq:correaltion_caseI}4\cos^2\theta_{13}\sin^2\theta_{12}=1,\quad \sin^2\theta_{23}=\frac{1}{3}+\frac{\tan{\theta_{13}}}{9}\left(\tan{\theta_{13}}\pm2\sqrt{6-2\tan^{2}{\theta_{13}}}\right)\,.
\end{equation}
The measured value of the reactor angle $\sin^{2}{\theta_{13}}=0.0234$~\cite{Capozzi:2013csa} can be reproduced for $\theta\simeq25.091^{\circ}$. With this value of $\theta$, $\sin^{2}{\theta_{12}}\simeq0.256$ and $\sin^2\theta_{23}\simeq0.420$ follow from Eq.~\eqref{eq:mixing_paramaters_nu1}. We see that $\theta_{23}$ in the experimentally preferred regions can be achieved while $\theta_{12}$ is predicted to be quite close to its $3\sigma$ lower bound~\cite{Capozzi:2013csa} due to the correlation with $\theta_{13}$ shown in Eq.~\eqref{eq:correaltion_caseI}. As has been emphasized in section~\ref{sec:2_derivation}, agreement with experimental data can be easily achieved after subleading corrections are included. A concrete model realization of this scenario is presented in the following. In order to see quantitatively to which extent this mixing pattern can accommodate the present experimental data, we perform a conventional $\chi^2$ analysis. The minimum of the $\chi^2$ is $\chi^2_{\text{min}}=9.865$ for NO and 10.454 for IO with
\begin{eqnarray}
\nonumber&\theta_{bf}=0.436 (0.434),\qquad \sin^2\theta_{12}(\theta_{bf})=0.256(0.256), \\
\label{eq:best_fit_Fir_caseIII}&\sin^2\theta_{13}(\theta_{bf})= 0.0238(0.0244),\qquad \sin^2\theta_{23}(\theta_{bf})=0.421(0.422)\,,
\end{eqnarray}
where the number before (in) the parenthesis denotes the best fitting value for NO (IO) spectrum. Obviously $\theta_{23}$ in the first octant is favored, and $\theta_{23}$ in the second octant can also be accommodated by interchanging the second and third rows of Eq.~\eqref{eq:UPMNS_one}. The resulting predictions for $\theta_{13}$, $\theta_{12}$ and CP phases remain the same as those given by Eq.~\eqref{eq:mixing_paramaters_nu1}, while $\theta_{23}$ becomes its complementary angle. The best fitting results are as follows:
\begin{eqnarray}
\nonumber&\hskip-0.3in\theta_{bf}=0.433(0.435),\qquad \sin^2\theta_{12}(\theta_{bf})=0.256(0.256),\\
\label{eq:best_fit_Sec_caseIII}&\hskip-0.3in\sin^2\theta_{13}(\theta_{bf})=0.0246(0.0242),~~ \sin^2\theta_{23}(\theta_{bf})=0.578(0.579),~~ \chi^2_{\mathrm{min}}=27.807(10.086)\,,
\end{eqnarray}
for NO (IO) neutrino mass spectrum. Furthermore, the predictions for the effective mass $|m_{ee}|$ is plotted in Fig.~\ref{fig:onubetabeta_caseI_new}. $\left|m_{ee}\right|$ is found to be around 0.049eV or 0.023eV for IO spectrum, which can be tested by forthcoming $0\nu\beta\beta$ decay experiments. Nevertheless the allowed regions of $|m_{ee}|$ are somewhat complex for NO spectrum, and the effective mass can be very small for certain values of the lightest neutrino mass. In concrete models where the mixing pattern in Eq.~\eqref{eq:mixing_paramaters_nu1} is produced at leading order, $|m_{ee}|$ could lie outside of the red and blue areas of Fig.~\ref{fig:onubetabeta_caseI_new} after possible subleading order corrections are considered. Depending on the specific form of the corrections and how large they are, different predictions for $|m_{ee}|$ can be obtained.

In the following we shall construct a model based on $S_{4}$ family symmetry and generalized CP symmetry $H_{CP}$. The auxiliary symmetry $Z_3\times Z_4$ is introduced to disentangle the flavon fields associated with the neutrino sector from those associated with the charged lepton sector and to eliminate unwanted dangerous operators. By a judicious choice of flavons, the above discussed symmetry breaking pattern of $S_4\rtimes H_{CP}$ into $Z^{ST^2SU}_2\times H^{\nu}_{CP}$ and $Z^{TST^{2}U}_4\rtimes H^{l}_{CP}$ is explicitly realized at leading order. As a result, the interesting mixing texture in Eq.~\eqref{eq:UPMNS_one} is reproduced exactly in this model, and realistic $\theta_{12}$ can be achieved after higher order corrections are included. This model is formulated in the context of supersymmetry. We assign the three generations of left-handed lepton doublets $l$ and of right-handed neutrino $\nu^{c}$ to $S_4$ triplet $\mathbf{3}$. The right-handed charged leptons $e^{c}$, $\mu^{c}$ and $\tau^{c}$ are singlet states of $S_{4}$, and they transform as $\mathbf{1}$, $\mathbf{1^{\prime}}$ and $\mathbf{1}$ respectively. The matter fields, flavon fields, driving fields and their transformation properties under the family symmetry $S_{4}\times Z_{3}\times Z_{4}\times U(1)_{R}$ are summarized in Table~\ref{tab:fields_model_one_column}.

\begin{table}[t!]
\renewcommand{\tabcolsep}{2.0mm}
\centering
\begin{tabular}{|c||c|c|c|c|c|c||c|c|c|c|c||c|c|c|c|}
\hline \hline
Field &   $l$    &  $\nu^{c}$     &  $e^{c}$     &   $\mu^{c}$    &    $\tau^{c}$  &  $h_{u,d}$   &  $\varphi_T$ & $\phi$ &  $\xi$ &  $\eta $ & $\varphi_{S}$     & $\rho^{0}$ & $\varphi^{0}_T$ &  $\eta^{0}$  &   $\varphi^{0}_S$     \\ \hline

$S_4$  &   $\mathbf{3}$  &  $\mathbf{3}$  &  $\mathbf{1}$  & $\mathbf{1}^\prime$   &  $\mathbf{1}$   & $\mathbf{1}$   &   $\mathbf{3}$ & $\mathbf{3}^\prime$ &   $\mathbf{1}$ &  $\mathbf{2}$  &  $\mathbf{3}$    &  $\mathbf{2}$  &  $\mathbf{3}$ &   $\mathbf{2}$ & $\mathbf{3}$     \\ \hline

$Z_{3}$  & $\omega_3$  & $1$  &   $\omega^2_3$   & $\omega^2_3$   &  $\omega^2_3$  &  $1$   &  $1$   & $1$  & $\omega^2_3$   &   $\omega^2_3$   & $\omega^2_3$   &    $1$  &  $1$   &  $\omega^2_3$ & $\omega^2_3$ \\ \hline

$Z_{4}$  & $-1$  & $1$  &  $-i$   & $1$   &  $i$  &  $1$   &  $i$   & $i$  &  $-1$   &   $-1$   & $-1$   &    $-1$  &  $-1$   &  $1$ & $1$ \\ \hline

$U(1)_R$  &  $1$  & $1$  & $1$  &   $1$  & $1$  &  $0$ &    $0$  &  $0$ & $0$  & $0$  & $0$   &  $2$ &  $2$ &   $2$ & $2$ \\ \hline \hline
\end{tabular}
\caption{\label{tab:fields_model_one_column}The field contents and their classification under the family symmetry $S_4\times Z_{3}\times Z_{4}$ and $U(1)_R$, where $\omega_{3}=e^{i2\pi/3}$.}
\end{table}

\subsection{\label{subsec:vacuum_alignment_one}Vacuum alignment}

The issue of vacuum alignment is handled with the help supersymmetric driving field mechanism~\cite{Altarelli:2005yx}. This approach utilises a global $U(1)_{R}$ continuous symmetry which contains the discrete $R-$parity as a subgroup.  The flavon and Higgs fields are uncharged under $U(1)_{R}$, the matter fields carry $R$ charge $+1$ and the driving fields $\rho^{0}$, $\varphi^{0}_{T}$, $\eta^{0}$ and $\varphi^{0}_{S}$ carry two units of $R$ charge. Consequently all terms in the superpotential  either contain two matter superfields or one driving field. The leading order (LO) driving superpotential $w_{d}$ invariant under the family symmetry $S_{4}\times Z_{3}\times Z_{4}$ is of the form
\begin{eqnarray}
\label{eq:wd_one} w_{d}=w^{l}_{d}+w^{\nu}_{d}\,,
\end{eqnarray}
where $w^{l}_{d}$ and $w^{\nu}_{d}$ are responsible for the LO vacuum alignment of the charged lepton sector and neutrino sector respectively, and they can be expressed as
\begin{eqnarray}
\nonumber&&  w^{l}_{d}=f_{1}(\rho^{0}(\varphi_{T}\varphi_{T})_{\bf2})_{\bf1}+f_{2}(\rho^{0}(\varphi_{T}\phi)_{\bf2})_{\bf1}+f_{3}(\rho^{0}(\phi\phi)_{\bf2})_{\bf1}+f_{4}(\varphi^{0}_{T}(\varphi_{T}\varphi_{T})_{\bf3})_{\bf1}\\
\label{eq:driving_one} &&\qquad+f_{5}(\varphi^{0}_{T}(\varphi_{T}\phi)_{\bf3})_{\bf1}+f_{6}(\varphi^{0}_{T}(\phi\phi)_{\bf3})_{\bf1},\\
\nonumber&&w^{\nu}_{d}=g_{1}\xi (\eta^{0}\eta)_{\bf1}+g_{2}(\eta^{0}(\eta\eta)_{\bf2})_{\bf1}+g_{3}(\eta^{0}(\varphi_{S}\varphi_{S})_{\bf2})_{\bf1}+g_{4}\xi (\varphi^{0}_{S}\varphi_{S})_{\bf1}+g_{5}(\varphi^{0}_{S}(\eta\varphi_{S})_{\bf3})_{\bf1}\\
\label{eq:driving_two}&&\qquad+g_{6}(\varphi^{0}_{S}(\varphi_{S}\varphi_{S})_{\bf3})_{\bf1}\,,
\end{eqnarray}
where the subscripts $\mathbf{1}$, $\mathbf{2}$, $\mathbf{3}$ etc stand for contractions into the corresponding $S_{4}$ irreducible representations. Note that the terms proportional to $f_4$, $f_6$ and $g_6$ give null contributions because of the antisymmetric contractions $\left(\mathbf{3}\otimes\mathbf{3}\right)_{\mathbf{3}}$ and $\left(\mathbf{3^{\prime}}\otimes\mathbf{3^{\prime}}\right)_{\mathbf{3}}$. As we require the theory invariant under the generalized CP transformations defined in Eq.~\eqref{eq:GCP_trans},  all couplings $f_{i}$ and $g_{i}$ would be real. The driving field is
assumed to have vanishing vacuum expectation value (VEV). In the limit of unbroken supersymmetry, the vacuum configuration is fixed by the vanishing $F-$term of the driving field. For the vacuum alignment of the charged lepton sector, we have
\begin{eqnarray}
\nonumber&&\frac{\partial w^{l}_{d}}{\partial\rho^{0}_{1}}=2f_{1}(\varphi^{2}_{T_{2}}-\varphi_{T_{1}}\varphi_{T_{3}})+\sqrt{3}f_{2}(\varphi_{T_{1}}\phi_{1}+\varphi_{T_{3}}\phi_{3})+2f_{3}(\phi^{2}_{2}-\phi_{1}\phi_{3})=0,\\
\nonumber&&\frac{\partial w^{l}_{d}}{\partial\rho^{0}_{2}}=\sqrt{3}f_{1}(\varphi^{2}_{T_{1}}+\varphi^{2}_{T_{3}})+f_{2}(\varphi_{T_{1}}\phi_{3}-2\varphi_{T_{2}}\phi _{2}+\varphi_{T_{3}}\phi_{1})+\sqrt{3}f_{3}(\phi^{2}_{1}+\phi^{2}_{3})=0,\\
\nonumber&&\frac{\partial w^{l}_{d}}{\partial\varphi^{0}_{T_{1}}}=f_{5}(\varphi_{T_{1}}\phi_{2}+\varphi_{T_{2}}\phi_{1})=0,\\
\nonumber&&\frac{\partial w^{l}_{d}}{\partial\varphi^{0}_{T_{2}}}=f_{5}(\varphi_{T_{1}}\phi_{1}-\varphi_{T_{3}}\phi_{3})=0,\\
&&\frac{\partial w^{l}_{d}}{\partial\varphi^{0}_{T_{3}}}=-f_{5}(\varphi_{T_{2}}\phi_{3}+\varphi_{T_{3}}\phi_{2})=0\,.
\end{eqnarray}
This set of equations are satisfied by the alignment:
\begin{equation}
\label{eq:vacuum_one}
\langle\varphi_{T}\rangle=\left(\begin{array}{l}
    1 \\
    0  \\
    0
\end{array}\right)v_{T}, \quad
\langle\phi\rangle =\left(\begin{array}{l}
   0 \\
   0  \\
   1
\end{array}\right)v_{\phi}, \quad
v_{\phi}=-\frac{f_{2}\pm\sqrt{f^{2}_{2}-12f_{1}f_{3}}}{2\sqrt{3}f_{3}}v_{T}\,.
\end{equation}
The VEVs $v_{\phi}$ and $v_{T}$ are naturally of the same order of magnitude, since they are related through the couplings $f_{1}$, $f_2$ and $f_{3}$ which are expected to have absolute values of order one. To reproduce the observed hierarchy among the charged lepton masses, we choose
\begin{equation}
\frac{v_{\phi}}{\Lambda}\sim\frac{v_{T}}{\Lambda}\sim\lambda^2\,,
\end{equation}
where $\lambda\approx 0.23$ is the Cabibbo angle~\cite{pdg}. The $F-$term conditions of the neutrino sector are
\begin{eqnarray}
\nonumber&&\frac{\partial w^{\nu}_{d}}{\partial\eta^{0}_{1}}=g_{1}\xi\eta_{1}+g_{2}(\eta^{2}_{2}-\eta^{2}_{1})+2g_{3}(\varphi^{2}_{S_{2}}-\varphi_{S_{1}}\varphi_{S_{3}})=0,\\
\nonumber&&\frac{\partial w^{\nu}_{d}}{\partial\eta^{0}_{2}}=g_{1}\xi\eta_{2}+2g_{2}\eta_{1}\eta_{2}+\sqrt{3}\,g_{3}(\varphi^{2}_{S_{1}}+\varphi^{2}_{S_{3}})=0,\\
\nonumber&&\frac{\partial w^{\nu}_{d}}{\partial\varphi^{0}_{S_{1}}}=g_{4}\xi\varphi_{S_{3}}+g_{5}(\sqrt{3}\,\eta_{2}\varphi_{S_{1}}-\eta_{1}\varphi_{S_{3}})=0,\\
\nonumber&&\frac{\partial w^{\nu}_{d}}{\partial\varphi^{0}_{S_{2}}}=g_{4}\xi\varphi_{S_{2}}+2g_{5}\eta_{1}\varphi_{S_{2}}=0,\\
&&\frac{\partial w^{\nu}_{d}}{\partial\varphi^{0}_{S_{3}}}=g_{4}\xi\varphi_{S_{1}}+g_{5}(\sqrt{3}\,\eta_{2}\varphi_{S_{3}}-\eta_{1}\varphi_{S_{1}})=0\,.
\end{eqnarray}
It is then straightforward to work out the most general solutions to these equations. Disregarding the ambiguity caused by $S_{4}$ family symmetry transformations we find three possible non-trivial solutions. The first one is given by
\begin{equation}
\label{eq:vacuum1_model1}\langle\xi\rangle=v_{\xi}, \qquad \langle\eta\rangle=\left(\begin{array}{c}
  1 \\
  0
\end{array}\right)v_{\eta}, \qquad
\langle\varphi_{S}\rangle=\left(\begin{array}{c}
  0  \\
  1  \\
 0
\end{array}\right)v_{S}\,,
\end{equation}
with
\begin{equation}
v_{\eta}=-\frac{g_{4}v_{\xi}}{2g_{5}},\qquad v^{2}_{S}=\frac{g_{4}(2g_{1}g_{5}+g_{2}g_{4})}{8g_{3}g^{2}_{5}}v^{2}_{\xi}\,,
\end{equation}
where $v_{\xi}$ is undetermined and generally complex. Given the representation matrices in Appendix~\ref{sec:appA}, it is easy to check that this vacuum breaks the $S_4$ family symmetry to $Z^{T^{2}STU}_{4}$. The second solution is
\begin{equation}
\label{eq:vacuum2_model1}\langle\xi\rangle=v_{\xi}, \qquad \langle\eta\rangle=\left(\begin{array}{c}
   1  \\
 \sqrt{3}
\end{array}\right)v_{\eta}, \qquad
\langle\varphi_{S}\rangle=\left(\begin{array}{c}
   1  \\
   0  \\
  -1
\end{array}\right)v_{S}\,,
\end{equation}
with
\begin{equation}
v_{\eta}=\frac{g_{4}v_{\xi}}{4g_{5}},\qquad v^{2}_{S}=-\frac{g_{4}(2g_{1}g_{5}+g_{2}g_{4})}{16g_{3}g^{2}_{5}}v^{2}_{\xi}\,.
\end{equation}
The residual family symmetry $Z^{T^{2}SU}_{4}$ is preserved by this alignment. The third takes the form
\begin{equation}
\label{eq:vacuum_two}
\langle\xi\rangle=v_{\xi}, \qquad  \langle\eta\rangle =\left(\begin{array}{c}
    1  \\
  \sqrt{3}
\end{array}\right)v_{\eta}, \qquad
\langle\varphi_{S}\rangle=\left(\begin{array}{c}
   1  \\
\sqrt{2}  \\
   1
\end{array}\right)v_{S}\,,
\end{equation}
where
\begin{equation}
\label{eq:relation_m1}v_{\eta}=-\frac{g_{4}v_{\xi}}{2g_{5}},\qquad v^{2}_{S}=\frac{g_{4}(g_{1}g_{5}-g_{2}g_{4})}{4g_{3}g^{2}_{5}} v^{2}_{\xi},\qquad v_{\xi}~~ \mathrm{undetermined} \,,
\end{equation}
We see that the two VEVs $v_{\eta}$ and $v_{\xi}$ share the same phase modulo $\pi$, while the phase difference between $v_{S}$ and $v_{\xi}$ is $0, \pi$ for $g_{3}g_{4}(g_{1}g_{5}-g_{2}g_{4})>0$ or $\pm\pi/2$ for $g_{3}g_{4}(g_{1}g_{5}-g_{2}g_{4})<0$. Since the phase of $v_{\xi}$ can always be absorbed by lepton fields, we could take $v_{\xi}$ to be real without loss of generality. Consequently $v_{\eta}$ is real, and $v_{S}$ is either real or pure imaginary depending on the combination $g_{3}g_{4}(g_{1}g_{5}-g_{2}g_{4})$ being positive or negative. We find that the symmetry $S_{4}\rtimes H_{CP}$ is broken to $Z^{ST^{2}SU}_{2}\times H^{\nu}_{CP}$ by the VEVs of $\xi$, $\eta$ and $\varphi_{S}$, where the remnant CP symmetry $H^{\nu}_{CP}=\{\rho_{\bf3}(1),\rho_{\bf3}(ST^{2}SU)\}$ for real $v_{S}$ and $H^{\nu}_{CP}=\{\rho_{\bf3}(T^{2}U),\rho_{\bf3}(TST^{2})\}$ for pure imaginary $v_{S}$. Only the third solution can allow us to derive the interesting mixing texture of Eq.~\eqref{eq:UPMNS_one} discussed above. Since the three vacuum configurations in Eq.~\eqref{eq:vacuum1_model1}, Eq.~\eqref{eq:vacuum2_model1} and Eq.~\eqref{eq:vacuum_two} are degenerate in the supersymmetric limit, supersymmetry breaking effects are needed to discriminate the last one as the the lowest minimum of the scalar potential. Here we consider the possibility of lifting the vacuum degeneracy by the soft supersymmetry breaking mass terms. The most general $Z_3-$breaking soft mass terms involving $\xi$, $\eta$ and $\varphi_{S}$ can be written as
\begin{equation}
\mathcal{L}^{m}_{soft}=m^2_{\xi}|\xi|^2+m^2_{\eta}|\eta|^2+m^2_{\varphi}|\varphi_S|^2+\widetilde{m}^2_{\xi}\xi^2+\widetilde{m}^2_{\eta}\left(\eta\eta\right)_{\mathbf{1}}+\widetilde{m}^2_{\varphi}\left(\varphi_{S}\varphi_{S}\right)_{\mathbf{1}}\,,
\end{equation}
where we assume $m^2_{\xi, \eta, \varphi}<0$ such that the first three terms stabilize the potential for all the three vacuum solutions. One can straightforwardly check that these soft mass terms take the same value for the first two vacuum in Eqs.~(\ref{eq:vacuum1_model1}, \ref{eq:vacuum2_model1}), and obtain another different value for the third vacuum in Eq.~\eqref{eq:vacuum_two}. With an appropriate choice of the
soft parameters, it is possible to distinguish the three configurations and assure the desired one in Eq.~\eqref{eq:vacuum_two} as the setting with the lowest minimum. Furthermore, the three VEVs $v_{\xi}$, $v_{\eta}$ and $v_{S}$ are expected to be of the same order of magnitude without fine tuning among the parameters $g_{i} (i=1,2,3,4,5)$. As usual, we shall take them to be of the same order as the VEVs of charged lepton sector flavons, i.e.
\begin{equation}
 \frac{v_{\xi}}{\Lambda}\sim\frac{v_{\eta}}{\Lambda}\sim\frac{v_{S}}{\Lambda}\sim\lambda^{2}\,.
\end{equation}

\subsection{\label{subsec:model_one}The structure of the model}

The superpotential for the charged lepton masses is
\begin{eqnarray}
\nonumber w_{l}&=&\frac{y_{\tau}}{\Lambda}\tau^{c}(l\varphi_{T})_{\bf1}h_{d}+\frac{y_{\mu_1}}{\Lambda^{2}}\mu^{c}(l(\varphi_{T}\varphi_{T})_{\bf3^{\prime}})_{\bf1^{\prime}}h_{d}
+\frac{y_{\mu_2}}{\Lambda^{2}}\mu^{c}(l(\varphi_{T}\phi)_{\bf3^{\prime}})_{\bf1^{\prime}}h_{d} \\
\label{eq:wl_one}&& +\frac{y_{\mu_3}}{\Lambda^{2}}\mu^{c}(l(\phi\phi)_{\bf3^{\prime}})_{\bf1^{\prime}}h_{d}+\sum^{4}_{i=1}\frac{y_{e_i}}{\Lambda^{3}}e^{c}(lO_{i})_{\bf1}h_{d}+\ldots\,,
\end{eqnarray}
where
\begin{equation}
O=\left\{\varphi_{T}\varphi_{T}\varphi_{T}, \varphi_{T}\varphi_{T}\phi, \varphi_{T}\phi\phi, \phi\phi\phi\right\}\,.
\end{equation}
Notice that all possible $S_4$ contractions should be considered. Dots stand for higher dimensional operators corrections which we will be discussed later. All the Yukawa couplings are real because of the generalized CP symmetry. Substituting the flavon VEVs in Eq.~\eqref{eq:vacuum_one}, we find the charged lepton mass matrix is diagonal with
\begin{equation}
m_{e}=\left|y_{e}\frac{v^{3}_T}{\Lambda^3}\right|v_{d},\qquad m_{\mu}=\left|y_{\mu_1}\frac{v^{2}_T}{\Lambda^2}-y_{\mu_2}\frac{v_{\phi}v_{T}}{\Lambda^2}-y_{\mu_3}\frac{v^{2}_{\phi}}{\Lambda^2}\right|v_{d},\qquad m_{\tau}=\left|y_{\tau}\frac{v_{T}}{\Lambda}\right|v_{d}\,,
\end{equation}
where $v_{d}=\langle h_d\rangle$, $y_{e}$ stands for the total result of all the different contributions of the $y_{e_i}$ terms. For $v_{\phi}\sim v_{T}\sim\lambda^2\Lambda$, the mass hierarchies of the charged lepton are obtained, i.e.
\begin{equation}
m_{e}:m_{\mu}:m_{\tau}\simeq\lambda^{4}:\lambda^{2}:1\,.
\end{equation}
As the representation matrix of the element $T^{2}STU$ is diagonal $\rho_{\mathbf{3}}(T^{2}STU)=\mathrm{diag}\left(-i, 1, i\right)$, we have $\rho^{\dagger}_{\bf3}(T^{2}STU)m^{\dagger}_{l}m_{l}\rho_{\bf3}(T^{2}STU)=m^{\dagger}_{l}m_{l}$. It is easy to check that the $Z^{(D)}_4$ subgroup is preserved by the vacuum of $\varphi_T$ and $\phi$, where $Z^{(D)}_4$ is the diagonal subgroup generated by $Z^{T^2STU}_4$ and the auxiliary $Z_4$ in usual way. Consequently the combination $m^{\dagger}_{l}m_{l}$ is predicted to be diagonal due to this residual $Z^{(D)}_4$ symmetry, and the lepton mixing arises from the neutrino sector.

The light neutrino masses are generated via type-I seesaw mechanism. The LO superpotential responsible for neutrino masses is
\begin{equation}
\label{eq:wnu_one}w_{\nu}=\frac{y_{1}}{\Lambda}\xi(\nu^{c}l)_{\mathbf{1}}h_{u}+\frac{y_{2}}{\Lambda}((\nu^{c}l)_{\mathbf{2}}\eta)_{\mathbf{1}}h_{u}+\frac{y_{3}}{\Lambda}((\nu^{c}l)_{\mathbf{3}}\varphi_{S})_{\mathbf{1}}h_{u}+M(\nu^{c}\nu^{c})_{\mathbf{1}}\;,
\end{equation}
where again all couplings are real due to the invariance under the generalized CP. The last term is the Majorana mass term for the right-handed neutrinos,
\begin{equation}
m_{M}=M\left(\begin{array}{ccc}
    0  &  0  &  1  \\
    0  &  1  &  0  \\
    1  &  0  &  0
\end{array}\right)\,.
\end{equation}
Hence the three right-handed neutrinos have a degenerate mass $M$. With the vacuum alignment of $\xi$, $\eta$ and $\varphi_{S}$ in Eq.~\eqref{eq:vacuum_two}, we find the Dirac mass matrix is of the following form,
\begin{equation}
m_{D}=y_{1}v_{u}\frac{v_{\xi}}{\Lambda}\left(\begin{array}{ccc}
    0  &  0  &  1  \\
    0  &  1  &  0  \\
    1  &  0  &  0
\end{array}\right)+y_{2}v_{u}\frac{v_{\eta}}{\Lambda}\left(\begin{array}{ccc}
    3  &  0  &  -1  \\
    0  &  2  &  0  \\
    -1  &  0  &  3
  \end{array}\right)+y_{3}v_{u}\frac{v_{S}}{\Lambda}\left(\begin{array}{ccc}
    0  &~  1  &  -\sqrt{2}  \\
    -1  &~  0  &  1  \\
    \sqrt{2}  &~  -1  &  0
  \end{array}\right)\;.
\end{equation}
The light neutrino mass matrix is given by the see-saw relation: $m_{\nu}=-m^{T}_{D}m^{-1}_{M}m_{D}$, we find that $m_{\nu}$ is of the same form as the one shown in Eq.~\eqref{eq:gen_mass_nu} with
\begin{equation}
\alpha=\big(\frac{8}{3}y^{2}-8x^{2}-1\big)m_{0}, \quad \beta=\big(2x-2x^{2}+\frac{1}{3}y^{2}\big)m_{0},\quad \gamma=-\sqrt{2}y^{2}m_{0}, \quad \epsilon=-6xym_{0}\,,
\end{equation}
where
\begin{equation}
x=\frac{y_{2}v_{\eta}}{y_{1}v_{\xi}}, \qquad  y=\frac{y_{3}v_{S}}{y_{1}v_{\xi}}, \qquad m_{0}=y^2_{1}\frac{v^2_{\xi}}{\Lambda^2}\frac{v^2_{u}}{M}\,.
\end{equation}
Note that the phase of $v_{\xi}$ can be factorized out as an overall phase of $m_{\nu}$ and therefore it can be absorbed by field redefinition. Accordingly Eq.~\eqref{eq:relation_m1} implies that the VEVs $v_{\xi}$ and $v_{\eta}$ are real while $v_{S}$ is real for $g_3g_4(g_1g_5-g_2g_4)>0$ and pure imaginary for $g_3g_4(g_1g_5-g_2g_4)<0$.

In case of real $v_{S}$, all the four parameters $\alpha$, $\beta$, $\gamma$ and $\epsilon$ are real. The VEVs of the flavon $\xi$, $\eta$ and $\varphi_{S}$ break the $S_{4}$ family symmetry to $Z^{ST^2SU}_2$ and break the generalized CP to $H^{\nu}_{CP}=\left\{\rho_{\mathbf{r}}(1), \rho_{\mathbf{r}}(ST^2SU)\right\}$ in the neutrino sector. Hence the desired symmetry breaking pattern discussed at the beginning of this section is exactly reproduced here. The lepton flavor mixing matrix is of the form shown in Eq.~\eqref{eq:UPMNS_one}, and the predictions for light neutrino masses and mixing parameters are presented in Eqs~(\ref{eq:neutrino_mass_caseI},\ref{eq:mixing_paramaters_nu1}) with $\tan2\theta=-\frac{12 x y}{1-2x+10x^2-4y^2}$.
Since the BM mixing has to undergo somewhat large corrections in order to be in accordance with experimental data, $\tan2\theta$ should be around 1.2, as shown in Eqs.~(\ref{eq:best_fit_Fir_caseIII}, \ref{eq:best_fit_Sec_caseIII}). This required value of $\theta$ can be naturally achieved in our model since both parameters $x$ and $y$ are of order one. On the other hand, if $v_{S}$ is pure imaginary, $\alpha$, $\beta$ and $\gamma$ are real while $\epsilon$ is an imaginary parameter. The remnant symmetry in the neutrino sector would be $Z^{ST^2SU}_2\times H^{\nu}_{CP}$ with $H^{\nu}_{CP}=\left\{\rho_{\mathbf{r}}(T^2U), \rho_{\mathbf{r}}(TST^2)\right\}$. However, the mixing pattern enforced by this residual symmetry can not fit the measured values of the mixing angles. Consequently we shall focus on the case of real $v_{S}$ henceforth.

It is useful to study the constraints on the model imposed by the observed values of the mass-squared splitting $\delta m^2\equiv m^2_2-m^2_1$, $\Delta m^2\equiv m^2_3-(m^2_1+m^2_2)/2$ and the reactor mixing angle $\theta_{13}$. As the light neutrino mass matrix effectively depends on three real (imaginary) parameters $x$, $y$ and $m_{0}$, their values can be completely fixed. Given the best fitting results $\delta m^2=7.54\times 10^{-5}\mathrm{eV}^2$, $\Delta m^2=2.43\times 10^{-3} (-2.38\times10^{-3}) \mathrm{eV}^2$ and $\sin^2\theta_{13}=0.0234 (0.0240)$ for NO (IO) neutrino mass spectrum from Ref.~\cite{Capozzi:2013csa}, the possible solutions for $x$, $y$ and the corresponding predictions for the light neutrino masses, the lepton mixing angles, CP phases and the effective mass $|m_{ee}|$ of neutrinoless double-beta decay are collected in Table~\ref{tab:effective5}. Note that there are other solutions predicting $\theta_{23}=30.137^{\circ}$ which is out of the $3\sigma$ range~\cite{Capozzi:2013csa}, and consequently they are not included in Table~\ref{tab:effective5}. It is remarkable that the absolute values of the light neutrino masses are fixed at leading order in the present model. We find that the light neutrino mass spectrum can be either NO or IO. Regarding the sum of the light neutrino masses, the latest Planck result is $\sum m_{\nu}<0.23\mathrm{eV}$ at $95\%$ confidence level~\cite{Ade:2013zuv}. This bound is saturated for all the solutions except the second one which gives $m_1+m_2+m_3\simeq0.238 \mathrm{eV}$ close to the upper bound. Furthermore, the effective mass $|m_{ee}|$ can take the values 12.650 meV, 33.044 meV, 22.821meV and 48.936meV in this model. The most stringent upper limit on $|m_{ee}|$ from GERDA~\cite{Agostini:2013mzu}, EXO-200~\cite{Auger:2012ar,Albert:2014awa} and KamLAND-ZEN~\cite{Gando:2012zm} is $|m_{ee}|<(120-250) \mathrm{meV}$. Hence our predictions for $|m_{ee}|$ are compatible with present experimental measurements. Our model could be directly tested by future neutrinoless double-beta decay experiments such as nEXO which is expected to have the mass sensitivity of $5\sim11$ meV~\cite{Tosi:2014zza}.

\begin{table}[t!]
\centering
\renewcommand{\tabcolsep}{1.55mm}
{\small
\begin{tabular}{|c|c|c|c|c|c|c|c|c|c|c|}
 \hline  \hline
 $\left(x,y\right)$  & $m_1$ & $m_2$   & $m_3$ & $|m_{ee}|$ &  $\alpha_{21}$ & $\alpha_{31}$ &  $\delta_{\rm{CP}}$ & $\theta_{23}/^\circ$  & $\theta_{12}/^\circ$  & \text{mass order}\\ \hline
 $(-0.109,-0.729)$ & 13.535 & 16.081 & 51.487 & 12.650 & $0$  & \multirow{4}{*}{$\pi$} & \multirow{4}{*}{$\pi$} & \multirow{2}{*}{40.392} & \multirow{2}{*}{30.395} & \multirow{2}{*}{NO} \\ \cline{1-6}
 $(0.855,-0.602)$ & 73.975 & 74.483  & 89.106 & 33.044 & $\pi$ &  & & & & \\ \cline{1-6} \cline{9-11}
 $(-0.057,0.468)$ & 48.529 & 49.300 & 3.569 & 22.821 &  $\pi$ &  & &\multirow{2}{*}{40.459} &\multirow{2}{*}{30.405} & \multirow{2}{*}{IO} \\ \cline{1-6}
 $(0.093,0.606)$ & 50.284 & 51.028 & 13.644 & 48.936 & $0$ &  & & & & \\ \hline \hline
\end{tabular}}
\caption{\label{tab:effective5}The predictions for light neutrino masses $m_i(i=1,2,3)$, the lepton flavor mixing parameters and the effective mass $|m_{ee}|$ of the neutrinoless double-beta decay, where the unit of mass is meV. }
\end{table}

Higher dimensional operators, suppressed by additional powers of the cutoff scale $\Lambda$, can be added to the leading terms studied above. As a result, the LO predictions would be modified. The subleading corrections to the driving superpotential are,
\begin{equation}
\Delta w^l_d=\frac{1}{\Lambda}(\rho^0\Psi^3_\nu)_{\mathbf{1}}+\frac{1}{\Lambda}(\varphi^0_T\Psi^3_\nu)_{\mathbf{1}}\,,\quad \Delta w^\nu_d=\frac{1}{\Lambda^4}(\eta^0\Psi^4_l\Psi^2_{\nu})_{\mathbf{1}}+\frac{1}{\Lambda^4}(\varphi^0_S\Psi^4_l\Psi^2_{\nu})_{\mathbf{1}}\,.
\end{equation}
where $\Psi_{\nu}=\{\xi, \eta, \varphi_{S}\}$, $\Psi_{l}=\{\phi, \varphi_{T}\}$ and the couplings in front of each operators are omitted. Notice that there are generally several independent $S_4$ contractions for each operator. The new VEV configuration is obtained by imposing the vanishing of the first derivative
of $w_d+\Delta w_d$ with respect to the driving fields $\rho^0$, $\varphi^0_T$, $\eta^0$ and $\varphi^0_S$. To the first order in the $1/\Lambda$ expansion, it is straightforward to find that the LO vacuum in Eq.~\eqref{eq:vacuum_one} and Eq.~\eqref{eq:vacuum_two} is modified into
\begin{eqnarray}
\nonumber&\langle\varphi_T\rangle=\left(v_{T},\delta v_{T_2}, \delta v_{T_3}\right),\qquad \langle\phi\rangle=\left(\delta v_{\phi_1}, \delta v_{\phi_2}, v_{\phi}+\delta v_{\phi_3}\right),\\
&\langle\eta\rangle=\left(v_{\eta}+\delta v_{\eta_1}, \sqrt{3}v_{\eta}+\delta v_{\eta_2}\right),\qquad \langle\varphi_{S}\rangle=\left(v_{S}+\delta v_{S_1}, \sqrt{2}v_{S}+\delta v_{S_2}, v_{S}+\delta v_{S_3}\right)\,.
\end{eqnarray}
Note that all components of $\langle\varphi_T\rangle$, $\langle\phi\rangle$, $\langle\eta\rangle$ and $\langle\varphi_{S}\rangle$ acquire different corrections so that their alignments are tilted. Moreover, Since $\Delta w^{l}_d$ and $\Delta w^{\nu}_{d}$ are suppressed by $1/\Lambda$ and $1/\Lambda^4$ respectively, the shifts $\delta v_{T_2}$, $\delta v_{T_3}$, $\delta v_{\phi_1}$, $\delta v_{\phi_2}$ and $\delta v_{\phi_3}$ are of relative order $\lambda^2$ with respect to the LO results, while the deviations $\delta v_{\eta_1}$, $\delta v_{\eta_2}$, $\delta v_{S_1}$, $\delta v_{S_2}$ and $\delta v_{S_3}$ in the neutrino sector are of relative order $\lambda^8$.

In the same fashion, the subleading terms of the Yukawa superpotential $w_{\nu}$ and $w_{l}$, which are invariant under the family symmetry $S_4\times Z_{3}\times Z_{4}$, are of the following form:
\begin{eqnarray}
\nonumber &&\Delta w_l=\frac{1}{\Lambda^5}\tau^c(l\Psi^5_l)_{\mathbf{1}}h_d+\frac{1}{\Lambda^3}\mu^c(l\Psi^3_\nu)_{\mathbf{1}^{\prime}}h_d+\frac{1}{\Lambda^4}e^c(l\Psi_l\Psi^3_\nu)_{\mathbf{1}}h_d\,,\\
&&w_\nu=\frac{1}{\Lambda^5}(l\nu^c\Psi^4_l\Psi_\nu)_{\bf1}h_u+\frac{1}{\Lambda^3}(\nu^c\nu^c\Psi^4_l)_{\bf1}\,.
\end{eqnarray}
The subleading corrections to the lepton mass and mixing matrices are obtained by inserting the corrected VEV alignment into the LO operators plus the contribution of the higher dimensional Yukawa operators evaluated with the unperturbed VEVs. It is easy to check that the neutrino mass matrix receives a relative corrections of order $\lambda^8$. As a result, the subleading corrections to lepton mixing of the neutrino sector are suppressed by $\langle\Phi_{l}\rangle^4/\Lambda^4\sim\lambda^8$ with respect to LO results and thus they can be ignored. In the charged lepton sector, all non-diagonal entries become non-vanishing after
the inclusion of the subleading contributions. Eventually the corrected charged lepton mass matrix has the following structure,
\begin{equation}
 m_l\sim\left(\begin{array}{ccc}
 m_e & \lambda^2m_e & \lambda^2m_e  \\
 \lambda^2m_\mu & m_\mu & \lambda^2m_\mu \\
 \lambda^2m_\tau & \lambda^2m_\tau & m_\tau
 \end{array}\right)\,.
\end{equation}
We can estimate the higher order corrections to the LO predictions for the lepton mixing angles in Eq.~\eqref{eq:mixing_paramaters_nu1} as follows,
\begin{equation}
\delta \sin^2\theta_{13}\sim\lambda^2,\qquad \delta \sin^2\theta_{12}\sim\lambda^2,\qquad \delta \sin^2\theta_{23}\sim\lambda^2\,.
\end{equation}
Therefore the LO relation $4\cos^2\theta_{13}\sin^2\theta_{12}=1$ is violated by small terms of order $\lambda^2$ when the subleading contributions are included. As a consequence, the observed value of $\theta_{12}$ can be achieved although a value of $\theta_{12}$ close to the present $3\sigma$ upper bound would be unnatural in our model.

\section{\label{sec:model_row}Model predicting one row of BM mixing with $S_4$ and generalized CP}
\cleqn

In this section, we shall present an explicit model realization for the mixing pattern investigated in section~\ref{sec:general_analysis_one_row}. The model is also based on $S_4$ family symmetry and generalized CP, which is supplemented by $Z_{5}\times Z_{6}$. The flavon fields and driving fields are properly arranged such that $S_4\rtimes H_{CP}$ is broken to $K^{(TST^2, T^2U)}_4\rtimes H^{\nu}_{CP}$ with $H^{\nu}_{CP}=\{$$\rho_{\bf r}(1)$, $\rho_{\bf r}(T^{2}U)$, $\rho_{\bf r}(TST^{2})$, $\rho_{\bf r}(ST^{2}SU)\}$ in the neutrino sector at leading order, and the flavor symmetry preserved by the charged lepton mass matrix $m^{\dagger}_{l}m_{l}$ is $K^{(S,U)}_4$. As a result, the lepton flavor mixing is predicted to be of the BM form at leading order. Furthermore, the next-to-leading-order (NLO) corrections break the remnant symmetry down to $Z^{SU}_2\times H^{l}_{CP}$ in the charged lepton sector. Consequently the resulting PMNS matrix has one row of the form $\left(1/2, 1/2, 1/\sqrt{2}\right)$
which is exactly the third row of the BM mixing pattern, and agreement with experimental data can be achieved. As we shall show below, the general model independent results of section~\ref{sec:general_analysis_one_row} can be naturally reproduced in this model. The involved fields and their transformation rules under the family symmetry are summarized in Table~\ref{tab:fields_model_one_row}. We start to explore the vacuum structure of the model in the following section.

\begin{table}[t!]
\renewcommand{\tabcolsep}{1.3mm}
\centering
\begin{tabular}{|c||c|c|c|c|c|c||c|c|c|c|c|c|c||c|c|c|c|c|c|c|c|c|}
\hline \hline
Field &   $l$    &  $\nu^{c}$     &  $e^{c}$     &   $\mu^{c}$    &    $\tau^{c}$  &  $h_{u,d}$ & $\xi$ &  $\eta$  &  $\varphi_T$ & $\phi$ &  $\rho$ &  $\sigma$ & $\varphi_{S}$      & $\xi^{0}$ & $\rho^{0}$ & $\zeta^{0}$ & $\eta^{0}$  & $\varphi^{0}_T$ & $\kappa^{0}$ & $\sigma^{0}$  &   $\varphi^{0}_S$  \\ \hline

$S_4$  &   $\mathbf{3}$  &  $\mathbf{3}$  &  $\mathbf{1}^{\prime}$  & $\mathbf{1}^\prime$   &  $\mathbf{1}$   & $\mathbf{1}$  & $\mathbf{1}$  & $\mathbf{2}$ &   $\mathbf{3}$ & $\mathbf{3}^\prime$ &   $\mathbf{1}$ &  $\mathbf{2}$  &  $\mathbf{3}^{\prime}$  & $\mathbf{1}$  & $\mathbf{1}$   & $\mathbf{1}^\prime$ & $\mathbf{2}$  &  $\mathbf{3}$ & $\mathbf{1}$ & $\mathbf{2}$ & $\mathbf{3}^{\prime}$     \\ \hline

$Z_5$ & $\omega^{3}_5$ & $\omega^2_5$ & $\omega^2_5$ & $\omega^2_5$ & $\omega^2_5$ & $1$ & $1$ & $1$ & $1$ & $1$ &$\omega_5$ & $\omega_5$ & $\omega_5$ & $1$ & $1$ & $1$ & $1$& $1$  & 1 & $\omega^3_5$ & $\omega^3_5$  \\ \hline

$Z_6$ & $1$ & $1$  & $\omega^{4}_{6}$ & $\omega^{5}_{6}$ &  $\omega^{4}_{6}$ & $1$ & $\omega^{4}_{6}$ & $\omega_{6}$ & $\omega^{2}_{6}$ & $\omega^{3}_{6}$ & $1$ & $1$ & $1$ & $\omega^{2}_{6}$ & $\omega^{4}_{6}$  & $\omega_{6}$ & $\omega_{6}$ & $\omega_{6}$ & 1 &   $1$ & $1$  \\ \hline

$U(1)_R$  &  $1$  & $1$  & $1$  &   $1$  & $1$  &  $0$ &  $0$ &  $0$ & $0$  &  $0$ & $0$  & $0$  & $0$ & $2$ & $2$ & $2$ & $2$ & $2$ &  $2$ & 2 & $2$  \\ \hline \hline
\end{tabular}
\caption{\label{tab:fields_model_one_row}The particle contents and their transformation properties under the family symmetry $S_{4}\times Z_{5}\times Z_{6}$ and $U(1)_R$, where $\omega_5=e^{2i\pi/5}$ and $\omega_{6}=e^{2i\pi/6}$.}
\end{table}

\subsection{\label{subsec:vacuum_alignment_model2}Vacuum alignment}

The most general flavon superpotential invariant under the symmetry of the model is
\begin{eqnarray}
\nonumber&&w_d=M_{\xi}\xi^{0}\xi+f_{1}\xi^{0}(\varphi_{T}\varphi_{T})_{\bf1}+f_{2}\rho^{0}\xi^{2}+f_{3}\rho^{0}(\eta\eta)_{\bf1}+f_{4}\zeta^{0}(\varphi_{T}\phi)_{\bf1^{\prime}}+f_{5}\xi(\eta^{0}\eta)_{\bf1}\\
\nonumber&&\qquad+f_{6}(\eta^{0}(\varphi_{T}\phi)_{\bf2})_{\bf1}+f_{7}(\varphi^{0}_{T}(\varphi_{T}\phi)_{\bf3})_{\bf1}+M^{2}_{\kappa}\kappa^{0}+f_{8}\kappa^{0}(\phi\phi)_{\bf1}+g_{1}\rho(\sigma^{0}\sigma)_{\bf1}+g_{2}(\sigma^{0}(\sigma\sigma)_{\bf2})_{\bf1}\\
\label{eq:wd_model2}&&\qquad+g_{3}(\sigma^{0}(\varphi_{S}\varphi_{S})_{\bf2})_{\bf1}+g_{4}\rho(\varphi^{0}_{S}\varphi_{S})_{\bf1}+g_{5}(\varphi^{0}_{S}(\sigma\varphi_{S})_{\bf3^{\prime}})_{\bf1}+g_{6}(\varphi^{0}_{S}(\varphi_{S}\varphi_{S})_{\bf3^{\prime}})_{\bf1}\,,
\end{eqnarray}
where all couplings $f_{i}$ and $g_{i}$ are real due to the imposed generalized CP symmetry. In the charged lepton sector, the equations for the vanishing of the derivatives of $w_{d}$ with respect to each component of the driving fields are as follows:
\begin{eqnarray}
\nonumber&&\frac{\partial w_{d}}{\partial\xi^{0}}=M_{\xi}\xi+f_{1}(2\varphi_{T_{1}}\varphi_{T_{3}}+\varphi^{2}_{T_{2}})=0,\\
\nonumber&&\frac{\partial w_{d}}{\partial\rho^{0}}=f_{2}\xi^{2}+f_{3}(\eta^{2}_{1}+\eta^{2}_{2})=0,\\
\nonumber&&\frac{\partial w_{d}}{\partial\zeta^{0}}=f_{4}(\varphi_{T_{1}}\phi_{3}+\varphi_{T_{2}}\phi_{2}+\varphi_{T_{3}}\phi_{1})=0,\\
\nonumber&&\frac{\partial w_{d}}{\partial\eta^{0}_{1}}=f_{5}\xi\eta_{1}+\sqrt{3}f_{6}(\varphi_{T_{1}}\phi_{1}+\varphi_{T_{3}}\phi_{3})=0,\\
\nonumber&&\frac{\partial w_{d}}{\partial\eta^{0}_{2}}=f_{5}\xi\eta_{2}+f_{6}(\varphi_{T_{1}}\phi_{3}-2\varphi_{T_{2}}\phi_{2}+\varphi_{T_{3}}\phi_{1})=0,\\
\nonumber&&\frac{\partial w_{d}}{\partial\varphi^{0}_{T_{1}}}=f_{7}(\varphi_{T_{1}}\phi_{2}+\varphi_{T_{2}}\phi_{1})=0,\\
\nonumber&&\frac{\partial w_{d}}{\partial\varphi^{0}_{T_{2}}}=f_{7}(\varphi_{T_{1}}\phi_{1}-\varphi_{T_{3}}\phi_{3})=0,\\
\nonumber&&\frac{\partial w_{d}}{\partial\varphi^{0}_{T_{3}}}=-f_{7}(\varphi_{T_{2}}\phi_{3}+\varphi_{T_{3}}\phi_{2})=0,\\
&&\frac{\partial\omega^{l}_{d}}{\partial\kappa^{0}}=M^{2}_{\kappa}+f_{8}(2\phi_{1}\phi_{3}+\phi^{2}_{2})=0\,.
\end{eqnarray}
We find one solution (up to $S_{4}$ transformations) for above equations: \begin{equation}
\label{eq:VEVs_five}
\langle\xi\rangle=v_{\xi}, \quad  \langle\eta\rangle=\left(\begin{array}{c}
    1 \\ 0
  \end{array}\right)v_{\eta},\quad \langle\varphi_{T}\rangle=\left(\begin{array}{c}
    1+i \\ 0  \\ i-1
  \end{array}\right)v_{T}, \quad \langle\phi\rangle=\left(\begin{array}{c}
    i-1 \\ 0  \\  1+i
  \end{array}\right)v_{\phi}\,,
\end{equation}
where the VEVs $v_{\xi}$, $v_{\eta}$, $v_{T}$ and $v_{\phi}$ are related by
\begin{equation}
\label{eq:VEV_relation_model2}v^{2}_{\eta}=-\frac{f_{2}}{f_{3}}v^{2}_{\xi}, \qquad v^{2}_{T}=\frac{M_{\xi}v_{\xi}}{4f_{1}}, \qquad v_{\phi}=\frac{f_{5}v_{\xi}v_{\eta}}{4\sqrt{3}f_{6}v_{T}}\,,
\end{equation}
with
\begin{equation}
v^{3}_{\xi}=-\frac{3f_{3}f^{2}_{6}M_{\xi}M^{2}_{\kappa}}{f_{1}f_{2}f^{2}_{5}f_{8}}\,.
\end{equation}
Hence the VEV $v_{\xi}$ is fixed to be
\begin{equation}
\label{eq:vxi_model2}v_{\xi}=-\left(\frac{3f_{3}f^{2}_{6}M_{\xi}M^{2}_{\kappa}}{f_{1}f_{2}f^{2}_{5}f_{8}}\right)^{1/3},\quad \left(\frac{3f_{3}f^{2}_{6}M_{\xi}M^{2}_{\kappa}}{f_{1}f_{2}f^{2}_{5}f_{8}}\right)^{1/3}e^{i\pi/3},~~\mathrm{or}~~
\left(\frac{3f_{3}f^{2}_{6}M_{\xi}M^{2}_{\kappa}}{f_{1}f_{2}f^{2}_{5}f_{8}}\right)^{1/3}e^{5i\pi/3}\,.
\end{equation}
In the present paper, we shall concentrate on the fist solution, i.e. the case of real $v_{\xi}$. The other two options of complex $v_{\xi}$ would not be considered. Accordingly the VEVs $v_{\eta}$, $v_{T}$ and $v_{\phi}$ would be real or pure imaginary. If $v_{\eta}$, $v_{T}$ and $v_{\phi}$ are all real parameters, this can be achieved for $f_2f_3<0$ and $f_1M_{\xi}v_{\xi}>0$, the residual CP symmetry preserved by the vacuum of Eq.~\eqref{eq:VEVs_five} is $H^{l}_{CP}=\{\rho_{\bf r}(TST^{2}),\rho_{\bf r}(TST^{2}U)\}$. If $v_{\eta}$ is real and $v_{T}$, $v_{\phi}$ are pure imaginary, this can be realized for $f_2f_3<0$ and $f_1M_{\xi}v_{\xi}<0$, another two of the 24 generalized CP symmetries are preserved
with $H^{l}_{CP}=\{\rho_{\bf r}(T^{2}ST),\rho_{\bf r}(T^{2}STU)\}$. On the other hand, the generalized CP symmetry $H^{l}_{CP}$ will be completely broken for imaginary $v_{\eta}$ no matter $v_{T}$, $v_{\phi}$ are real or imaginary. It is easy to check that the determined vacuum in Eq.~\eqref{eq:VEVs_five} breaks $S_{4}$ family symmetry to $Z^{SU}_{2}$ subgroup. Furthermore, since the different VEVs are related via dimensionless couplings in Eq.~\eqref{eq:VEV_relation_model2}, these VEVs are expected to have the same order of magnitude which we choose to be $\lambda^2\Lambda$.

In the neutrino sector, the vacuum is determined by $F-$term conditions associated with the driving fields $\sigma^{0}$ and $\varphi^{0}_{S}$,
\begin{eqnarray}
\nonumber&&\frac{\partial w_{d}}{\partial\sigma^{0}_{1}}=g_{1}\rho\sigma_{1}+g_{2}(\sigma^{2}_{2}-\sigma^{2}_{1})+2g_{3}(\varphi^{2}_{S_{2}}-\varphi_{S_{1}}\varphi_{S_{3}})=0,\\
\nonumber&&\frac{\partial w_{d}}{\partial\sigma^{0}_{2}}=g_{1}\rho\sigma_{2}+2g_{2}\sigma_{1}\sigma_{2}+\sqrt{3}g_{3}(\varphi^{2}_{S_{1}}+\varphi^{2}_{S_{3}})=0,\\
\nonumber&&\frac{\partial w_{d}}{\partial\varphi^{0}_{S_{1}}}=g_{4}\rho\varphi_{S_{3}}+g_{5}(\sqrt{3}\sigma_{2}\varphi_{S_{1}}-\sigma_{1}\varphi_{S_{3}})+2g_{6}\varphi_{S_{1}}\varphi_{S_{2}}=0,\\
\nonumber&&\frac{\partial w_{d}}{\partial\varphi^{0}_{S_{2}}}=g_{4}\rho\varphi_{S_{2}}+2g_{5}\sigma_{1}\varphi_{S_{2}}+g_{6}(\varphi^{2}_{S_{1}}-\varphi^{2}_{S_{3}})=0,\\
&&\frac{\partial w_{d}}{\partial\varphi^{0}_{S_{3}}}=g_{4}\rho\varphi_{S_{1}}+g_{5}(\sqrt{3}\sigma_{2}\varphi_{S_{3}}-\sigma_{1}\varphi_{S_{1}})-2g_{6}\varphi_{S_{2}}\varphi_{S_{3}}=0\,.
 \end{eqnarray}
A solution to this equation with each flavon acquiring non-zero VEV is given by
\begin{equation}
\label{eq:VEVs_seven}
\langle\rho\rangle=v_{\rho}, \qquad \langle\sigma\rangle=\left(\begin{array}{c}
    1 \\ \sqrt{3}
  \end{array}\right)v_{\sigma}, \qquad  \langle\varphi_{S}\rangle=\left(\begin{array}{c}
    1 \\ 0 \\ -1
  \end{array}\right)v_{S}\,,
\end{equation}
where the VEVs obey the relations
\begin{equation}
\label{eq:VEV_relations_m2}v_{\sigma}=\frac{g_{4}v_{\rho}}{4g_{5}}, \qquad v_{S}=\frac{v_{\rho}}{4g_{5}}\sqrt{-\frac{g_{4}(2g_{1}g_{5}+g_{2}g_{4})}{g_{3}}}\,,
\end{equation}
with $v_{\rho}$ undetermined. The vacuum alignment in Eq.~\eqref{eq:VEVs_seven} is invariant under the action of both the $TST^2$ and $T^2U$ elements of $S_{4}$, consequently it breaks the $S_4$ family symmetry to Klein four $K^{(TST^2, T^2U)}_4$ subgroup. Furthermore, since all couplings $g_{i}$ are real, then Eq.~\eqref{eq:VEV_relations_m2} implies that $v_{\sigma}$ and $v_{\rho}$ have the same phase up to $\pi$, and the phase difference between $v_{\rho}$ and $v_{S}$ is $0$, $\pi$ or $\pm\frac{\pi}{2}$ determined by the sign of $g_{3}g_{4}(2g_{1}g_{5}+g_{2}g_{4})$. Similar to previous model, we expect a common order of magnitude for all the VEVs which is taken to be $\lambda^2\Lambda$.

\subsection{\label{subsec:LO_model2}Leading order results}
The charged lepton masses are described by the following superpotential
\begin{equation}
\label{eq:charged_superpotential_three}
w_{l}=\frac{y_{\tau}}{\Lambda}\tau^{c}(l\varphi_{T})_{\bf1}h_{d}+\frac{y_{\mu}}{\Lambda^{2}}\mu^{c}\xi(l\phi)_{\bf1^{\prime}}h_{d}+...\,,
\end{equation}
where dots represent higher dimensional operators which we will consider later. After the electroweak and flavor symmetries breaking by the VEVs shown in Eq.~\eqref{eq:VEVs_five}, we obtain a charged lepton mass matrix as follows
\begin{equation}
m_{l}=\left(\begin{array}{ccc}
   0  &  ~0~  &  0  \\
\frac{(1+i)y_{\mu}v_{\xi}v_{\phi}}{\Lambda^{2}}  &  ~0~  & \frac{(i-1)y_{\mu}v_{\xi}v_{\phi}}{\Lambda^{2}} \\
\frac{(i-1)y_{\tau}v_{T}}{\Lambda}  &  ~0~  &  \frac{(1+i)y_{\tau}v_{T}}{\Lambda}
\end{array}\right)v_{d}\,.
\end{equation}
As a consequence the unitary matrix $U_{l}$, which corresponds to the transformation of the charged leptons used to diagonalize $m^{\dagger}_{l}m_{l}$, is of the following form:
\begin{equation}
\label{eq:UL_model2}U_{l}=\frac{1}{\sqrt{2}}\left(\begin{array}{ccc}
0 ~&~ e^{\frac{i \pi }{4}}  ~&~  e^{-\frac{i\pi }{4}} \\
-\sqrt{2}  & 0  &  0  \\
0  ~&~  e^{-\frac{i\pi }{4}} ~&~ e^{\frac{i \pi }{4}}
\end{array}\right)\,.
\end{equation}
The charged lepton masses are given by,
\begin{equation}
m^{2}_{e}=0, \quad m^{2}_{\mu}=4y^{2}_{\mu}\frac{|v_{\xi}v_{\phi}|^{2}}{\Lambda^{4}}\;v^2_{d}, \quad m^{2}_{\tau}=4y^{2}_{\tau}\frac{|v_{T}|^{2}}{\Lambda^{2}}v^2_{d}\,.
\end{equation}
Note that the correct mass hierarchy between muon and tau is generated for $v_{\xi}/\Lambda\sim v_{T}/\Lambda\sim v_{\phi}/\Lambda\sim\lambda^{2}$. The electron is massless at LO and its mass is generated by higher dimensional operators, which will be studied in section~\ref{subsec:NLO_model2}. From the view of symmetry and its breaking, although the VEVs of $\xi$, $\eta$, $\varphi_{T}$ and $\phi$ leave $Z^{SU}_2$ invariant, the remnant flavor symmetry of  $m^{\dagger}_{l}m_{l}$ is $K^{(S,U)}_{4}$. In other words, we have $\rho^{\dagger}_{\mathbf{3}}(S)m^{\dagger}_{l}m_{l}\rho_{\mathbf{3}}(S)=m^{\dagger}_{l}m_{l}$ and $\rho^{\dagger}_{\mathbf{3}}(U)m^{\dagger}_{l}m_{l}\rho_{\mathbf{3}}(U)=m^{\dagger}_{l}m_{l}$. The enhancement of the remnant flavor symmetry from $Z^{SU}_2$ to $K^{(S,U)}_{4}$ is because that $\left|v_{T}\right|^2$ and $\left|v_{\phi}\right|^2$ instead of $v_{T}$ and $v_{\phi}$ are involved in $m^{\dagger}_{l}m_{l}$. Moreover, it is straightforward to check that the residual CP symmetry preserved by the combination $m^{\dagger}_{l}m_{l}$ is $H^{l}_{CP}=\big\{$$\rho_{\mathbf{r}}(TST^2)$, $\rho_{\mathbf{r}}(TST^2U)$, $\rho_{\mathbf{r}}(T^2ST)$, $\rho_{\mathbf{r}}(T^2STU)\big\}$.

Now we come to the neutrino sector. The LO superpotential of for the neutrino masses is
\begin{equation}
\label{eq:wnu_model2}w_{\nu}=y(\nu^{c}l)_{\bf1}h_{u}+y_{1}\rho(\nu^{c}\nu^{c})_{\bf1}+y_{2}((\nu^{c}\nu^{c})_{\bf2}\sigma)_{\bf1} +y_{3}((\nu^{c}\nu^{c})_{\bf3^{\prime}}\varphi_{S})_{\bf1}\,,
\end{equation}
where the first term is Dirac mass term and the last three are Majorana mass terms. The generalized CP symmetry constrains all the couplings to be real. The flavons $\rho$, $\sigma$ and $\varphi_{S}$ get VEVs shown in Eq.~\eqref{eq:VEVs_seven}, and then the Dirac and right-handed Majorana neutrino mass matrices read as
\begin{equation}
m_{D}=yv_{u}\left(\begin{array}{ccc}
  0  &  0  &  1  \\
  0  &  1  &  0  \\
  1  &  0  &  0
\end{array}\right),\quad m_{M}=y_{1}v_{\rho}\left(
\begin{array}{ccc}
 0 & 0 & 1 \\
 0 & 1 & 0 \\
 1 & 0 & 0
\end{array}
\right)+y_{2}v_{\sigma}\left(
\begin{array}{ccc}
 3 & 0 & -1 \\
 0 & 2 & 0 \\
 -1 & 0 & 3
\end{array}
\right)+y_{3}v_{S}\left(
\begin{array}{ccc}
 0 & 1 & 0 \\
 1 & 0 & 1 \\
 0 & 1 & 0
\end{array}
\right)\,.
\end{equation}
The light neutrino mass matrix is given by the seesaw relation $m_{\nu}=-m^{T}_{D}m^{-1}_{M}m_{D}$, and we find $m_{\nu}$ is of the same form as that in Eq.~\eqref{eq:neutrino_matrix_reK4} with
\begin{eqnarray}
\nonumber && a=\frac{\left[-3y^2_{1}v^2_{\rho}+2\left(6y^2_{2}v^2_{\sigma} +y^2_{3}v^2_{S}\right)\right]y^2v^2_{u}}{3(y_{1}v_{\rho}-4y_{2}v_{\sigma}) \left[(y_{1}v_{\rho}+2y_{2}v_{\sigma})^2-2y^2_{3}v^2_{S}\right]}, \\ \nonumber&&b=\frac{\left[3y_{2}v_{\sigma}(y_{1}v_{\rho}+2y_{2}v_{\sigma})-y^2_{3}v^2_{S}\right]y^2v^2_{u}}{3 (y_{1}v_{\rho}-4y_{2}v_{\sigma})\left[(y_{1}v_{\rho}+2y_{2}v_{\sigma})^2-2y^2_{3}v^2_{S}\right]}, \\
&&c=\frac{y_{3}y^2v_{S}v^2_{u}}{(y_{1}v_{\rho}+2y_{2}v_{\sigma})^2-2y^2_{3}v^2_{S}}\,.
\end{eqnarray}
Hence $m_{\nu}$ is exactly diagonalized by the unitary transformation $U_{\nu}$ shown in Eq.~\eqref{eq:Unu:reK4}, and the resulting mass eigenvalues are $a+2b-\sqrt{2}\,c$, $a+2b+\sqrt{2}\,c$ and $-a+4b$. As shown in Eq.~\eqref{eq:wnu_model2}, here the VEVs of $\rho$, $\sigma$ and $\varphi_S$ breaks both $S_4$ family symmetry and generalized CP in the neutrino sector. From the vacuum alignment of section~\ref{subsec:vacuum_alignment_model2}, we know that the remnant family symmetry is $K^{(TST^2, T^2U)}_4$. Since the phase of $v_{\rho}$ can be factored out from $m_{\nu}$, $v_{\rho}$ can be taken to be real. As a consequence, $v_{\sigma}$ is real and $v_{S}$ can be real or purely imaginary. If $v_{S}$ is imaginary, this can be realized for $g_{3}g_{4}(2g_{1}g_{5}+g_{2}g_{4})>0$. We find the generalized CP symmetry $H_{CP}$ is broken to $H^{\nu}_{CP}=\{\rho_{\bf r}(S)$, $\rho_{\bf r}(ST^{2}U)$, $\rho_{\bf r}(T^{2}ST),\rho_{\bf r}(T^{2}SU)\}$ in the neutrino sector. The parameters $a$, $b$ are real while $c$ is purely imaginary. Accordingly the light neutrino masses would be partially degenerate with $m_{1}=m_{2}$ which is not viable.
If $v_{S}$ is real, this scenario can be achieved for $g_{3}g_{4}(2g_{1}g_{5}+g_{2}g_{4})<0$. The residual CP symmetry would be $H^{\nu}_{CP}=\{\rho_{\bf r}(1), \rho_{\bf r}(T^{2}U), \rho_{\bf r}(TST^{2}), \rho_{\bf r}(ST^{2}SU)\}$ which has been discussed in section~\ref{sec:general_analysis_one_row}. Then all the three parameters $a$, $b$ and $c$ are real. The phenomenological constraints of $\delta m^2\equiv m^2_2-m^2_1$ and $\Delta m^2\equiv m^2_3-(m^2_1+m^2_2)/2$ can be easily satisfied by properly choosing the values of $a$, $b$ and $c$. Either NO or IO neutrino mass spectrum is allowed. Henceforth $v_{S}$ will be considered as real.

In the end, combining the unitary transformation $U_{l}$ and $U_{\nu}$ from the charged lepton and the neutrino sectors, we obtain the lepton mixing matrix
\begin{equation}
U_{PMNS}=U^{\dagger}_{l}U_{\nu}=\frac{1}{2}\left(\begin{array}{ccc}
\sqrt{2}  ~&~  -\sqrt{2}   ~&~  0 \\
1   ~&~  1  ~&~   \sqrt{2}\,i \\
1   ~&~  1  ~&~   -\sqrt{2}\,i
\end{array}
\right)\,.
\end{equation}
Therefore the lepton flavor mixing is the BM pattern at LO. In the following section, we shall analyze the higher order corrections needed to modify the BM mixing in order to obtain an acceptable lepton mixing pattern.

\subsection{\label{subsec:NLO_model2}Next-to-leading-order corrections}

In brief, at leading order the model gives rise to a vanishing electron mass $(m_{e}=0)$ and the BM mixing pattern leading to $\theta_{13}=0^{\circ}$ and $\theta_{12}=\theta_{23}=45^{\circ}$ which obviously don't match with the experimental measurements. Therefore the next-to-leading-order (NLO) corrections are crucial to achieve agreement with the present data. We will demonstrate in the following that a non-zero electron mass and realistic mass hierarchies among the charged lepton are obtained after the NLO contributions are included. In addition, the LO remnant symmetry $K^{(S,U)}_{4}$ of $m^{\dagger}m_{l}$ is further broken down to $Z^{SU}_{2}$ such that the symmetry breaking patterns discussed in section~\ref{sec:general_analysis_one_row} are realized and the resulting PMNS matrix is of the form of Eq.~\eqref{eq:PMNS_caseIII_1oct}. We first start with the corrections to the flavon superpotential $w_{d}$ in Eq.~\eqref{eq:wd_model2} which determines the vacuum alignment. The symmetry allowed NLO terms including the driving fields $\xi^{0}$, $\rho^{0}$, $\zeta^{0}$, $\eta^{0}$, $\varphi^{0}_{T}$ and $\kappa^{0}$ are
\begin{eqnarray}
\nonumber&& \Delta w^{l}_{d}=f_{9}\xi^{0}\xi(\phi\phi)_{\bf1}/\Lambda+f_{10}\rho^{0}\xi(\varphi_{T}\varphi_{T})_{\bf1}/\Lambda+f_{11}\rho^{0}(\varphi_{T}(\phi\phi)_{\bf3})_{\bf1}/\Lambda+f_{12}\zeta^{0}(\eta(\varphi_{T}\varphi_{T})_{\bf2})_{\bf1^{\prime}}/\Lambda\\
\nonumber&&\qquad+f_{13}(\eta^{0}\eta)_{\bf1}(\varphi_{T}\varphi_{T})_{\bf1}/\Lambda+f_{14}((\eta^{0}\eta)_{\bf2}(\varphi_{T}\varphi_{T})_{\bf2})_{\bf1}/\Lambda+f_{15}((\varphi^{0}_{T}\eta)_{\bf3}(\eta\phi)_{\bf3})_{\bf1}/\Lambda\\
\nonumber&&\qquad+f_{16}((\varphi^{0}_{T}\eta)_{\bf3^{\prime}}(\eta\phi)_{\bf3^{\prime}})_{\bf1}/\Lambda+f_{17}((\varphi^{0}_{T}\eta)_{\bf3}(\varphi_{T}\varphi_{T})_{\bf3})_{\bf1}/\Lambda+f_{18}((\varphi^{0}_{T}\eta)_{\bf3^{\prime}}(\varphi_{T}\varphi_{T})_{\bf3^{\prime}})_{\bf1}/\Lambda\\
&&\qquad+f_{19}\kappa^{0}\xi^{3}/\Lambda+f_{20}\kappa^{0}\xi(\eta\eta)_{\bf1}/\Lambda+f_{21}\kappa^{0}(\eta(\varphi_{T}\phi)_{\bf2})_{\bf1}/\Lambda+f_{22}\kappa^{0}(\varphi_{T}(\varphi_{T}\varphi_{T})_{\bf3})_{\bf1}/\Lambda\,.
\end{eqnarray}
We see that they are suppressed by one of power of $1/\Lambda$ with respect to the LO terms in Eq.~\eqref{eq:wd_model2}. The new vacuum configuration is obtained by searching for the zeros of the $F-$terms of $w_{d}+\Delta w^{l}_d$ with respect to the driving fields $\xi^{0}$, $\rho^{0}$, $\zeta^{0}$, $\eta^{0}$, $\varphi^{0}_{T}$ and $\kappa^{0}$. To the first order in $1/\Lambda$ expansion, the LO vacuum alignment of the charged lepton sector is modified into
\begin{eqnarray}
\nonumber& \langle\xi\rangle=v_{\xi}+\delta v_{\xi}, \qquad \langle\eta\rangle=\left(\begin{array}{c}
    v_{\eta}+\delta v_{\eta_{1}} \\ \delta v_{\eta_{2}}
  \end{array}\right), \\
\label{eq:correct_vacuum_five}& \langle\varphi_{T}\rangle =\left(\begin{array}{c}
    (1+i)(v_{T}+\delta v_{T_{1}}) \\  \delta v_{T_{2}}  \\  (i-1)(v_{T}+\delta v_{T_{3}})
  \end{array}\right), \qquad
    \langle\phi\rangle =\left(\begin{array}{c}
    (i-1)(v_{\phi}+\delta v_{\phi_{1}}) \\  -i\delta v_{\phi_{2}}  \\  (1+i)(v_{\phi}+\delta v_{\phi_{3}})
  \end{array}\right)\,.
\end{eqnarray}
The shifts $\delta v_{\xi}$, $\delta v_{\eta_{i}}$, $\delta v_{T_{i}}$ and $\delta v_{\phi_{i}}$ are solved to be
\begin{eqnarray}\label{eq:correct_vacuum_seven}
\nonumber&\delta v_{\xi}=X\frac{v_{\xi}}{\Lambda}, \qquad \delta v_{\eta_{1}}=(X-\frac{f_{10}}{2f_{1}f_{2}})\frac{M_{\xi}v_{\eta}}{\Lambda},\\
\nonumber & \delta v_{\eta_{2}}=\delta v_{T_{2}}=0, \qquad  \delta v_{T_{1}}=\delta v_{T_{3}}=(X-\frac{f_{9}M^{2}_{\kappa}}{2f_{8}M^{2}_{\xi}})\frac{M_{\xi}v_{T}}{2\Lambda},\\
&\delta v_{\phi_{1}}=\delta v_{\phi_{3}}=\frac{3f_{6}\left[f_{2}(f_{6}f_{20}-f_{5}f_{21})-f_{3}f_{6}f_{19}\right]}{2f_{1}f_{2}f^{2}_{5}f_{8}}\frac{M_{\xi}v_{\phi}}{\Lambda},\qquad \delta v_{\phi_{2}}=-\frac{\sqrt{3}(f_{15}+f_{16})v^{2}_{\eta}v_{\phi}}{f_{7}v_{T}\Lambda}\,,
\end{eqnarray}
where $X$ is a real parameter of order one with
\begin{eqnarray}
\hskip-0.3in  X=\frac{\big[f_{2}f_{5}\big(2f_{8}(f_{13}+f_{14})-3f_{6}f_{21}\big)+f^{2}_{5}f_{8}f_{10}+3f^{2}_{6}(f_{2}f_{20}-f_{3}f_{19})\big]M^{2}_{\xi}-f_{1}f_{2}f^{2}_{5}f_{9}M^{2}_{\kappa}}{3f_{1}f_{2}f^{2}_{5}f_{8}M^{2}_{\xi}}\,.
\end{eqnarray}
Notice that the shifts of the vacuum are suppressed by $\lambda^2$ compared with the LO VEVs, and the structure of the LO vacuum of the flavons $\eta$ and $\varphi_{T}$ is unchanged by the NLO corrections. Because the NLO driving superpotental $\Delta w^{l}_d$ only contain the charged lepton flavon fields $\xi$, $\eta$, $\varphi_{T}$ and $\phi$, hence their VEVs still preserve the $Z^{SU}_2$ subgroup even at NLO. Indeed the vacuum in Eq.~\eqref{eq:correct_vacuum_five} is the most general form which is compatible with the residual family symmetry $Z^{SU}_2$ in the charged lepton sector.

In the same way, the subleading corrections to the flavon superpotential of $\rho$, $\sigma$ and $\varphi_{S}$ are of the form
\begin{equation}
\Delta w^{\nu}_{d}=(\sigma^{0}\xi\varphi_{T}\Psi^{2}_{\nu})_{\bf1}/\Lambda^{2}+(\sigma^{0}\phi^{2}\Psi^{2}_{\nu})_{\bf1}/\Lambda^{2}
+(\varphi^{0}_{S}\xi\varphi_{T}\Psi^{2}_{\nu})_{\bf1}/\Lambda^{2}+(\varphi^{0}_{S}\phi^{2}\Psi^{2}_{\nu})_{\bf1}/\Lambda^{2}\,.
\end{equation}
where $\Psi_{\nu}=\{\rho,\sigma,\varphi_{S}\}$ denotes the neutrino flavon fields, and the real coupling constant in front of each term has been omitted. The resulting contributions to the $F-$terms of the driving fields $\sigma^{0}$ and $\varphi^{0}_{S}$ are suppressed by $\langle\xi\rangle\langle\varphi_T\rangle/\Lambda^2\sim\langle\phi\rangle^2/\Lambda\sim\lambda^4$ with respect to the LO terms in Eq.~\eqref{eq:wd_model2}. Hence they induce shifts in the VEVs of $\rho$, $\sigma$ and $\varphi_{S}$ at relative order $\lambda^4$. After some straightforward algebra, the new VEVs can be written as
\begin{eqnarray}
\label{eq:correct_vacuum_seven}
\langle\rho\rangle=v_{\rho}, \quad \langle\sigma\rangle=\left(\begin{array}{c}
    1+\epsilon_{1}\lambda^{4}  \\  \sqrt{3}+\epsilon_{2}\lambda^{4}
  \end{array}\right)v_{\sigma}\;,\quad  \langle\varphi_{S}\rangle=\left(\begin{array}{c}
    1+\epsilon_{3}\lambda^{4} \\ \epsilon_{4}\lambda^{4}  \\ -1+\epsilon_{5}\lambda^{4}
  \end{array}\right)v_{S} \,,
\end{eqnarray}
where $v_{\rho}$ remains undetermined, and the coefficients $\epsilon_{i}(i=1,2,...5)$ are unspecified constants with absolute value of order one. In the following we study the subleading corrections to the LO mass matrices from both the modified vacuum and higher dimensional operators in the Yukawa superpotential $w_{l}$ and $w_{\nu}$.

In the neutrino sector, the subleading operators are obtained by adding to each term of $w_{\nu}$ the factor of $\xi\varphi_{T}$ or $\phi^2$ in all possible ways, i.e.
\begin{equation}
\Delta w_{\nu}=(\nu^{c}l\xi\varphi_{T})_{\bf1}h_{u}/\Lambda^{2}+(\nu^{c}l\phi^2)_{\bf1}h_{u}/\Lambda^{2}
+(\nu^{c}\nu^{c}\xi\varphi_{T}\Psi_{\nu})_{\mathbf{1}}/\Lambda^2+(\nu^{c}\nu^{c}\phi^2\Psi_{\nu})_{\mathbf{1}}/\Lambda^2\,.
\end{equation}
In addition to these corrections, we have to consider the ones from $w_{\nu}$ in Eq.~\eqref{eq:wnu_model2} with the deviations of the VEVs at NLO, as shown in Eq.~\eqref{eq:correct_vacuum_seven}. Eventually we find that the neutrino mass matrix is corrected by terms of relative order $\lambda^4$ in every entry. As a result, the lepton mixing parameters acquire corrections of order $\lambda^4$ which can be safely neglected.

The NLO operators contributing to the charged lepton masses are given by
\begin{eqnarray}\label{eq:charged_superpotential_four}
\nonumber&& \Delta w_{l}=y_{e_1}e^{c}\xi(l(\eta\phi)_{\bf3^{\prime}})_{\bf1^{\prime}}h_{d}/\Lambda^{3}+y_{e_2}e^{c}\xi(l(\varphi_{T}\varphi_{T})_{\bf3^{\prime}})_{\bf1^{\prime}}h_{d}/\Lambda^{3}+y_{e_3}e^{c}((l\varphi_{T})_{\bf2}(\phi\phi)_{\bf2})_{\bf1^{\prime}}h_{d}/\Lambda^{3}\\
\nonumber &&\qquad+y_{e_4}e^{c}((l\varphi_{T})_{\bf3}(\phi\phi)_{\bf3^{\prime}})_{\bf1^{\prime}}h_{d}/\Lambda^{3}+y_{e_5}e^{c}((l\varphi_{T})_{\bf3^{\prime}}(\phi\phi)_{\bf3})_{\bf1^{\prime}}h_{d}/\Lambda^{3}+y_{\mu_1}\mu^{c}\xi(l(\eta\varphi_{T})_{\bf3^{\prime}})_{\bf1^{\prime}}h_{d}/\Lambda^{3}\\
\nonumber&&\qquad+y_{\mu_2}\mu^{c}((l\eta)_{\bf3}(\phi\phi)_{\bf3^{\prime}})_{\bf1^{\prime}}h_{d}/\Lambda^{3}+y_{\mu_3}\mu^{c}((l\eta)_{\bf3^{\prime}}(\phi\phi)_{\bf3})_{\bf1^{\prime}}h_{d}/\Lambda^{3}+y_{\mu_4}\mu^{c}(l\phi)_{\bf1^{\prime}}(\varphi_{T}\varphi_{T})_{\bf1}h_{d}/\Lambda^{3}\\
\nonumber&&\qquad+y_{\mu_5}\mu^{c}((l\phi)_{\bf2}(\varphi_{T}\varphi_{T})_{\bf2})_{\bf1^{\prime}}h_{d}/\Lambda^{3}+y_{\mu_6}\mu^{c}((l\phi)_{\bf3}(\varphi_{T}\varphi_{T})_{\bf3^{\prime}})_{\bf1^{\prime}}h_{d}/\Lambda^{3}\\
\nonumber&&\qquad+y_{\mu_7}\mu^{c}((l\phi)_{\bf3^{\prime}}(\varphi_{T}\varphi_{T})_{\bf3})_{\bf1^{\prime}}h_{d}/\Lambda^{3}+y_{\tau_1}\tau^{c}\xi(l(\eta\phi)_{\bf3})_{\bf1}h_{d}/\Lambda^{3}+y_{\tau_2}\tau^{c}\xi(l(\varphi_{T}\varphi_{T})_{\bf3})_{\bf1}h_{d}/\Lambda^{3}\\
\nonumber&&\qquad+y_{\tau_3}\tau^{c}(l\varphi_{T})_{\bf1}(\phi\phi)_{\bf1}h_{d}/\Lambda^{3}+y_{\tau_4}\tau^{c}((l\varphi_{T})_{\bf2}(\phi\phi)_{\bf2})_{\bf1}h_{d}/\Lambda^{3}\\
&&\qquad+y_{\tau_5}\tau^{c}((l\varphi_{T})_{\bf3}(\phi\phi)_{\bf3})_{\bf1}h_{d}/\Lambda^{3}+y_{\tau_6}\tau^{c}((l\varphi_{T})_{\bf3^{\prime}}(\phi\phi)_{\bf3^{\prime}})_{\bf1}h_{d}/\Lambda^{3}\;.
\end{eqnarray}
The charged lepton mass matrix is obtained by inserting the shifted vacuum alignment of Eq.~\eqref{eq:correct_vacuum_five} into the
LO operators plus the contribution of these higher dimensional operators evaluated with the LO VEVs of Eq.~\eqref{eq:VEVs_five}. We find that the charged lepton mass matrix including NLO corrections takes the following form
\begin{eqnarray}
m_{l}\simeq\left(\begin{array}{ccc}
-(1+i)a_{1}v_{T}v^{2}_{\phi}/\Lambda^{3}  & ~ 4iy_{e_2}v_{\xi}v^{2}_{T}/\Lambda^{3}~  & (1-i)a_{1}v_{T}v^{2}_{\phi}/\Lambda^{3} \\
(1+i)y_{\mu}v_{\xi}v_{\phi}/\Lambda^{2}   &~  -ib_1v_{\xi}v_{\phi}v^2_{\eta}/(\Lambda^{3}v_{T}) ~ &  (i-1)y_{\mu}v_{\xi}v_{\phi}/\Lambda^{2}  \\
(i-1)y_{\tau}v_{T}/\Lambda       &              ~  0 ~                      &  (1+i)y_{\tau}v_{T}/\Lambda
\end{array}\right)v_{d}\,,
\end{eqnarray}
where both $a_1$ and $b_1$ are real parameters,
\begin{eqnarray}
\nonumber&a_1=4(\sqrt{3}\,y_{e3}+y_{e4})+y_{e_1}\frac{v_{\xi}v_{\eta}}{v_{T}v_{\phi}}=4(\sqrt{3}\,y_{e3}+y_{e4})+4\sqrt{3}y_{e_1}\frac{f_6}{f_5},\\
\label{eq:parameter_one}&b_1=\frac{v_{T}}{v_{\xi}v_{\phi}v^2_{\eta}}\left(y_{\mu}v_{\xi}\delta v_{\phi_{2}}\Lambda+8y_{\mu2}v_{\eta}v^{2}_{\phi}\right)=\frac{2y_{\mu_2}f_5}{\sqrt{3}f_{6}}-\frac{\sqrt{3}y_{\mu}(f_{15}+f_{16})}{f_7}\,.
\end{eqnarray}
In order to diagonalize the charged lepton mass matrix $m^{\dagger}_{l}m_{l}$, it is helpful to apply the LO unitary transformation $U_{l}$ in Eq.~\eqref{eq:UL_model2} first, i.e.
\begin{eqnarray}
\hskip-0.2in
\label{eq:charg_mass_matr_refine_model2}U^{\dagger}_{l}m^{\dagger}_{l}m_{l}U_{l}\simeq\left(\begin{array}{ccc}
    16y^2_{e_2}\frac{|v_{\xi}|^2|v_T|^4}{\Lambda^6}+b^2_1\frac{|v_{\xi}|^2|v_{\phi}|^2|v_{\eta}|^4}{|v_T|^2\Lambda^6} & ~ 2b_1y_{\mu}\frac{|v_{\xi}|^2|v_{\phi}|^2v^{*2}_{\eta}}{v^{*}_T\Lambda^5} & 0 \\
 2b_1y_{\mu}\frac{|v_{\xi}|^2|v_{\phi}|^2v^{2}_{\eta}}{v_T\Lambda^5} & ~4y^2_{\mu}\frac{|v_{\xi}|^2|v_{\phi}|^2}{\Lambda^4} & 0  \\
   0  & ~ 0  &  4y^2_{\tau}\frac{|v_{T}|^{2}}{\Lambda^{2}}
  \end{array}\right)v^{2}_{d}\,,
\end{eqnarray}
which can be easily diagonalized by a rotation in the (1, 2) sector. From Eq.~\eqref{eq:VEV_relation_model2} and Eq.~\eqref{eq:vxi_model2}, we see that $v^2_{\eta}$ is real since $v_{\xi}$ is chosen to be real, while the VEV $v_{T}$ can be real or pure imaginary depending on the sign of the product $f_{2}f_{3}f_{8}$. In case of $f_{2}f_{3}f_{8}<0$, $v_{T}$ is real. The combination $m^{\dagger}_{l}m_{l}$ is invariant under $Z^{SU}_2\times H^{l}_{CP}$ with $H^{l}_{CP}=\left\{\rho_{\mathbf{r}}(TST^2), \rho_{\mathbf{r}}(TST^2U) \right\}$. As a consequence, the scenario analyzed in section~\ref{sec:general_analysis_one_row} is realized. Then the lepton mixing matrix is of the form
\begin{equation}
\label{eq:mixing1_m2}U_{PMNS}=\frac{1}{2}\left(
\begin{array}{ccc}
 \sin\theta+\sqrt{2}\cos\theta ~&~ \sin\theta-\sqrt{2}\cos\theta ~&~ i\sqrt{2}\sin\theta \\
 \cos\theta-\sqrt{2}\sin\theta ~&~ \cos\theta+\sqrt{2}\sin\theta ~&~ i\sqrt{2}\cos\theta \\
 1 ~&~ 1 ~&~ -i\sqrt{2} \\
\end{array}
\right)\,,
\end{equation}
where the parameter $\theta$ is
\begin{equation}
\tan2\theta=-\frac{b_1}{y_{\mu}}\frac{v^2_{\eta}}{v_{T}\Lambda}\,.
\end{equation}
The predictions for lepton mixing parameters are given in Eq.~\eqref{eq:mix_parameters_caseIII_1oct}. In this case both Dirac CP phase and Majorana CP phases are trivial, and very good agreement with the experimental data can be achieved for appropriate values of the parameter $\theta$, as shown in Eq.~\eqref{eq:bf_caseIII_1oct} and Eq.~\eqref{eq:bf_caseIII_2oct}. In order to achieve the correct size of $\theta\sim\lambda$, an accidental enhancement of the combination $\frac{b_1}{y_{\mu}}=\frac{2y_{\mu_2}f_5}{\sqrt{3}y_{\mu}f_{6}}-\frac{\sqrt{3}(f_{15}+f_{16})}{f_7}$ of order $1/\lambda$ is required. If the two terms $\frac{2y_{\mu_2}f_5}{\sqrt{3}y_{\mu}f_{6}}$ and $\frac{\sqrt{3}(f_{15}+f_{16})}{f_7}$ are of opposite sign, then the two factors sum up and the required values can be easily explained. The charged lepton masses are determined to be
\begin{equation}
m_{e}\simeq4\left|y_{e_2}\frac{v_{\xi}v^2_{T}}{\Lambda^3}\right|v_d,\qquad m_{\mu}\simeq2\left|y_{\mu}\frac{v_{\xi}v_{\phi}}{\Lambda^2}\right|v_{d},\qquad m_{\tau}\simeq2\left|y_{\tau}\frac{v_{T}}{\Lambda}\right|v_{d}\,.
\end{equation}
The electron mass is generated at NLO level, and realistic charged lepton mass hierarchy $m_{e} : m_{\mu} : m_{\tau}\simeq\lambda^4 : \lambda^2 : 1$ is produced.

For the mixing pattern shown in Eq.~\eqref{eq:mixing1_m2}, the atmospheric mixing angle $\theta_{23}$ fulfills
\begin{equation}
\sin^2\theta_{23}=\frac{1+\cos2\theta}{3+\cos2\theta}=\frac{1}{1+\sec^2\theta}\leq\frac{1}{2}\,.
\end{equation}
As a result, $\theta_{23}$ deviates from maximal mixing and it lies in the first octant in this model.

Since the octant of $\theta_{23}$ is not known so far, we would like to minimally modify this model to accommodate the situation of $\theta_{23}>45^{\circ}$. The family symmetry is still $S_4\times Z_5\times Z_6$. For the assignment of the fields, only the right-handed charged leptons $\mu^{c}$ and $\tau^{c}$ are changed to be in $\left(\mathbf{1}, \omega^2_5, \omega^3_{6}\right)$ and $\left(\mathbf{1}^{\prime}, \omega^2_5, \omega^3_{6}\right)$ under $S_4\times Z_5\times Z_6$. Because both flavon fields and driving fields are kept intact, the vacuum is unchanged. Then the LO vacuum configuration is still given in Eqs.~\eqref{eq:VEVs_five} and \eqref{eq:VEVs_seven}, and the NLO VEVs are given by Eqs.~\eqref{eq:correct_vacuum_five} and \eqref{eq:correct_vacuum_seven}. After including the subleading contributions in the same manner described in  previous paragraphs, we find that the PMNS matrix is related to the corresponding one of previous model by exchanging its second and third rows. As a consequence, the atmospheric angle $\theta_{23}$ is in the second octant.

\section{\label{sec:conclusion}Summary and conclusions}

Although the BM mixing pattern has already been ruled out by experiment data, the scheme of keeping one column or one row of BM mixing is viable. We perform a comprehensive analysis of how to naturally realize this scheme from $S_4$ family symmetry and generalized CP symmetry in this paper. Furthermore, two models with $S_4$ family symmetry and generalized CP are constructed to implement the model independent results enforced by remnant symmetry.

We firstly study the deviation from BM mixing which originates from a rotation between two generation of neutrinos or charged leptons. The phenomenological predictions for the lepton mixing angles and Dirac CP phase are discussed in detail. In this approach, all mixing parameters depend on two real parameters $\theta$ and $\delta$ while the Majorana CP phases are indeterminate. For an additional rotation of 1-2 or 1-3 generation of charged leptons in the BM basis, good agreement with experiment data can be achieved, and the Dirac CP phase $\delta_{CP}$ is constrained to be in the range of $\pm\left[2.52,\pi\right]$ or $\left[-0.62, 0.62\right]$ respectively, after the present $3\sigma$ bounds of mixing angles from global data analysis are taken into account. For rotations in the neutrino sector, the measured values of the lepton mixing angles can not be accommodated. With the help of independent permutations of rows and columns of the PMNS matrix, interesting mixing patterns shown in Eq.~\eqref{eq:neutrino_perturbation} is found. The Dirac CP phase is in the range of $\pm\left[2.04, \pi\right]$ or $\left[-1.10, 1.10\right]$. Note that $\delta_{CP}$ can vary within a quite wide range.

Since the BM mixing can be derived if we impose $S_4$ family symmetry and spontaneously break it down to $G_{\nu}=K^{(TST^2,T^2U)}_4$ in the neutrino sector and to $G_{l}=Z^{TST^2U}_4$ or $G_{l}=K^{\left(S,U\right)}_4$ in the charged lepton sector. It is easy to see that one column of the BM matrix would be retained if we degrade $G_{\nu}$ from $K_4$ to $Z_2$ subgroup, and one row of the BM mixing would be preserved once $G_{l}$ is degraded from $K_4$ (or $Z_4$) to $Z_2$. In order to have definite predictions for the leptonic CP violating phases, we extend the $S_4$ family symmetry to include generalized CP symmetry. The phenomenological implications of the symmetry breaking of $S_4\rtimes H_{CP}$ into $Z^{ST^2SU}_2\times H^{\nu}_{CP}$ in the neutrino sector and $Z^{TST^{2}U}_4\rtimes H^{l}_{CP}$ in the charged lepton sector have been discussed by Feruglio et al~\cite{Feruglio:2012cw}. The resulting PMNS matrix is found to have one column of the form $\left(1/2, 1/\sqrt{2}, 1/2\right)^{T}$ or $\left(1/2, 1/2, 1/\sqrt{2}\right)^{T}$, and the Dirac CP phase $\delta_{CP}$ as well as the Majorana CP phases are predicted to be conserved to account for the measured values of the mixing angles. In this work, the predictions for $0\nu2\beta$ decay are studied. The effective mass $\left|m_{ee}\right|$ is predicted to be around 0.049 eV and 0.023 eV for inverted ordering spectrum. Hence this mixing pattern can be tested by future $0\nu2\beta$ experiments.

It is usually assumed the remnant symmetry in the neutrino sector is $Z_2\times CP$ in the context of family symmetry combined with generalized CP. In this work, we also consider another situation that $Z_2\times CP$ is preserved in the charged lepton sector instead of in the neutrino sector.  The lepton flavor mixing arising from the remnant symmetry $K^{(TST^2, T^2U)}_4\rtimes H^{\nu}_{CP}$ in the neutrino sector and $Z^{SU}_{2}\times H^{l}_{CP}$ in the charged lepton sector is explored in a model independent way. One row of PMNS matrix is determined to be $\left(1/2, 1/2, -i/\sqrt{2}\right)$, and both Dirac CP and Majorana CP are fully conserved as well to fit the data on mixing angles. In this case, The effective mass $\left|m_{ee}\right|$ is determined to be around the $3\sigma$ upper or lower limit for inverted hierarchy. This prediction can also be tested by future $0\nu2\beta$ experiments. Furthermore, our above prediction for $\delta_{CP}$ can be directly tested by forthcoming long baseline neutrino oscillation experiments LBNE, LBNO and Hyper-Kamiokande. If signal of leptonic CP violation is discovered, our proposal would be ruled out.

Inspired by the above fascinating results, we construct a model based on $S_4\rtimes H_{CP}$ which is spontaneously broken down to $Z^{ST^2SU}_2\times H^{\nu}_{CP}$ in the neutrino sector and $Z^{TST^{2}U}_4\rtimes H^{l}_{CP}$ in the charged lepton sector by the VEVs of flavons. The PMNS matrix is really found to be of the form predicted in Ref.~\cite{Feruglio:2012cw}. At leading order, the light neutrino mass matrix effectively contains only three real parameters which can be fixed by the measured values of the mass-squared difference $\delta m^2\equiv m^2_2-m^2_1$ and $\Delta m^2\equiv m^2_3-(m^2_1+m^2_2)/2$ and the reactor angle $\theta_{13}$. As a consequence, the light neutrino masses are completely determined. The predictions for the effective mass $|m_{ee}|$ are safely below the present upper limit, and yet they are within the future sensitivity of planned neutrinoless double-beta decay experiments. Although $\theta_{12}$ is slightly smaller than its $3\sigma$ lower bound at leading order, agreement with experimental data can be achieved after subleading corrections are included.

Moreover, we present another model and its variant where the BM mixing is realized at LO. After the NLO corrections are included, the charged lepton mass hierarchy is obtained and the BM mixing is corrected by the effect of charged lepton diagonalization. One row of PMNS matrix is determined to be $\left(1/2, 1/2, 1/\sqrt{2}\right)$ or $\left(1/2, 1/2, -i/\sqrt{2}\right)$, and all the general model independent predictions for lepton flavor mixing in section~\ref{sec:general_analysis_one_row} are naturally reproduced. The Dirac CP phase $\delta_{CP}$ is trivial 0 or $\pi$ for $f_2f_3f_8<0$.

In the past years, family symmetry and generalized CP symmetry has been shown to be a very powerful and promising framework to predict lepton mixing angles and CP violating phases. It is intriguing to extend this approach to the quark sector to understand the established CP violation at $B-$factory and strong CP problem.

\section*{Acknowledgements}

This work is supported by the National Natural Science Foundation of China under Grant Nos. 11275188 and 11179007.

\vfill
\newpage

\section*{Appendix}

\begin{appendix}

\section{\label{sec:appA}Group theory of $S_4$ and Clebsch-Gordan coefficients}
\cleqn

$S_4$ is a symmetric group of degree four, and it is a good candidate for a family symmetry to realize the tri-bimaximal and BM mixing. Hence $S_4$ has been widely studied in the literature. For the sake of being self-contained, in the following we shall present our convention for the $S_4$ group, the working basis and the associated Clebsch-Gordan coefficients. $S_4$ group can be generated by three generators $S$, $T$ and $U$ obeying the relations~\cite{Li:2013jya}
\begin{equation}
S^{2}=T^{3}=U^{2}=(ST)^3=(SU)^{2}=(TU)^{2}=(STU)^{4}=1\,.
\end{equation}
Note that the chosen generators $\widetilde{S}$ and $\widetilde{T}$ of Ref.~\cite{Altarelli:2009gn} are related to our generators $S$, $T$ and $U$ via $\widetilde{S}=ST^{2}SU$ and $\widetilde{T}=T^{2}STU$ or vice versa $S=\widetilde{T}^{2}$, $T=\widetilde{T}\widetilde{S}$, $U=\widetilde{S}\widetilde{T}^{2}\widetilde{S}\widetilde{T}$. It is straightforward to check that the multiplication rules $\widetilde{T}^4=\widetilde{S}^2=\left(\widetilde{S}\widetilde{T}\right)^3=\left(\widetilde{T}\widetilde{S}\right)^3=1$ are satisfied. The 24 group elements can be divided into the five conjugacy classes as follows:
\begin{eqnarray}
\nonumber 1C_1 &=& \left\{1 \right\}, \\
\nonumber 3C_2 &=& \left\{S, TST^2, T^2ST \right\},  \\
\nonumber 6C_2^{\prime} &=& \left\{U,TU,SU,T^2U,STSU,ST^2SU\right\},  \\
\nonumber 8C_3 &=& \left\{T,ST,TS,STS,T^2,ST^2,T^2S,ST^2S\right\},  \\
6C_4 &=& \left\{STU,TSU,T^2SU,ST^2U,TST^2U,T^2STU\right\}\,,
\end{eqnarray}
where $kC_{n}$ denotes a conjugacy class with $k$ elements and the subscript $n$ is the order of its elements. Since the number of conjugacy class is equal to the number the number of irreducible representation, $S_4$ has five irreducible representations: two
singlet representations $\mathbf{1}$ and $\mathbf{1}^{\prime}$, one doublet representation $\mathbf{2}$ and two triplet representations $\mathbf{3}$ and $\mathbf{3}^{\prime}$. Note that both $\mathbf{3}$ and $\mathbf{3^{\prime}}$ are faithful representations of $S_4$. Our choice for the representation matrices of the generators $S$, $T$ and $U$ are listed in Table~\ref{tab:representation}. For the three-dimensional representation $\mathbf{3}$, the representation matrices for the elements are as follows:
\begin{table}[t!]
\begin{center}
\begin{tabular}{|c|c|c|c|}\hline\hline
 ~~  &  $S$  &   $T$    &  $U$  \\ \hline
~~~$\mathbf{1}$, $\mathbf{1^{\prime}}$ ~~~ & 1   &  1  & $\pm1$  \\ \hline
   &   &    &    \\ [-0.16in]
$\mathbf{2}$ &  $\begin{pmatrix}
    1&0 \\
    0&1
    \end{pmatrix}$
    & $\frac{1}{2}\begin{pmatrix}
    -1 & ~\sqrt{3} \\
    -\sqrt{3} & ~-1
    \end{pmatrix}$
    & $\begin{pmatrix}
    1 & 0 \\
    0 & ~-1
    \end{pmatrix}$\\
    &   &   &  \\ [-0.16in] \hline
&   &   &  \\ [-0.16in]
$\mathbf{3}$, $\mathbf{3^{\prime}}$ & $ \begin{pmatrix}
    -1& 0  & 0  \\
    0  & 1 & 0 \\
    0 & 0 & -1
    \end{pmatrix}$
    & $\frac{1}{2}\begin{pmatrix}
     i &  -\sqrt{2}\;i  & ~-i \\
   \sqrt{2}  & 0  & \sqrt{2} \\
   i & \sqrt{2}\;i   & -i
   \end{pmatrix} $
    & $\mp\left( \begin{array}{ccc}
    0 & 0 & -i \\
    0 & 1 & 0 \\
    i & 0 & 0
    \end{array}\right)$\\
&   &   &  \\ [-0.16in]\hline\hline
\end{tabular}
\caption{\label{tab:representation} The representation matrices of the generators $S$, $T$ and $U$ for the five irreducible representations of $S_4$ in our working basis.}
\end{center}
\end{table}

\begin{eqnarray*}
\nonumber&&1C_1: 1=\begin{pmatrix}
 1 & 0 & 0 \\
 0 & 1 & 0 \\
 0 & 0 & 1
\end{pmatrix},\\
\nonumber&&3C_2: S=\begin{pmatrix}
 -1 & 0 & 0 \\
 0 & 1 & 0 \\
 0 & 0 & -1
\end{pmatrix},\quad  TST^{2}=\begin{pmatrix}
 0 & 0 & -1 \\
 0 & -1 & 0 \\
 -1 & 0 & 0
\end{pmatrix},\quad T^{2}ST=\begin{pmatrix}
 0 & 0 & 1 \\
 0 & -1 & 0 \\
 1 & 0 & 0
\end{pmatrix},\\
\nonumber&&6C_2^{\prime}: U=\begin{pmatrix}
 0 & 0 & i \\
 0 & -1 & 0 \\
 -i & 0 & 0
\end{pmatrix},\quad TU=\frac{1}{2}\begin{pmatrix}
 -1 & \sqrt{2}\,i & -1 \\
 -\sqrt{2}\,i & 0 & \sqrt{2}\,i \\
 -1 & -\sqrt{2}\,i & -1
\end{pmatrix}, \quad SU=\begin{pmatrix}
 0 & 0 & -i \\
 0 & -1 & 0 \\
 i & 0 & 0
\end{pmatrix},\\
\nonumber&&\qquad T^{2}U=\frac{1}{2}\begin{pmatrix}
 -1 & -\sqrt{2} & 1 \\
 -\sqrt{2} & 0 & -\sqrt{2} \\
 1 & -\sqrt{2} & -1
\end{pmatrix},\quad STSU=\frac{1}{2}\begin{pmatrix}
 -1 & -\sqrt{2}\,i & -1 \\
 \sqrt{2}\,i & 0 & -\sqrt{2}\,i \\
 -1 & \sqrt{2}\,i & -1
\end{pmatrix},\\
\nonumber&&\qquad ST^{2}SU=\frac{1}{2}\begin{pmatrix}
 -1 & \sqrt{2} & 1 \\
 \sqrt{2} & 0 & \sqrt{2} \\
 1 & \sqrt{2} & -1
\end{pmatrix},\\
\nonumber&& 8C_3: T=\frac{1}{2}\begin{pmatrix}
 i & ~-\sqrt{2}\,i & ~-i \\
 \sqrt{2} & 0 & \sqrt{2} \\
 i & \sqrt{2}\,i & -i
\end{pmatrix},\quad ST=\frac{1}{2}\begin{pmatrix}
 -i & \sqrt{2}\,i & i \\
 \sqrt{2} & 0 & \sqrt{2} \\
 -i & ~-\sqrt{2}\,i & i
\end{pmatrix},\quad TS=\frac{1}{2}\begin{pmatrix}
 -i & -\sqrt{2}\,i & i \\
 -\sqrt{2} & 0 & -\sqrt{2} \\
 -i & \sqrt{2}\,i & i
\end{pmatrix},\\
\nonumber&&\qquad STS=\frac{1}{2}\begin{pmatrix}
 i & \sqrt{2}\,i & -i \\
 -\sqrt{2} & 0 & -\sqrt{2} \\
 i & -\sqrt{2}\,i & -i
\end{pmatrix},~~ T^{2}=\frac{1}{2}\begin{pmatrix}
 -i & \sqrt{2} & -i \\
 \sqrt{2}\,i & 0 & -\sqrt{2}\,i \\
 i & \sqrt{2} & i
\end{pmatrix},~~ ST^{2}=\frac{1}{2}\begin{pmatrix}
 i & -\sqrt{2} & i \\
 \sqrt{2}\,i & 0 & -\sqrt{2}\,i \\
 -i & -\sqrt{2} & -i
\end{pmatrix},\\
\nonumber&&\qquad T^{2}S=\frac{1}{2}\begin{pmatrix}
 i & \sqrt{2} & i \\
 -\sqrt{2}\,i & 0 & \sqrt{2}\,i \\
 -i & \sqrt{2} & -i
\end{pmatrix},\quad ST^{2}S=\frac{1}{2}\begin{pmatrix}
 -i & -\sqrt{2} & -i \\
 -\sqrt{2}\,i & 0 & \sqrt{2}\,i \\
 i & -\sqrt{2} & i
\end{pmatrix},\\
\nonumber&&6C_4: STU=\frac{1}{2}\begin{pmatrix}
 1 & -\sqrt{2}\,i & 1 \\
 -\sqrt{2}\,i & 0 & \sqrt{2}\,i \\
 1 & \sqrt{2}\,i & 1
\end{pmatrix},\quad TSU=\frac{1}{2}\begin{pmatrix}
 1 & \sqrt{2}\,i & 1  \\
 \sqrt{2}\,i & 0 & -\sqrt{2}\,i \\
 1 & -\sqrt{2}\,i & 1
\end{pmatrix},\\
\nonumber&&\qquad T^{2}SU=\frac{1}{2}\begin{pmatrix}
 1 & -\sqrt{2} & -1 \\
 \sqrt{2} & 0 & \sqrt{2} \\
 -1  & -\sqrt{2} & 1
\end{pmatrix},\quad ST^{2}U=\frac{1}{2}\begin{pmatrix}
 1 & \sqrt{2}  & -1 \\
 -\sqrt{2} & 0 & -\sqrt{2} \\
 -1 &  \sqrt{2}  &  1
\end{pmatrix},\quad TST^{2}U=\begin{pmatrix}
 i & 0 & 0 \\
 0 & 1 & 0 \\
 0 & 0 & ~-i
\end{pmatrix},\\
\nonumber&&\qquad T^{2}STU=\begin{pmatrix}
 -i & 0 & 0 \\
 0 & 1 & 0 \\
 0 & 0 & i
\end{pmatrix}\,.
\end{eqnarray*}
For the $\mathbf{3}^{\prime}$ representation, the matrices representing the elements of $1C_1$, $3C_2$ and $8C_3$ are the same as those listed above for the representation $\mathbf{3}$, while they are the opposite for $6C^{\prime}_2$ and $6C_{4}$. The reason is that the generator $U$ changes its sign in $\mathbf{3}$ and $\mathbf{3}^{\prime}$ representations,  the elements in $1C_1$, $3C_2$ and $8C_3$ contain an even number of $U$, while those in $6C^{\prime}_2$ and $6C_{4}$ contain an odd number of $U$. Character of an element is the trace of its representation matrix. The character table of $S_4$ group can be easily obtained, as shown in Table~\ref{tab:S4Classes}. The Kronecker products between various irreducible representations follow immediately:
\begin{table}[t!]
\begin{center}
\begin{tabular}{|c|c|c|c|c|c||c|}
  \hline
  & $\chi_{\mathbf{1}}$ & $\chi_{\mathbf{1^\prime}}$ & $\chi_{\mathbf{2}}$ & $\chi_{\mathbf{3}}$ & $\chi_{\mathbf{3^\prime}}$ & Example\\
  \hline
  $1C_1$ & $1$ & $1$ & $2$ & $3$ & $3$  &  1  \\
  $3C_2$ & $1$ & $1$ & 2 & $-1$ & $-1$   & $S$  \\
  $6C^{\prime}_2$ & $1$ & $-1$ & $0$ & $-1$ & $1$  &  $U$  \\
  $8C_3$ & $1$ & $1$ & $-1$ & $0$ & $0$  &   $T$ \\
  $6C_4$ & $1$ & $-1$ & $0$ & $1$ & $-1$ & $STU$  \\
  \hline
  \end{tabular}
\end{center}
\begin{center}
\caption{\label{tab:S4Classes}Character table of $S_4$. We give an example of the elements for each class in the last column.}
\end{center}
\end{table}

\begin{eqnarray}
\nonumber&& \mathbf{1}\otimes R=R\otimes\mathbf{1}=R,\quad \mathbf{1^{\prime}}\otimes\mathbf{1^{\prime}}=\mathbf{1},\quad \mathbf{1^{\prime}}\otimes\mathbf{2}=\mathbf{2},\quad \mathbf{1^{\prime}}\otimes\mathbf{3}=\mathbf{3^{\prime}},\quad \mathbf{1^{\prime}}\otimes\mathbf{3^{\prime}}=\mathbf{3},\\
\nonumber&& \mathbf{2}\otimes\mathbf{2}=\mathbf{1}\oplus\mathbf{1^{\prime}}\oplus\mathbf{2},\quad \mathbf{2}\otimes\mathbf{3}=\mathbf{2}\otimes\mathbf{3^{\prime}}=\mathbf{3}\oplus\mathbf{3^{\prime}},\\
&&\mathbf{3}\otimes\mathbf{3}=\mathbf{3^{\prime}}\otimes\mathbf{3^{\prime}}=\mathbf{1}\oplus\mathbf{2}\oplus\mathbf{3}\oplus\mathbf{3^{\prime}},\quad
\mathbf{3}\otimes\mathbf{3^{\prime}}=\mathbf{1^{\prime}}\oplus\mathbf{2}\oplus\mathbf{3}\oplus\mathbf{3^{\prime}}\,.
\end{eqnarray}
where $R$ denotes any $S_4$ irreducible representation. In the following, we shall present the Clebsch-Gordan (CG) coefficients in our basis. we use $\alpha_{i}$ to indicate the elements of the first representation of the product and $\beta_{i}$ to indicate those of the second representation. We first report the CG coefficients associated with the singlet representation $\mathbf{1^{\prime}}$:
\begin{equation}
\begin{array}{l}
\mathbf{1^{\prime}}\otimes\mathbf{1^{\prime}}~=\mathbf{1}~\sim\alpha\beta\\[-7pt]
            \\[4pt]
\mathbf{1^{\prime}}\otimes\mathbf{2}~=\mathbf{2}~\sim\begin{pmatrix}
                    \alpha\beta_2 \\
                    -\alpha\beta_1 \\
            \end{pmatrix}\\[-7pt]
                  \\[4pt]
\mathbf{1^{\prime}}\otimes\mathbf{3}~=\mathbf{3^{\prime}}\sim\begin{pmatrix}
                    \alpha\beta_1 \\
                    \alpha\beta_2 \\
                    \alpha\beta_3\\
                    \end{pmatrix}\\[-7pt]
                   \\[4pt]
\mathbf{1^{\prime}}\otimes\mathbf{3^{\prime}}~=\mathbf{3}~\sim\begin{pmatrix}
                            \alpha\beta_1 \\
                            \alpha\beta_2 \\
                            \alpha\beta_3\\
                    \end{pmatrix}
\end{array}
\end{equation}
The CG coefficients for the products involving the doublet representation $\mathbf{2}$ are the following ones:
\begin{equation}
\begin{array}{ll}
\mathbf{2}\otimes\mathbf{2}~=\mathbf{1}\oplus\mathbf{1^{\prime}}\oplus\mathbf{2} &\qquad\quad
\text{with}\qquad\quad\left\{\begin{array}{l}
                    \mathbf{1}~\sim\alpha_1\beta_1+\alpha_2\beta_2\\[-7pt]
                    \\[4pt]
                     \mathbf{1^{\prime}}\sim\alpha_1\beta_2-\alpha_2\beta_1\\[-7pt]
                    \\[4pt]
                    \mathbf{2}~\sim\begin{pmatrix}
                        \alpha_2\beta_2-\alpha_1\beta_1 \\
                        \alpha_1\beta_2+\alpha_2\beta_1\\
                    \end{pmatrix}
                  \end{array}
            \right.\\[-7pt]
                    \\[2pt]
\mathbf{2}\otimes\mathbf{3}~=\mathbf{3}\oplus\mathbf{3^{\prime}}& \qquad\quad
\text{with}\qquad\quad\left\{\begin{array}{l}
                    \mathbf{3}~\sim\begin{pmatrix}
                        \sqrt{3}\alpha_{2}\beta_{3}-\alpha_1\beta_1\\
                        2\alpha_1\beta_2 \\
                        \sqrt{3}\alpha_2\beta_1-\alpha_1\beta_3 \\
                    \end{pmatrix}\\[-7pt]
                    \\[4pt]
                   \mathbf{3^{\prime}}~\sim\begin{pmatrix}
                        \sqrt{3}\alpha_1\beta_3+\alpha_2\beta_1\\
                        -2\alpha_2\beta_2 \\
                        \sqrt{3}\alpha_1\beta_1+\alpha_2\beta_3 \\
                    \end{pmatrix}\\
                    \end{array}
            \right.\\[-7pt]
                    \\[2pt]
\mathbf{2}\otimes\mathbf{3^{\prime}}~=\mathbf{3}\oplus\mathbf{3^{\prime}}& \qquad\quad
\text{with}\qquad\quad\left\{\begin{array}{l}
                    \mathbf{3}~\sim\begin{pmatrix}
                        \sqrt{3}\alpha_1\beta_3+\alpha_2\beta_1\\
                        -2\alpha_2\beta_2 \\
                        \sqrt{3}\alpha_1\beta_1+\alpha_2\beta_3 \\
                    \end{pmatrix}\\[-7pt]
                    \\[4pt]
                    \mathbf{3^{\prime}}\sim\begin{pmatrix}
                        \sqrt{3}\alpha_{2}\beta_{3}-\alpha_1\beta_1\\
                        2\alpha_1\beta_2 \\
                        \sqrt{3}\alpha_2\beta_1-\alpha_1\beta_3 \\
                    \end{pmatrix}
                    \end{array}
            \right.
\end{array}
\end{equation}
Finally the CG coefficients involving the three-dimensional representations $\mathbf{3}$ and $\mathbf{3^{\prime}}$ are as follows:
\begin{equation}
\begin{array}{ll}
\mathbf{3}\otimes\mathbf{3}~=\mathbf{3^{\prime}}\otimes\mathbf{3^{\prime}}=\mathbf{1}\oplus\mathbf{2}\oplus\mathbf{3}\oplus\mathbf{3^{\prime}}& \qquad\quad
\text{with}\qquad\quad\left\{
\begin{array}{l}
\mathbf{1}~\sim\alpha_1\beta_3+\alpha_2\beta_2+\alpha_3\beta_1\\[-7pt]
                    \\[4pt]
\mathbf{2}~\sim\begin{pmatrix}
       2\alpha_2\beta_2-\alpha_1\beta_3-\alpha_3\beta_1\\
      \sqrt{3}(\alpha_1\beta_1+\alpha_3\beta_3)\\
     \end{pmatrix}\\[-7pt]
                  \\[4pt]
\mathbf{3}~\sim\begin{pmatrix}
         \alpha_1\beta_2-\alpha_2\beta_1\\
         \alpha_3\beta_1-\alpha_1\beta_3\\
         \alpha_2\beta_3-\alpha_3\beta_2\\
        \end{pmatrix}\\[-7pt]
                    \\[4pt]
\mathbf{3^{\prime}}\sim\begin{pmatrix}
         -\alpha_2\beta_3-\alpha_3\beta_2\\
         \alpha_1\beta_1-\alpha_3\beta_3\\
         \alpha_1\beta_2+\alpha_2\beta_1\\
    \end{pmatrix}
\end{array}\right.\\[-7pt]
                    \\[2pt]
\mathbf{3}\otimes\mathbf{3^{\prime}}~=\mathbf{1^{\prime}}\oplus\mathbf{2}\oplus\mathbf{3}\oplus\mathbf{3^{\prime}}& \qquad\quad
\text{with}\qquad\quad\left\{
\begin{array}{l}
\mathbf{1^{\prime}}\sim\alpha_1\beta_3+\alpha_2\beta_2+\alpha_3\beta_1\\[-7pt]
                    \\[4pt]
\mathbf{2}~\sim\begin{pmatrix}
     \sqrt{3}(\alpha_1\beta_1+\alpha_3\beta_3)\\
     \alpha_1\beta_3+\alpha_3\beta_1-2\alpha_2\beta_2\\
     \end{pmatrix}\\[-7pt]
                    \\[4pt]
\mathbf{3}~\sim\begin{pmatrix}
         -\alpha_2\beta_3-\alpha_3\beta_2\\
         \alpha_1\beta_1-\alpha_3\beta_3\\
         \alpha_1\beta_2+\alpha_2\beta_1\\
    \end{pmatrix}\\[-7pt]
                    \\[4pt]
\mathbf{3^{\prime}}\sim\begin{pmatrix}
         \alpha_1\beta_2-\alpha_2\beta_1\\
         \alpha_3\beta_1-\alpha_1\beta_3\\
         \alpha_2\beta_3-\alpha_3\beta_2\\
        \end{pmatrix}
\end{array}\right.
\end{array}
\end{equation}
Note that all the CG coefficients are real. The group structure of $S_4$ has been studied comprehensively in Ref.~\cite{S4_group}. It has nine $Z_2$ subgroups, four $Z_3$ subgroups, three $Z_4$ subgroups, four $K_4\cong Z_2\times Z_2$ subgroups, four $S_3$ subgroups, three $D_4$ subgroups~\footnote{Here $D_4$ is the symmetry group of the square, and its order is eight. Its mathematical definition is $D_4=\langle r,s|r^4=s^2=(rs)^2=1\rangle$. } and the alternating group $A_4$ as a subgroup. In the present work, we focus on the Abelian subgroups as the remnant symmetry, which can be expressed in terms of the generators $S$, $T$ and $U$ as follows:

\begin{itemize}[leftmargin=1.5em]

\item{$Z_2$ subgroups}
\begin{equation}
\label{eq:Z2-subgroups}
\begin{array}{lll}
Z_2^{ST^{2}SU}=\{1,ST^{2}SU\},& ~~~ Z_2^{TU}=\{1,TU\},& ~~~ Z_2^{STSU}=\{1,STSU\},\\
Z_2^{T^2U}=\{1,T^2U\},&~~~ Z_2^{U}=\{1,U\}, &~~~ Z_2^{SU}=\{1,SU\}, \\
Z_2^{S}=\{1,S\},&~~~ Z_2^{T^2ST}=\{1,T^2ST\}, &~~~ Z_2^{TST^{2}}=\{1,TST^{2}\}\,.
\end{array}
\end{equation}
The former six $Z_{2}$ subgroups are related to each other by group conjugation, and the latter three subgroups are conjugate to each other as well.
\item{$Z_3$ subgroups}
\begin{equation}
\label{eq:Z3-subgroups}
\begin{array}{ll}
Z_3^{ST}=\{1,ST,T^{2}S\},&\qquad Z_3^{T}=\{1,T,T^{2}\},\\
Z_3^{STS}=\{1,STS,ST^2S\},&\qquad Z_3^{TS}=\{1,TS,ST^{2}\}\,.
\end{array}
\end{equation}
All the above $Z_3$ subgroups are conjugate to each other.
\item{$Z_4$ subgroups}
\begin{eqnarray}\label{eq:Z4-subgroups}
\nonumber&&Z_4^{TST^{2}U}=\{1,TST^{2}U,S,T^{2}STU\},\quad Z_4^{ST^2U}=\{1,ST^2U,TST^{2},T^2SU\},\\
&&Z_4^{TSU}=\{1,TSU,T^2ST,STU\}\,,
\end{eqnarray}
which are related with each under group conjugation.
\item{$K_4$ subgroups}
\begin{equation}\label{eq:K4-subgroups}
\begin{array}{l}
K^{(S, TST^{2})}_4\equiv Z_2^{S}\times Z_2^{TST^{2}}=\{1,S,TST^{2},T^{2}ST\},\\
K^{(S,U)}_4\equiv Z_2^{S}\times Z_2^{U}=\{1,S,U,SU\}, \\
K^{(TST^{2}, T^{2}U)}_4\equiv Z_2^{TST^{2}}\times Z_2^{T^{2}U}\equiv \{1,TST^{2},T^2U,ST^{2}SU\}, \\
K^{(T^{2}ST, TU)}_4\equiv Z_2^{T^{2}ST}\times Z_2^{TU}=\{1,T^{2}ST,TU,STSU\}\,,
\end{array}
\end{equation}
where $K^{(S, TST^{2})}_4$ is a normal subgroup of $S_4$, and the other three $K_4$ subgroups are conjugate to each other.

\end{itemize}

\section{\label{sec:appB}The general analysis of $S_4$ breaking to $Z_2$ in neutrino sector and to $K_4$ in charged lepton sector with remnant CP}
\cleqn

In this appendix, we shall analyze the last scenario in which $S_4\rtimes H_{CP}$ is broken down to $Z_2\times H^{\nu}_{CP}$ in the neutrino sector and $K_{4}\rtimes H^{l}_{CP}$ in the charged lepton sector. Since $S_4$ has nine $Z_2$ subgroups given by Eq.~\eqref{eq:Z2-subgroups} and four $K_4$ subgroups in Eq.~\eqref{eq:K4-subgroups}, there are $9\times4=36$ possible preserved remnant family symmetry in the neutrino and the charged lepton sectors, the remnant CP symmetry is fixed by the consistency condition.  Reminding that the different residual subgroups $\left(Z_2, K_4\right)$ related by group conjugation lead to the same prediction for lepton flavor mixing, it is sufficient to only consider five representative cases: $\left(G_{\nu}, G_{l}\right)=\big(Z^{ST^2SU}_2, K^{\left(S,U\right)}_4\big)$, $\big(Z^{ST^2SU}_2, K^{(S, TST^2)}_4\big)$, $\big(Z^{T^2ST}_2, K^{(S,U)}_4\big)$, $\big(Z^{S}_2, K^{(S, TST^2)}_4\big)$ and $\big(Z^{U}_2, K^{(S, U)}_4\big)$. Here the unbroken $Z_2$ flavor symmetry in the neutrino sector fixes only one column of the mixing matrix. For the last two cases $\left(G_{\nu}, G_{l}\right)=\big(Z^{S}_2, K^{(S, TST^2)}_4\big)$ or $\big(Z^{U}_2, K^{(S, U)}_4\big)$, one column of the PMNS matrix turns out to be $\left(0, 0, 1\right)^{T}$ which leads to a vanishing $\theta_{13}$. Hence this case can not be accommodated by the present data. For the centered two cases $\left(G_{\nu}, G_{l}\right)=\big(Z^{ST^2SU}_2, K^{(S, TST^2)}_4\big)$, $\big(Z^{T^2ST}_2, K^{(S,U)}_4\big)$, the PMNS matrix would have one column of form $(0,1/\sqrt{2}, 1/\sqrt{2})^{T}$ which also give rise to a zero $\theta_{13}$. For the remaining case $\left(G_{\nu}, G_{l}\right)=\big(Z^{ST^2SU}_2, K^{\left(S,U\right)}_4\big)$, one column of $U_{PMNS}$ would be $\left(1/\sqrt{2}, 1/2, 1/2\right)^{T}$ up to permutation.  We shall demonstrate that the corresponding lepton flavor mixing matrix is of the same form as that predicted by $\left(G_{\nu}, G_{l}\right)=\big(Z^{ST^2SU}_2, Z^{TST^{2}U}_4\big)$ in section~\ref{sec:model_column} after the residual CP symmetry is further included.

Following the same steps listed in section~\ref{sec:general_analysis_one_row}, we can straightforwardly find that the most general mass matrix $m^{\dagger}_{l}m_{l}$, which is invariant under residual family symmetry $G_{l}=K^{\left(S,U\right)}_4$, is of the following form:
\begin{equation}
\label{eq:ch_mass_K4Gl}m^{\dagger}_{l}m_{l}=\left(\begin{array}{ccc}
m_{11}  &  0  &  im_{13}  \\
  0  &  m_{22}  &  0  \\
-im_{13}  &  0 &  m_{11}
\end{array}\right)\,,
\end{equation}
where $m_{11}$, $m_{13}$ and $m_{22}$ are real parameters. This charged lepton mass matrix is diagonalized by the transformation
\begin{eqnarray}
\label{eq:eigen_Glk4}V^{\dagger}_{l}m^{\dagger}_{l}m_{l}V_{l}=\mathrm{diag}(m_{22}, m_{11}-m_{13}, m_{11}+m_{13})\,,
\end{eqnarray}
with
\begin{equation}
\label{eq:VL_appB}V_{l}=\frac{1}{\sqrt{2}}\left(
\begin{array}{ccc}
 0 ~&~ e^{-\frac{i \pi }{4}} ~&~ e^{\frac{i \pi }{4}} \\
 \sqrt{2} ~&~ 0 ~&~ 0 \\
 0 ~&~ e^{\frac{i \pi }{4}} ~&~ e^{-\frac{i \pi }{4}} \\
\end{array}
\right)\,.
\end{equation}
Because the order of the charged lepton masses is undetermined in our framework, $V_{l}$ can undergo rephasing and permutations of its column vectors. The mass matrix $m^{\dagger}_{l}m_{l}$ in Eq.~\eqref{eq:ch_mass_K4Gl} is further constrained by the residual CP symmetry $H^{l}_{CP}$ which should be consistent with the residual family symmetry $K^{\left(S,U\right)}_4$,
\begin{eqnarray}
\label{eq:consistence_appendix}X_{l\mathbf{r}}\rho^{*}_{\mathbf{r}}(S)X^{-1}_{l\mathbf{r}}=\rho_{\mathbf{r}}(S^{\prime}),\quad X_{l\mathbf{r}}\rho^{*}_{\mathbf{r}}(U)X^{-1}_{l\mathbf{r}}=\rho_{\mathbf{r}}(U^{\prime}),\quad S^{\prime}, U^{\prime}\in K^{\left(S,U\right)}_4\,.
\end{eqnarray}
Then we find the residual CP symmetry $H^l_{CP}$ can take the following values,
\begin{equation}
H^{l}_{CP}=\{\rho_{\mathbf{r}}(T^{2}ST), \rho_{\mathbf{r}}(TST^{2}), \rho_{\mathbf{r}}(TST^{2}U), \rho_{\mathbf{r}}(T^{2}STU)\}\,.
\end{equation}
We can straightforwardly check that the remnant CP invariance condition is automatically fulfilled by the charged lepton mass matrix $m^{\dagger}_{l}m_{l}$ in Eq.~\eqref{eq:ch_mass_K4Gl}. Hence no new constraints are generated. Note that $X_{l\mathbf{r}}=\rho_{\mathbf{r}}(1)$, $\rho_{\mathbf{r}}(S)$, $\rho_{\mathbf{r}}(U)$, $\rho_{\mathbf{r}}(SU)$ also satisfy the consistence equation of Eq.~\eqref{eq:consistence_appendix}. However, the symmetric requirement of section 2 is not fulfilled since both $\rho_{\mathbf{r}}(U)$ and $\rho_{\mathbf{r}}(SU)$ are not symmetric matrices.

Now we come to the neutrino sector with the remnant symmetry $Z^{ST^{2}SU}_2\times H^{\nu}_{CP}$. From section~\ref{sec:model_column}, we know that the most general neutrino mass matrix invariant under $G_{\nu}=Z^{ST^{2}SU}_2$  is of the form
\begin{equation}
 m_{\nu} = \alpha\left(
\begin{array}{ccc}
 0 & 0 & 1 \\
 0 & 1 & 0 \\
 1 & 0 & 0
\end{array}
\right)+\beta \left(
\begin{array}{ccc}
 -3 & 0 & 1 \\
 0 & -2 & 0 \\
 1 & 0 & -3
\end{array}
\right)+\gamma\left(
\begin{array}{ccc}
 0 & 1 & 0 \\
 1 & 0 & 1 \\
 0 & 1 & 0
\end{array}
\right)+\epsilon\left(
\begin{array}{ccc}
 \sqrt{2} & ~-1 & 0 \\
 -1 & ~0 & 1 \\
 0 & ~1 & -\sqrt{2}
\end{array}
\right)\,.
\end{equation}
The residual CP symmetry is $H^{\nu}_{CP}=\{\rho_{\mathbf{r}}(1)$, $\rho_{\mathbf{r}}(ST^{2}SU)$, $\rho_{\mathbf{r}}(T^{2}U)$, $\rho_{\mathbf{r}}(TST^{2})\}$. In the following, we shall present the predictions for the lepton mixing  matrix for different residual CP transformations.
\begin{itemize}[leftmargin=1.5em]
\item{$X_{\nu\mathbf{r}}=\rho_{\mathbf{r}}(1),\rho_{\mathbf{r}}(ST^{2}SU)$}

The four parameters $\alpha$, $\beta$, $\gamma$ and $\epsilon$ are real, neutrino mass matrix $m_{\nu}$ is diagonalized by the unitary matrix $U_{\nu}$ given in Eq.~\eqref{eq:UPMNS_one}. Including the contribution $V_{l}$ in Eq.~\eqref{eq:VL_appB} from the charged lepton sector, the lepton mixing matrix $U_{PMNS}$ is of the form
\begin{equation}
\hskip-0.4in U_{PMNS}=V^{\dagger}_{l}{U_{\nu}}=
\frac{1}{2}\left(
\begin{array}{ccc}
 -\sqrt{2} \sin\theta &~ \sqrt{2} &~ -\sqrt{2} \cos\theta
   \\
 \sin\theta+i \sqrt{2} \cos\theta &~ 1 &~ \cos\theta-i
   \sqrt{2} \sin\theta \\
 \sin \theta-i \sqrt{2} \cos\theta &~ 1 &~ \cos\theta+i
   \sqrt{2} \sin\theta \\
\end{array}
\right)\,.
\end{equation}
It turns out to be identical with the lepton mixing matrix originating from the residual symmetries $Z^{ST^2SU}_2\times H^{\nu}_{CP}$ in neutrino sector and $Z^{TST^2U}_4\rtimes H^{l}_{CP}$ in charged lepton sector with $H^{\nu}_{CP}=$$\{\rho_{\mathbf{r}}(T^{2}U)$,$\rho_{\mathbf{r}}(TST^{2})\}$. This mixing pattern can not fit the measured mixing angles well.

\item{$X_{\nu\mathbf{r}}=\rho_{\mathbf{r}}(T^{2}U),\rho_{\mathbf{r}}(TST^{2})$}

Residual CP invariance condition implies that $\alpha$, $\beta$ and $\gamma$ are real while $\epsilon$ is pure imaginary.
In the same fashion, the PMNS matrix is determined to be of the form
\begin{equation}
U_{PMNS}=\frac{1}{2}\left(
\begin{array}{ccc}
 -\sqrt{2} \sin\theta &~ \sqrt{2} &~ -\sqrt{2} \cos\theta
   \\
 \sin\theta-\sqrt{2} \cos\theta  &~ 1 &~ \cos\theta
 +\sqrt{2} \sin\theta \\
 \sin\theta+\sqrt{2} \cos\theta &~ 1 &~ \cos\theta
   -\sqrt{2} \sin\theta  \\
\end{array}
\right)\,.
\end{equation}
Obviously this PMNS matrix coincides with Eq.~\eqref{eq:UPMNS_one} up to exchange of the rows.

\end{itemize}
In short, the symmetry breaking pattern of $S_4\rtimes H_{CP}$ into $Z^{ST^2SU}_2\times H^{\nu}_{CP}$ in the neutrino sector and $K^{(S,U)}_4\rtimes H^{l}_{CP}$ in the charged lepton sector leads to the same predictions for the lepton flavor mixing as the remnant symmetries $Z^{ST^2SU}_2\times H^{\nu}_{CP}$ of neutrino sector and $Z^{TST^2U}_4\rtimes H^{l}_{CP}$ of charged lepton sector. The same conclusion is obtained in Ref.~\cite{Feruglio:2012cw}.

\end{appendix}

\end{document}